\documentclass[12pt, a4paper]{article}
\usepackage{titlesec,titletoc}
\usepackage[export]{adjustbox}

\usepackage[mode=buildnew]{standalone}
 \usepackage[margin=3.2cm]{geometry}

\usepackage[
			anchorcolor=black,
			citecolor=black,
			colorlinks=true,
			filecolor=black,
			linkcolor=black,
			menucolor=black,
			runcolor=black,
			urlcolor=black,
			hyperfootnotes=false
]{hyperref}
\usepackage{amsmath,amsthm}
\usepackage[shrink=25]{microtype}
\usepackage[noabbrev]{cleveref}
\crefname{appendix}{Appendix}{Appendices}

\theoremstyle{plain}

\newtheorem{lemma}{Lemma}
\newtheorem{proposition}{Proposition}
\newtheorem{corollary}{Corollary}

\theoremstyle{definition}
\newtheorem{definition}{Definition}

\newtheorem{observation}{Observation}

\theoremstyle{remark}

\crefname{observation}{Observation}{Observation}
\crefname{theorem}{Theorem}{Theorems}
\crefname{lemma}{Lemma}{Lemmas}
\crefname{proposition}{Proposition}{Propositions}
\crefname{corollary}{Corollary}{Corollaries}
\crefname{definition}{Definition}{Definitions}
\crefname{assumption}{Assumption}{Assumptions}
\crefname{section}{Section}{Sections}
\crefname{figure}{Figure}{Figures}
\crefname{axiom}{Axiom}{Axioms}
\crefname{modification}{Modification}{Modifications}
\crefname{example}{Example}{Examples}

\usepackage[font={small}]{caption}[2008/08/24]

\usepackage{enumitem}

\usepackage{setspace}
\usepackage{dsfont}

\usepackage{amsmath,amsfonts,amssymb, mathrsfs}
\usepackage[T1]{fontenc}
\usepackage[]{lmodern}
\usepackage[english]{babel}
\usepackage{csquotes}
\usepackage[subrefformat=parens,labelformat=parens]{subfig}
\usepackage{multirow,array}

\usepackage{pgfplots}
\pgfplotsset{compat=1.7}
\usepackage{sidecap}
\usetikzlibrary{patterns,decorations.pathreplacing,positioning,external,shapes.geometric,arrows,calc,fillbetween}

\usepackage{pstricks}
\usepackage{pst-3d}
\usepackage{sgamevar, egameps}
\usepackage{graphicx}
\usepackage{epstopdf}

\usepackage{verbatim}
\usepackage{booktabs}
\usepackage[multiple]{footmisc}

\usepackage{placeins}

\definecolor{Vermilion}{RGB}{227, 85, 10}
\definecolor{BluishGreen}{RGB}{0,158,115}
\definecolor{ReddishPurple}{RGB}{204,121,167}
\definecolor{Teal}{RGB}{0,128,128}
\definecolor{Amber}{HTML}{FFBF00}
\usepackage[backend=biber,
			sorting=ynt,
			sortcites=true,
			bibencoding=inputenc,
			bibstyle=authoryear -comp,
			citestyle=authoryear -comp,
			doi=false,
			giveninits=true,
			isbn=false,
			uniquename=false,
			uniquelist=minyear,
			maxcitenames=2,
			maxnames=2,
			maxbibnames=99,
			natbib=true,
			uniquename=false,
			url=false,
			date=year
]{biblatex}
\bibliography{lobby}
	\AtEveryBibitem{\clearlist{language}}
	\AtEveryBibitem{\clearfield{number}}
	\AtEveryBibitem{\clearlist{month}}
	\AtEveryBibitem{\clearfield{Month}}
	\AtEveryBibitem{\clearfield{day}}
	\AtEveryBibitem{\clearfield{eprint}}
	\AtEveryBibitem{\clearfield{Volume}}
	\AtEveryBibitem{\clearfield{volume}}

\renewbibmacro{in:}{}
	\DeclareSourcemap{
  \maps[datatype=bibtex,overwrite=true]{
    \map{
      \step[fieldsource=pages,
            match=\regexp{pp?\.?(.+)},
           replace=\regexp{$1}]
    }
  }
}
\DeclareSourcemap{
  \maps[datatype=bibtex]{
    \map{
      \step[fieldsource=journal, match={RAND Journal of Economics},
      replace={RAND J of Economics}]
    }
  }
}

\DeclareSourcemap{
  \maps[datatype=bibtex]{
    \map{
      \step[fieldsource=journal, match={Proceedings of the National Academiy of Sciences},
      replace={PNAS}]
    }
  }
}

\DeclareSourcemap{
  \maps[datatype=bibtex]{
    \map{
      \step[fieldsource=journal, match={\regexp{Available at SSRN?\.?(.+)}},
      replace={mimeo}]
    }
  }
}

\DeclareSourcemap{
  \maps[datatype=bibtex]{
    \map{
      \step[fieldsource=journal, match={Working Paper},
      replace={mimeo}]
    }
  }
}

\DeclareSourcemap{
  \maps[datatype=bibtex]{
    \map{
      \step[fieldsource=journal, match={\regexp{arXiv?\.?(.+)}},
      replace={mimeo}]
    }
  }
}

\DeclareSourcemap{
  \maps[datatype=bibtex]{
    \map{
      \step[fieldsource=journal, match={NBER working paper},
      replace={mimeo}]
    }
  }
}

\DeclareSourcemap{
  \maps[datatype=bibtex]{
    \map{
      \step[fieldsource=journal, match={working paper},
      replace={mimeo}]
    }
  }
}

\DeclareSourcemap{
  \maps[datatype=bibtex]{
    \map{
      \step[fieldsource=journal, match={Econometrica: Journal of the Econometric Society},
      replace={Econometrica}]
    }
  }
}

\DeclareSourcemap{
  \maps[datatype=bibtex]{
    \map{
      \step[fieldsource=journal, match={The Review of Financial Studies},
      replace={RFS}]
    }
  }
}

\DeclareSourcemap{
  \maps[datatype=bibtex]{
    \map{
      \step[fieldsource=journal, match={The Review of Economic Studies},
      replace={ReStud}]
    }
  }
}

\DeclareSourcemap{
  \maps[datatype=bibtex]{
    \map{
      \step[fieldsource=journal, match={Review of Economic Studies},
      replace={ReStud}]
    }
  }
}

\DeclareSourcemap{
  \maps[datatype=bibtex]{
    \map{
      \step[fieldsource=journal, match={The Journal of Economic Theory},
      replace={JET}]
    }
  }
}

\DeclareSourcemap{
  \maps[datatype=bibtex,overwrite=true]{
    \map{
      \step[fieldsource=journal, match={American Economic Journal: Microeconomics},
      replace={AEJ:Micro}]
    }
  }
}
\DeclareSourcemap{
  \maps[datatype=bibtex,overwrite=true]{
    \map{
      \step[fieldsource=journal, match={American Economic Journal:Microeconomics},
      replace={AEJ:Micro}]
    }
  }
}

\DeclareSourcemap{
  \maps[datatype=bibtex,overwrite=true]{
    \map{
      \step[fieldsource=journal, match={American Economic Journal: Applied Economics},
      replace={AEJ:Applied}]
    }
  }
}

\DeclareSourcemap{
  \maps[datatype=bibtex,overwrite=true]{
    \map{
      \step[fieldsource=journal, match={Journal of Economic Theory},
      replace={JET}]
    }
  }
}

\DeclareSourcemap{
  \maps[datatype=bibtex,overwrite=true]{
    \map{
      \step[fieldsource=journal, match={American Journal of Political Science},
      replace={AJPS}]
    }
  }
}

\DeclareSourcemap{
  \maps[datatype=bibtex,overwrite=true]{
    \map{
      \step[fieldsource=journal, match={American Political Science Review},
      replace={APSR}]
    }
  }
}

\DeclareSourcemap{
  \maps[datatype=bibtex,overwrite=true]{
    \map{
      \step[fieldsource=journal, match={American Economic Review},
      replace={AER}]
    }
  }
}

\DeclareSourcemap{
  \maps[datatype=bibtex,overwrite=true]{
    \map{
      \step[fieldsource=journal, match={Journal of Economic Perspectives},
      replace={JEP}]
    }
  }
}

\DeclareSourcemap{
  \maps[datatype=bibtex,overwrite=true]{
    \map{
      \step[fieldsource=journal, match={Journal of Political Economy},
      replace={JPE}]
    }
  }
}

\DeclareSourcemap{
  \maps[datatype=bibtex,overwrite=true]{
    \map{
      \step[fieldsource=journal, match={The Quarterly Journal of Economics},
      replace={QJE}]
    }
  }
}

\DeclareSourcemap{
  \maps[datatype=bibtex,overwrite=true]{
    \map{
      \step[fieldsource=journal, match={Quarterly Journal of Political Science},
      replace={QJPS}]
    }
  }
}
\usetikzlibrary{automata, positioning}
\usepackage{soul}
\usepackage{datetime,titlesec}
\hypersetup{colorlinks=true,citecolor=blue}
\setlist{noitemsep,topsep=0pt}
\assignrefcontextentries[]{*}

\usepackage{booktabs,xcolor,colortbl,makecell,array,tabularx}
 \definecolor{cgreen}{HTML}{3B6D11}
 \definecolor{cgreenlight}{HTML}{EAF3DE}
 \definecolor{cred}{HTML}{8F1D1D}
 \definecolor{credlight}{HTML}{FCEBEB}
 \definecolor{cblue}{HTML}{184E9E}
 \definecolor{cbluelight}{HTML}{E6F1FB}
 \definecolor{cgray}{HTML}{5F5E5A}
 \definecolor{cgraylight}{HTML}{F1EFE8}
 \definecolor{camber}{HTML}{854F0B}
 \definecolor{camberlight}{HTML}{FAEEDA}
 \newcommand{\yes}[1]{\cellcolor{cgreenlight}\textcolor{cgreen}{\textbf{#1\;\checkmark}}}
  \newcommand{\no}[1]{\cellcolor{cgraylight}\textbf{#1\;\texttimes}}
 
 \newcommand{\sideleft}{\cellcolor{cbluelight}\textcolor{cblue}{\textbf{Left}}}
 \newcommand{\sideright}{\cellcolor{cbluelight}\textcolor{cblue}{\textbf{Right}}}
 \newcommand{\sideneither}{\cellcolor{cgraylight}\textcolor{cgray}{\textbf{Neither}}}
 \newcommand{\sidelean}{\cellcolor{credlight}\textcolor{cred}{\textbf{Lean left}}}
 \newcommand{\sideend}{\cellcolor{camberlight}\textcolor{camber}{\textbf{Neither viable}}}
 \newcommand{\elecrow}[1]{\midrule\multicolumn{5}{@{}l@{}}{\cellcolor{cgraylight}\small\textit{#1}}\\[2pt]\midrule}
 \renewcommand{\sideleft}{\cellcolor{credlight}\textcolor{cred}{\textbf{Left}}}
 \renewcommand{\sidelean}{\cellcolor{credlight}\textcolor{cred}{\textbf{Lean left}}}

 \newcommand{\sidefarright}{\cellcolor{camberlight}\textcolor{camber}{\textbf{Far right}}}
 \newcommand{\chosenL}[1]{\cellcolor{credlight}\textcolor{cred}{\textbf{#1\;$\leftarrow$}}}
 \newcommand{\chosenR}[1]{\cellcolor{cbluelight}\textcolor{cblue}{\textbf{#1\;$\leftarrow$}}}
 \newcolumntype{Y}{>{\raggedright\arraybackslash}X}

\makeatletter
\g@addto@macro\normalsize{
  \setlength\abovedisplayskip{3pt}
  \setlength\belowdisplayskip{8pt}
  \setlength\abovedisplayshortskip{3pt}
  \setlength\belowdisplayshortskip{8pt}
}
\makeatother

\titlespacing*{\part}
{0pt}{0ex}{0ex}
\titlespacing*{\subsection}
{0pt}{0.25ex plus .3ex minus .1ex}{0.25ex plus .15ex minus .15ex}
\titlespacing*{\section}
{0pt}{0.25ex plus .3ex minus .1ex}{0.25ex plus .15ex minus .15ex}
\titlespacing*{\subsubsection}
{0pt}{0.2ex plus .25ex minus .1ex}{0.2ex plus .15ex minus .1ex}
\titlespacing*{\paragraph}
{0pt}{0.2ex plus .2ex minus .1ex}{.15em plus .1ex}
\titlespacing*{\subparagraph}
{\parindent}{0.05ex plus .2ex minus .1ex}{.35em plus .1ex}

\theoremstyle{definition}

\theoremstyle{plain}

\crefname{result}{Result}{Results}

\title{ \textbf{Embracing the Enemy}\thanks{We are grateful to Antoine Loeper for inspiring discussions early on. We thank  Antonio Cabrales, Pawe\l~Doligalski, Dana Foarta, Gabriele Gratton, Elliot Lipnowski, Chara Papioti, Spencer Pantoja, Harry Pei, Elia Sartori, Christoph Schottm\"uller, Jonathan B. Slapin, Takuo Sugaya, Heidi Thysen, Jan Zapal, and audiences at various places. Johannes Schneider acknowledges financial support from PID-2024-149885OB-I00. We thank the institutions we spent time at during this research for their hospitality and inspiration: Schneider was at uc3m, UBern, and NHH (Bergen); Delgado-Vega was at the Chair of Political Economy and eDemocracy (ETH Zurich) and a Doctoral Scholar (scholarship PRE2019-089897) at U. Carlos III de Madrid.}}
\author{  \'Alvaro Delgado-Vega\thanks{University of Chicago; \url{adelgadovega@uchicago.edu }} \and Johannes Schneider\thanks{Johannes Kepler University; \url{johannes.schneider@jku.at}}}

\makeatletter
\def\pgfplots@stacked@diff{}
\makeatother

\date{\monthname~\the\year}
\tikzset{font=\footnotesize}
\pgfplotsset{
    ,tick label style = {font=\footnotesize}
    ,every axis label = {font=\footnotesize}
    ,legend style = {font=\footnotesize}
    ,label style = {font=\footnotesize}
}
\usepgfplotslibrary{fillbetween}

\begin{document}
\maketitle

\begin{abstract}

A principal can partially influence the allocation of power between two competing parties. The principal is closer to one party, the ``friend'', than to the other, the ``enemy''. The principal's optimal contract initially seeks to exclude the enemy. However, once the enemy gains power, the principal embraces him in exchange for policy moderation. Moderation also disciplines the friend, inducing him to move closer to the principal's preferred policy. Principals close to the friend fully embrace the enemy; more centrist principals divide their support. Commitment benefits the principal only if she is close to the friend and parties value power little.

\end{abstract}
\onehalfspacing
\textsc{JEL Codes:} D23, D74, D86

\textsc{Keywords:} Relational Contracts, Organizational Economics, Political Economy, Horizontal Moral Hazard, Coalition Formation

\begin{quote}
  \emph{Politics is the art of the possible, the attainable---the art of the next best.}\\ \phantom{A large number of characters teaching latex what to do here} {Otto von Bismarck}
\end{quote}
\section{Introduction}

All organizations must grapple with disagreement, especially when a discordant actor is in a position to shape collective decisions. One common strategy for limiting such influence is exclusion. Economists formalize this logic as the \emph{Ally Principle}: whenever possible, the principal delegates authority to aligned actors, thereby preventing the discordant actor from exercising power. Its political counterpart, the \emph{cordon sanitaire}, follows the same logic: mainstream parties collaborate to isolate radical parties and deny them meaningful influence.

Yet permanent exclusion proves challenging in practice. Consider, for instance, a parliament witnessing the emergence of an anti-establishment party. This newcomer competes directly with established parties, disrupting traditional governance patterns and reshaping the distribution of political power. Centrist parties---strategically positioned between ideological extremes---may become power brokers deciding, for example, who obtains cabinet posts. Their choices, however, are constrained: parliamentary arithmetic or shifts in public sentiment limit their options. A similar problem arises for centrist interest groups allocating electoral support---think of campaign contributors, major newspapers, or even groups of (perhaps organized) voters.\footnote{Voters and political parties can be thought of as having relational contracts that, even if sketchy, may be fully understood and relevant for politicians' decision-making. Although some features of our equilibrium are complex and may require too much from a voter, we cannot rule it out; we leave it to the reader to interpret in each application whether traces of our results can be observed in voters' behavior.} By endorsing aligned parties, they may bring them to power more often, but they cannot avoid the risk that a different party ultimately wins office. In this paper, we study optimal power brokerage in such scenarios. Our power broker, though influential, is not all-powerful.  What strategy should she pursue? Should she impose a cordon sanitaire around a discordant agent? Or should she, on the contrary, back him in exchange for his moderation?

We consider an infinite-horizon setting in which two political parties (both ``he'') repeatedly compete for the right to set policy. Both are power-hungry and agenda-driven, vying over \textit{who} leads and \textit{which} policy is made. A third player, the principal (``she''), is the power broker. She has an agenda, too, and only cares about that. The principal cannot choose the policy, but influences who gets the right to choose. She can increase a party's chances of obtaining that right, yet only to a limited extent. The principal's ideal policy is between the ideals of the parties but closer to one (her \textit{friend}) than the other (her \textit{enemy}).

We study the dynamics of the principal's optimal contract and derive three predictions: (i) at first, setting up a cordon sanitaire around the enemy is optimal; (ii) eventually, that cordon sanitaire will fall; and (iii) once it falls, it is not optimal to restore it but to embrace the enemy.

Our analysis unveils an inherently dynamic logic. If the game were played only once, whichever party received the right to act (the \textit{leading party}) would simply choose his preferred policy. Anticipating this polarization, the principal would \emph{fully endorse} the friend, that is, she would do everything in her power to maximize the friend's likelihood of taking the lead. However, if the interaction occurs repeatedly, fully endorsing the friend severely undermines the principal's influence. To see why, suppose the enemy becomes the leading party today. The principal can offer him the following contract: ``If, today, you choose a policy close to my bliss point, I will endorse you tomorrow and ever after.''\footnote{This contract, of course, hinges on the principal's ability to credibly promise that endorsement. Whether and when she can do that is a core aspect we will discuss in what follows.}

We show that this exchange---trading future endorsement for present policy moderation---is effective in disciplining the enemy but especially relevant if the friend and the principal are closely aligned. Suppose, for example, that the friend and the principal have identical preferences over policy and assume the principal has commitment power. In that case, the principal endorses the friend, upholding the cordon sanitaire, as long as the enemy is kept from leading. But once the enemy leads for the first time, the principal flips and persistently embraces the enemy in exchange for his moderation. Meanwhile, the friend continues to implement the principal's preferred policy whenever he leads.

If the principal and the friend are not fully aligned, the principal faces another challenge: to incentivize the friend to moderate his policy, too. Naturally, the principal can promise to endorse the friend in exchange. But then, her embrace of the enemy decreases, and the enemy moderates less. For that reason, the principal prefers to incentivize the friend indirectly: she leverages the fact that the friend benefits from the enemy's moderation, too. By moderating the enemy whenever he is in charge, the principal improves the friend's expected payoffs. Threatening to remove that moderation is enough to get the friend to choose the principal's preferred policy. Yet, if they disagree more, direct endorsement of the friend becomes necessary.

Principals with more centrist bliss points have higher ex-ante payoffs if they can incentivize the friend---either directly or indirectly---to choose their preferred policy. Intuitively, if the principal can drag the friend's policy choice to her bliss point, she is mechanically better off the closer she is to the enemy. However, if the principal is too distant from the friend, she cannot count on a fully complicit friend. Then, being more centrist makes her worse off: neither the friend's nor the enemy's policy can be pulled more to the center. Yet, the principal is further from the friend, whose policy choices are more relevant because of the initial cordon sanitaire.

A natural concern is the principal's ability to uphold her promises. To address this, we devote the second part of the paper to the case in which the principal lacks \emph{commitment power}, so all promises need to be self-enforcing. Commitment power has two advantages for the principal. First, in the optimal contract, the principal promises to endorse the enemy in future periods in exchange for his moderation today. With commitment, this promise is trivially \emph{credible}. Second, a committed principal can \emph{credibly threaten} to punish a deviating friend. Both threat and promise are curtailed if the principal lacks commitment because she genuinely aligns more with the friend and therefore, irrespective of her endorsement, always hopes for the friend to take the lead next.

We show that in some parameter regions, whether the principal can indeed commit bears no practical relevance. We can replicate the commitment outcome if (i) agents attach sufficient intrinsic value to leading or (ii) the principal is sufficiently balanced. Does cooperation unravel completely otherwise? We show that this depends on the level of alignment between principal and friend. If the principal and the friend fully align, the results are drastic: either the principal can implement the commitment outcome, or she cannot make use of dynamic incentives at all. Outside this extreme case, the unraveling is gradual. Even if the commitment contract cannot be implemented, the principal can often promise some endorsement to the enemy, thereby maintaining some moderation.

Our predictions relate to the Ally Principle mentioned in the beginning \citep[see also][]{bendor2001theories, bendor2004spatial, callander2008theory}. In our model, the Ally Principle is not invalidated in general, but should be abandoned as soon as the enemy gains agency. Thus, we explain \emph{when} the Ally Principle holds, and when and why it gets discarded. Indeed, the behavior we outlined resembles what we can observe when parliamentary democracies deal with a new extremist party. At first, mainstream parties try to exclude it through a cordon sanitaire. However, once the newcomer attains agency, they switch to an embracing strategy to moderate it. Often, the extremist then integrates into the political establishment. As examples, we document this pattern both in Weimar Germany and postwar Italy.

One may wonder if the principal benefits from the existence of a competitive setting. Absent the enemy, the principal would lose the disciplining effect of the power struggle on the friend, but also dispose of the enemy's undesired policy choices. The friend would choose the policy always, but the principal would lack any means of pulling his choice toward the center. We show that if the principal is non-extreme in her relative position and interactions are frequent enough, the benefits of competition are larger.

The paper proceeds as follows. After a brief review of the literature, we present the model in \cref{sec:model}. \cref{sec: Example} analyzes a special case, and \cref{sec: Analysis} extends the analysis to the general case. \cref{sec:implications} discusses the implications of the results and their connection to real-world cases. We conclude in \cref{sec:conclusion}. Proofs are in the Appendix.

\paragraph{Related Literature.}\addcontentsline{toc}{subsection}{Related Literature} Allocation of power is central in organizational economics and political economy \citep[see][for a survey]{bolton2013authority}. We focus on repeated interactions. Much of that literature considers settings where players fully control power allocation \citep[e.g.,][]{li2017power,lipnowski2020repeated,acharya2021optimal,luo2025relational} or where that allocation is stochastic and exogenous \citep[e.g.,][]{dixit2000dynamics}. We adopt a middle ground; our principal partly influences the power allocation.

To our knowledge, \citet{delgado2023which} is the only paper with a similar setup. However, while that model hinges on a common resource pool---which agents use either to provide public goods or to bribe the principal for private gain---our model focuses on agents in pure conflict over a fixed pie. Moreover, the principal's role differs: In \citet{delgado2023which} the principal extracts common resources for personal benefit. Our principal is solely concerned with distribution. She leverages the agents' conflict to implement her preferred allocation. Conceptually, that means we face---unlike in the literature cited above---a setting of horizontal moral hazard: shirking is not the main incentive problem, but the direction of the action is. Consequently, the differentiation between friend and enemy becomes a key feature of our model.

Our model connects to the relational contracting literature \citep[see, e.g., the surveys of][]{macleod2014economics,MaclomsonHandbook,watson2021theoretical}. Other papers within that literature consider relationships with multiple competing agents.\footnote{Examples include \citet{rayo2007relational,levin2002multilateral,board2011relational,andrews2016allocation,calzolari2017relational,nocke2023collective,barron2019policies}.} The closest among these are \citet{board2011relational,calzolari2017relational, barron2019policies}, who, like us, study direct and persistent competition between agents. In contrast to these papers, we analyze a non-transferable utility (NTU) setting that precludes front-loading payments. This NTU framework and our view of organizations as communities of fate---with persistent internal disagreements---connect our paper to the delegation literature \citep[e.g.,][]{Alonso:08,dessein2002authority,alonso2008coordination}. Different from us, that delegation literature focuses on information use. In our model of complete and perfect information, delegation has other inherent frictions: the principal is not only unable to censor the agents' action spaces, but also only partially controls \emph{which} agent to entrust with power. We explore how the principal optimally uses that limited delegation power.

The idea of trading power for policy is central to both political economy \citep[see, e.g.,][]{anesi2022making,acharya2021optimal,invernizzi2024institutions} and organizational economics \citep[see, e.g.,][]{gibbons2020marching,gibbons1992,li2017power}. We contribute by studying relational incentives in a setting of persistent partisan disagreement and stochastically shifting authority.

Our setting is a repeated game with the particular feature that a subset of the players (the two parties) engages in a constant-sum game while a third player (the principal) acts as a power broker to her own advantage. As we will see, non-stationary optimal contracts appear because this power broker lacks full control over delegation. Moreover, the game is asymmetric because the principal is more closely aligned with one party, the friend. We show that, in this setting, the principal's value for commitment depends non-trivially on her relative position. This affects both her ability to punish the friend after a deviation and the credibility of her promises to embrace the enemy on the equilibrium path. Unlike \citet{abreau86}, our punishments combine grim trigger (always for the enemy, sometimes for the principal) and sticks-and-carrots (for the friend). The carrot is returning to the on-path continuation contract. The constant-sum property allows us to derive penal codes and thereby characterize the equilibrium payoff set.\footnote{In a stochastic game, \citet{deb2023fostering} study a principal who, like ours, exploits the constant-sum competition between two agents. However, their research question, model, and analysis differ substantially.}

\section{Model}\label{sec:model}

Time is discrete and indexed by $t=1, 2, \ldots$ There are three risk-neutral players: one principal, $P$; and two parties, the friend, $F$, and the enemy, $E$. Players discount the future exponentially with discount factor $\beta \in (0,1)$. Throughout, information is perfect, and our players engage in the following repeated game.\footnote{In line with the repeated games literature, we assume that the principal can use a public randomization device that---although not strictly necessary---simplifies some proofs.}

    \paragraph{Stage Game.} At the beginning of every period $t$, $P$ provides a level of endorsement $s_t \in [-m,m]$, where $s_t=-m$ implies full endorsement of $F$, and $s_t=m$ implies full endorsement of $E$. Then, nature tosses a biased coin that yields $k_t \in \{F, E\}$. Endorsement $s_t$ affects the probabilities of this biased coin. The probability of $k_t=E$ is
    \(p(s_t) =1/2+s_t,\)
     and $k_t{=}F$ realizes with the complementary probability, $1-p(s_t)$. The principal's power is limited---that is, $m \in (0,1/2)$. If $k$ realizes, we say party $k$ is the \emph{leading party}. The leading party chooses a policy $y_t \in [0,1]$. The stage game ends and all players collect their within-period payoffs.

    \begin{figure}[t]
        \centering
        \begin{tikzpicture}[
            scale=1.05,
            every node/.style={font=\small}
        ]

        \def\X{8}
        \def\T{2.6}

        \draw[->, thick] (-0.3,0) node[left] {policy} -- (\X+0.45,0) node[right] {$y$};

        \draw[thick] (0,0.09) -- (0,-0.09);
        \draw[thick] (\T,0.09) -- (\T,-0.09);
        \draw[thick] (\X,0.09) -- (\X,-0.09);

        \node[above=15pt] at (0,0) {$\theta_F=$};
        \node[above=15pt] at (\T,0) {$\theta_P=$};
        \node[above=15pt] at (\X,0) {$\theta_E=$};

        \begin{scope}[yshift=0cm]

        \draw[very thick] (0,0) -- (\X,-1.6);
        \node[right] at (\X+.35,-1.6) {$u_F(y)=-|y|$};

        \draw[very thick, dashed] (0,-0.52) -- (\T,0) -- (\X,-1.08);
        \node[right] at (\X+.35,-1.08) {$u_P(y)=-|y-\theta|$};

        \draw[very thick, dotted] (0,-1.6) -- (\X,0);
        \node[left] at (0,-1.6) {$u_E(y)=-|y-1|$};

        \node[above=5pt] at (0,0) {$0$};
        \node[above=5pt] at (\T,0) {$\theta$};
        \node[above=5pt] at (\X,0) {$1$};

        \end{scope}

        \end{tikzpicture}
        \caption{Bliss points and policy payoffs.
        }
        \label{fig:bliss-points-payoffs}
    \end{figure}

    \paragraph{Flow Payoffs.} If the leader's policy is $y$, player $i \in \{F,E,P\}$ receives a stage payoff from that policy given by
    \[u_{i,t}(y):= -|y-\theta_i|,\]
    where $\theta_i$ denotes player $i$'s persistent bliss point in the policy space,
    \[\theta_F\equiv 0 \qquad \theta_P\equiv \theta \in [0,1/2) \qquad \theta_E\equiv 1.\]
    Only the party who leads in the current period, $k$, receives the leadership rent, $b>0$, in addition to $u_{k,t}$. Notationally, it is useful to distinguish between a player's continuation payoff at the beginning of a period, $w_i(\cdot)$, and \emph{after selection}, $v_i(\cdot)$.

    \paragraph{Solution Concept and Strategies.} We are looking for an ex-ante principal-preferred subgame-perfect Nash equilibrium (SPE) of the repeated stage game.

    We consider two distinct cases: The first is the \emph{commitment case}. Here, the principal chooses her strategy at the beginning of the game and publicly commits to it. The parties then play the principal-preferred SPE, taking $P$'s strategy as given. The second is the \emph{no-commitment case}. Here, the principal cannot commit a priori. Instead, we look for the principal-optimal SPE of the three-player repeated stage game.

    A \emph{strategy} describes a player's full contingent plan of action. We describe it by the function $s(\cdot)$ for the principal, $y_F(\cdot)$ for party $F$, and $y_E(\cdot)$ for party $E$. We let $\sigma:=(s(\cdot),y_F(\cdot),y_E(\cdot))$ describe a strategy profile, and we call a strategy profile that constitutes an SPE a \emph{contract}, using $C$ to denote a generic one. Our object of interest, the \emph{optimal contract}, $C^\ast$, is the principal's ex-ante preferred contract. The only difference between the commitment and the no-commitment case is that in the former, the principal's strategy $s(\cdot)$ need not be incentive compatible, whereas in the latter it must be.

    \subsection*{Preliminaries}
    Before moving to the analysis, we make a few preliminary observations that are useful to keep in mind.
    \paragraph{No Transfers.} Contrary to standard principal-agent models, our setting is one of ``horizontal disagreement.'' Thus, we implicitly assume our players have a strong mission-driven agenda, making a setting without monetary transfers salient (see, e.g., the discussions in \citet{Tirole:94,march1962business,gibbons2020marching}). There is one caveat to that assumption, however: If a principal holds decision-making power herself (and cares about it), we may think of $b$, the leadership rent, as the number of decisions she has to delegate to an agent. But then, the principal may want to increase that number (at her own expense). One could model that through direct utility transfers. Whether utility transfers benefit the principal or not depends on the parameters. A committed principal only benefits from utility transfers when she is sufficiently centrist. Without commitment, she always benefits from transfers when trying (and needing) to punish the friend. We provide a formal discussion in Online Appendix \ref{sub:transfers}.

        \paragraph{Implementable Contracts.} A contract describes an equilibrium of the game. It must, therefore, be incentive compatible. Parties should prefer choosing the policy prescribed by the contract over their best deviation. If party $k$ complies with contract $C$'s recommendation, $y_k(h)$, at a history $h$, he will choose policy $y_k(h)$, and then expect an on-path continuation payoff $w_k(C,h)$. If, on the contrary, he deviates---e.g., by picking his preferred policy, $y_k=\theta_k$---we switch to his \text{penal code} \citep{abreu1988theory}. Party $k$'s continuation payoff from entering his own penal code is $\underline{w}_k$.\footnote{Note that $k$'s penal code is a contract, $C^k$, in itself. We use the shorthand $w^k\equiv w(C^k)$ to describe $C^k$'s payoff vector. Each player has their own penal code and $j$'s expected payoff from entering $k$'s penal code is $w^k_j$. To emphasize that $w_k^k$ is indeed $k$'s worst continuation payoff we use $w_k^k\equiv\underline{w}_k$. Simple penal codes that only condition on the (last) deviator's identity suffice because information is perfect.} Hence, a policy $y$ is enforceable for $k$ within contract $C$ at history $h$ if and only if
        \begin{equation*}\label{Eq Agent DEC}
        -|y-\theta_k| + \beta w_k(C,h) \geq \beta \underline{w}_k. \tag{DEC}
        \end{equation*}
        Following \citet{levin2003relational}, we refer to \eqref{Eq Agent DEC} as $k$'s \emph{dynamic enforcement constraint}. A necessary condition for a strategy profile to be a contract is that \eqref{Eq Agent DEC} holds at all histories. If the principal has commitment power, the condition is also sufficient.

        \paragraph{Parties' Disagreement.} Parties disagree in two dimensions: First, they disagree along the policy dimension. There, they distribute a constant utility of $-1$ across them. Second, they compete over the per-period leadership rent, $b$. Taken together, parties' utility in every period sums to
        \(b+u_{F,t}(y)+u_{E,t}(y)\equiv b-1.\) We make two observations based on this \emph{constant-sum property.}
        \begin{observation}[\textbf{Pareto Efficiency}]\label{obs:Pareto}
            A principal-optimal contract is Pareto optimal.
        \end{observation}
        The observation follows as \emph{any} equilibrium is Pareto optimal across parties by the constant-sum property. A principal-optimal equilibrium maximizes total welfare.\footnote{Because direct utility transfers are not possible, the reverse is not true: there may be Pareto-optimal equilibria that do not maximize total welfare and are thus not principal-optimal.}

        Our second observation concerns the role of the principal, which is grounded in her power, $m$, to influence the selection process. Consider a contract that replicates the Nash equilibrium of the stage game (NES hereafter) in every period. In this contract, the principal endorses her friend, $s=-m$, and any party $k$ chooses their bliss point $y_k {=} \theta_k$.

        \begin{observation}[\textbf{No $P$-Power}]\label{lem:noP}
           NES is the (essentially) unique contract if $m{=}0.$
        \end{observation}
This is the classical min-max result that follows from the constant-sum property. If $m=0$, the principal has no power. Thus, to achieve moderation, parties have to coordinate on concessions. But then, any policy concession by one party must be compensated through future policy concessions by the other. Because parties are impatient, future concessions must exceed current ones, which in turn require even larger concessions after---as this is impossible, parties can only agree to polarize.

Note, however, that if $m>0$ but $b\to 0$, the principal can use endorsement to her advantage---especially in the commitment case. By threatening parties to support the opponent unconditionally, the principal can extract policy concessions even when leadership rents vanish. For instance, by moderating the enemy, she can demand concessions from the friend under the threat of withdrawing moderation once the enemy returns to leadership.
In the first part of the analysis, we will shut down this cross-subsidization channel and focus on an example in which the principal fully aligns with the friend, $\theta=\theta_F=0$. As we see, a committed principal still strictly improves upon NES for $\beta>0$ and $m>0$ even if $b \to 0$. The picture changes absent commitment. In the second part, we analyze the full model. We show that there, cross-subsidization interacts with the leadership rent, $b$, both with and without commitment.

\section{Example with full alignment}\label{sec: Example}

 We begin by analyzing the special case of a fully aligned friend, $\theta=\theta_F=0$.

\subsection{Example: Commitment}

    \paragraph{The NES contract.} A candidate for the principal would be the NES contract. This strategy maximizes the probability that the friend takes the lead, and he then chooses the principal's bliss point, $y_F=0$. However, the enemy, whenever he leads, has no incentive to choose anything but his bliss point, $y_E=1$. But is persistent NES ever optimal?

    \paragraph{The Optimal Contract.} It turns out that NES is never optimal. Instead, (almost) the opposite is true: the principal continuously {embraces the enemy} once an initial attempt to exclude him has failed. More formally, the optimal contract has two phases on the equilibrium path: an initial \emph{exclusion phase} that lasts \emph{until the first time the enemy leads}. Then it switches to a stationary \emph{embracing phase}.
    To characterize the optimum, we specify the contract in each phase and determine how deviations are punished. With commitment and $\theta=0$, the relevant punishment is that of a deviating enemy, who is punished by NES.\footnote{\label{fn:unpunished}The friend can be left unpunished as he chooses his static optimum on-path.}
    \begin{proposition}\label{prop:theta0}
        Suppose $\theta=0$ and a principal with commitment. For any $(m,\beta,b)$, the optimal contract has two phases.

        \textbf{Exclusion Phase.} The principal fully endorses the friend, $s^0=-m$, and the friend chooses $y_F^\ast=0$. The exclusion phase lasts until the enemy leads for the first time.

        \textbf{Embracing Phase.} From that point on, the contract is stationary. The principal fully endorses the enemy, $s^\ast=m$, the friend chooses $y_F^\ast=0$, and the enemy moderates to
        \[y^\ast_E=\max \left\{1- \frac{2 \beta m (b+1)}{1-\beta (1/2-m)},0\right\}.\]
    \end{proposition}

    Proposition \ref{prop:theta0} shows that the principal switches (on-path) strategies midgame. The game begins with an exclusion phase, which lasts until nature selects $E$ for the first time. Now, we transition irreversibly to the embracing phase. The principal asks the enemy to moderate and promises full endorsement in all future periods in return. The enemy complies to secure that endorsement.

    The intuition behind this result comes from two observations: the constant-sum property across parties and the fact that the principal's payoff equals the friend's payoff net of his expected leadership rents (i.e., $b$ in periods when he leads and $0$ otherwise). The principal can increase the enemy's continuation payoff by promising him marginally more endorsement in the future. This promise gives the enemy two benefits: First, by being in power more often, the average policy in the future is closer to his bliss point. For the sake of the argument, say the change in the expected policy increases the continuation utility of the enemy by $\gamma>0$. Second, the enemy's average leadership rent in the future increases, resulting in a continuation utility gain of, say, $\omega>0$. Thus, the enemy's total continuation payoff increases by $\gamma{+}\omega$.

    The principal only suffers from the enemy's future policy choices, but not from distributing the leadership rent toward him. Her continuation utility, therefore, decreases only by $\gamma$. However, she can ask the enemy to pay back his \emph{total} utility gains ($\gamma{+}\omega$) through \emph{today's policy choice}, which fully enters the principal's utility. Thus, the principal gains $\gamma + \omega$ today and loses $\gamma$ in the future---a profitable trade-off.\footnote{Policy $y_E^\ast < 1$ even when $b\to 0$. This is because even in that limit, the principal could trade policy for office. However, since $\omega\to 0$, all such trades would be almost 1-to-1. The equilibrium would give the principal the same payoffs as NES. Therefore, whenever $b>0$, the principal strictly improves over NES.}

    Naturally, these considerations only become relevant once the enemy leads for the first time. Past endorsements are sunk for the leading party and, as long as there are no incentives to provide to the enemy, the principal aims to exclude him.

    \subsection{Example: No Commitment}

    The previous results rely on the principal's promise to endorse the enemy in the future if he moderates today. However, this promise may not be credible if the principal lacks commitment power. After all, she always prefers her friend to lead. So, she may renege right after the enemy has taken his action and endorse her friend instead.

   Note that this decision relates to the principal's \emph{interim} continuation payoff, that is, \emph{after} the enemy chose his first policy. Although we know from \cref{prop:theta0} that more embracing of the enemy is always better \emph{ex ante} for the principal, she may renege on her promised endorsement if: (i) her \emph{interim} continuation payoff is not attractive enough or (ii) the consequences of reneging are not severe enough. As in any repeated game, the principal's dynamic enforcement constraint captures both issues: any contract $C$ at any history $h$ must satisfy
    \begin{equation*}\label{Eq:PrincipalDEC}
        w_P(C,h) \geq \underline{w}_P, \tag{DEC'}
\end{equation*}
    where $\underline{w}_P$ is the principal's payoff from her penal code. Although \cref{sec: nocommitment} gives more details on the construction of penal codes, it is instructive to see that NES is not the harshest credible punishment for $P$.  Recall that NES is the enemy's worst continuation play and thus---by the constant-sum property---the friend's best. Since friend and principal are aligned in policies, such a penal code is too mild. Instead, to punish the principal most effectively, the friend chooses a policy $y_F>0$, i.e., to the right of the principal's bliss point. This is demanding for the friend, who needs to be (at least) indifferent between continuing the principal's punishment and deviating to his bliss point $y_F=0$, which triggers his own penal code. As a result, another complexity appears:  the penal codes of friend and principal are interconnected. We have to characterize the friend's punishment although it is irrelevant on path. It turns out that a worst punishment for the friend has two phases: upon deviation, the principal fully embraces the enemy, demanding less moderation from him (the punishment phase), and, upon the friend's return to power, continuation play switches back to the on-path embracing phase (the back-to-business phase).\footnote{We provide further details in \cref{sec: nocommitment}.}

    Using our construction, we find that if the principal lacks commitment power, either the optimal contract is observationally identical to the optimal commitment contract, or NES is the optimal contract.

    \begin{proposition}\label{prop:theta0nocommitment}
        Assume a principal without commitment, and suppose $\theta=0$. The optimal contract is observationally equivalent to the optimal commitment contract if and only if \[b\geq\overline{b}_0:=\frac{(1 - \beta)^2}{\beta  (2 - \beta  (1 + 1/2 - m))  (1/2 + m)}.\]
        Otherwise, NES is the only available contract.

        For $b\geq \overline{b}_0$, the following \emph{penal codes} support the equilibrium: \begin{itemize}
            \item the NES contract for the enemy;
            \item a stationary contract with $y_E{=}1{>}y_F{=}\hat{y}_F{>}0$ and $s{=}-m$ for the principal;
            \item a punishment phase with $\hat{y}_E{\geq} y_E^\ast$ and $s{=}m$ that lasts until $F$ leads again; followed by a return to the on-path embracing phase for the friend.
         \end{itemize}

    \end{proposition}

    For extreme values of $b$, the results of \Cref{prop:theta0nocommitment} are straightforward. If $b\to \infty$, the policy is second-order to the enemy. He is willing to choose $y^\ast_E=0$ in exchange for endorsement, so endorsing him is costless for the principal.  If $b\to 0$, the game between principal and enemy becomes almost constant-sum, leaving no room for dynamic incentives.

    For intermediate values of $b$, however, two opposing effects are interacting. On the one hand, there is a marginal cost: the greater the endorsement for the enemy, the lower the friend's chances of leading and choosing $y_F^\ast =0$. On the other hand, there is an inframarginal gain: the greater the endorsement promised to the enemy, the more the principal can ask for his moderation.

  Together, these opposing effects imply that the principal's \emph{interim} continuation payoff is (strictly) convex in the level of endorsement she promises to the enemy. When endorsement is low, the enemy's policy moderation is small. Thus, by increasing the enemy's chances of leading, undesired policies become more likely---the marginal cost is large. However, as endorsement rises, so does the enemy's moderation, reducing the marginal cost. The inframarginal gain of policy moderation in exchange for endorsement is often constant in the level of endorsement. Therefore, there is no interior optimum: in the embracing phase, the principal's optimal strategy is either NES or full embracement.

\Cref{prop:theta0nocommitment} establishes, furthermore, that if $\theta=0$, then \textit{once full embracing ceases to be optimal, the only option left is NES}---in other words, the principal entirely loses her ability to make credible promises. To see why, we need to walk through the following (slightly technical) argument.

The key observation is that the friend's payoff consists of both leadership rent and policy payoff. The latter is identical to the principal's payoff because $\theta=0$. Moreover, full embracing of the enemy is strictly better \emph{ex ante} for the principal. Therefore, no embracing can only be optimal if full embracing violates the principal's \eqref{Eq:PrincipalDEC}.

Now, consider the knife-edge case in which full embracing makes \eqref{Eq:PrincipalDEC} hold with equality. This means the principal's \textit{interim} continuation payoff---at the embracing phase---is identical to her worst (ex-ante) payoff, $\underline{w}_P$. Consequently, the friend's interim \textit{policy} continuation payoff is identical and thus also $\underline{w}_P$ \emph{because} his policy preferences equal those of the principal. Moreover, since the enemy is fully embraced, the friend's expected leadership rent in the continuation game is minimized. Hence, in the embracing phase, the friend obtains his worst continuation payoff, that is, his penal code. Crucially, this implies that the friend's threat of punishing the principal harder than NES (which, recall, depends on him choosing $y_F >0$) becomes inseparable from the implementability of full embracing. Once the principal can no longer credibly promise to fully embrace the enemy, the friend can no longer credibly threaten the principal with policies to the right of their common bliss point. In short, the collapse of full embracing brings down both the commitment optimal contract and the principal's penal code. We revert to NES for $b<\bar{b}_0$.

Key to \cref{prop:theta0,prop:theta0nocommitment} is that the friend needs no incentives to choose the principal's bliss point. As we will see next, once there is (sufficient) disagreement between friend and principal, our results change.

\section{Analysis}\label{sec: Analysis}

\subsection{General Model: Commitment}
\begin{figure}
    \centering
    \includegraphics[width=0.49\textwidth,valign=t]{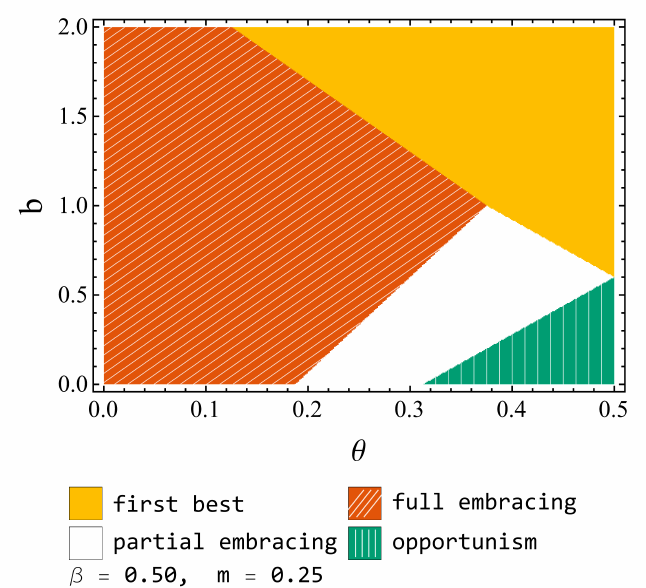}
    \includegraphics[width=0.49\textwidth,valign=t]{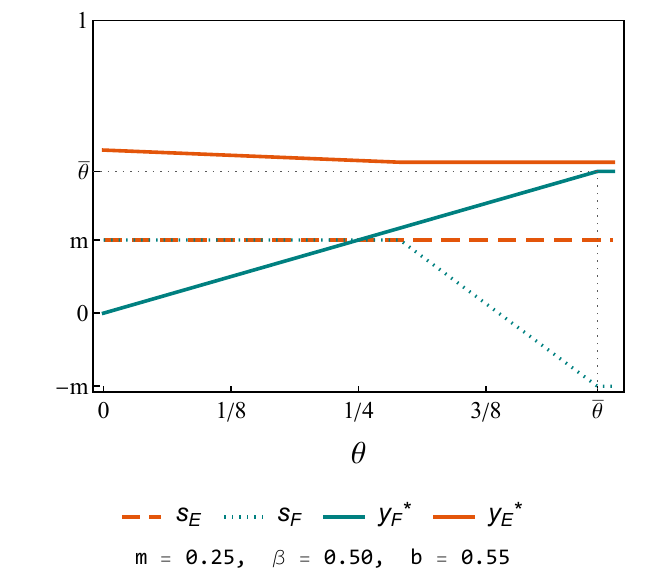}
    \caption{\textbf{Embracing Phase under Commitment.} \emph{Left:} Regimes in the $(\theta,b)$ space. Outside the first best, \emph{full embracing} ($s_F=s_E=m$) is optimal for low $\theta$, \emph{partial embracing} ($s_E=m>s_F>-m$) for intermediate $\theta$, and \emph{opportunism} ($s_E=m,s_F=-m$) for large $\theta$.\\
    \emph{Right:} Equilibrium strategies as a function of $\theta$ for given $(m,\beta,b)$. Dashed/dotted are the principal's strategies, solid those of the agents. The friend's strategy $y_F^\ast$ follows the 45-degree line until $\theta=\overline{\theta}$, then stays constant at $\overline{\theta}$.}\label{fig:commitment}
    \end{figure}

    We now turn to the general case, $\theta \in [0,1/2)$, in which there may also be disagreement between friend and principal. Then, the friend would only choose the principal's bliss point if incentivized accordingly. We begin with the embracing phase. \Cref{fig:commitment} illustrates the findings we develop next.

    \paragraph{Embracing.} As in any dynamic setting, the principal can incentivize the friend by (i) increasing the friend's on-path utility upon compliance, or (ii) threatening punishment if he deviates. In particular, a principal with commitment power can always threaten to min-max the friend, for example, through promising unconditional and full endorsement to the enemy, who then chooses his bliss point $y_E = 1$.

    If $\theta$ is close enough to $0$, the punishment threat---option (ii)---is sufficient to incentivize the friend. The reason is intuitive: Both principal and friend benefit from the enemy's moderation. Conditional on the enemy leading, their incentives are aligned (as long as $y_E\geq \theta$). The friend fears the enemy choosing his bliss point $1$ instead of the on-path $y_E^\ast<1$. Hence, when he leads, the friend has an incentive to choose $y_F=\theta$ even absent any promise of future endorsement. Moreover, since moderating the friend's choices benefits the enemy, the principal can also ask the enemy to moderate further. This \emph{cross-subsidization effect} ensures that if $\theta$ is close to $0$, the principal fully embraces the enemy in the embracing phase (i.e., after the enemy leads for the first time).

    More formally, the optimal contract requires the enemy to moderate until his \eqref{Eq Agent DEC} binds. The constant-sum property then implies that the losses of the friend when the enemy leads are bounded. In fact, the friend attains his largest possible continuation payoff \emph{conditional on the enemy leading}. If $\theta$ is small, that promised utility is enough for the friend's \eqref{Eq Agent DEC} to hold with slack if he chooses $y_F=\theta$. The remaining economic forces are identical to the example.
    Yet, there exists a threshold value
       \[\begin{split}\underline{\theta}&:=2\frac{\beta}{1-\beta}(b+1)m\beta \left(\frac{1}{2}+m\right),\end{split}\]
    such that, if the principal's bliss point $\theta>\underline{\theta}$, cross-subsidization is no longer enough to incentivize the friend to choose $y_F=\theta$. Then, the principal needs to provide additional, direct incentives for the friend to make him moderate to $y_F=\theta$. The optimal way to provide such incentives is to offer a leading friend to tilt the endorsement strategy in his favor in the \emph{next} period. Thus, during the embracing phase, the principal selects different endorsement strategies, $s_F, s_E$, depending on who has led last. Not surprisingly, it remains optimal to exploit the cross-subsidization effect and require the enemy to moderate until his \eqref{Eq Agent DEC} binds. Since---by the constant-sum property---the friend is indifferent between all continuation plays that give the enemy a certain payoff, the principal proposes the principal-optimal continuation contract that puts the enemy at his \eqref{Eq Agent DEC}. Using our reasoning from the example, it is thus optimal to set $s_E=m$ in that case.

    We also know from the example that, absent the problem of incentivizing the friend, the principal would like to incentivize the enemy's moderation by setting $s_F=m$. Thus, the only reason to lower $s_F$ is that, otherwise, the friend would never choose $y_F=\theta$ (i.e., the friend's \eqref{Eq Agent DEC} fails for $y_F=\theta$ and $s=m$). It is therefore optimal to set $s_F$ such that $F$'s \eqref{Eq Agent DEC} binds conditional on playing $y_F=\theta$. This leaves the enemy's incentives for moderation unchanged because, invoking the constant-sum property once more, if $F$'s \eqref{Eq Agent DEC} binds then $E$'s continuation payoff is maximal conditional on the friend leading.

    Quite intuitively, the relationship between $\theta$ and the optimal $s^\ast_F$ is monotone. The larger $\theta$, the more incentives the friend needs to select $y^\ast_F= \theta$, so $s^\ast_F$ declines. Depending on parameters, it may be the case that there is a second threshold
       \[\begin{split}\overline{\theta}&:=\frac{1-\beta(1/2+m)}{\beta(1/2+m)} \underline{\theta}\end{split}\]
   in the domain of $\theta$ such that $s^\ast_F$ hits the boundary $s^\ast_F{=}{-}m$. Before turning to this case, we formalize the discussion until now and characterize the optimal contract when $\theta<\overline{\theta}$.

    \begin{proposition}\label{prop:embraceGenCom}
        Suppose $\theta<\overline{\theta}$ and parameters are such that the first best is not feasible. The optimal contract has two phases.

        \textbf{Exclusion Phase}. The principal fully endorses the friend, $s^0{=}-m$, and the friend chooses $y_F^\ast{=}\theta$. The friend's \eqref{Eq Agent DEC} has slack. This phase ends with the enemy's first lead.

        \textbf{Embracing Phase.} From this point on, the contract is stationary. The friend chooses $y_F^\ast=\theta$, and the enemy, $y_E^\ast\in(\theta,1)$. The principal's endorsement strategy depends on $\theta$:
        \begin{itemize}
                \item If $\theta\leq \underline{\theta}$, she fully endorses the enemy, $s^\ast_E=s^\ast_F=s^\ast=m$.
                \item If, instead, $\theta \in (\underline{\theta},\overline{\theta})$, she fully endorses the enemy if he led last, $s^\ast_E=m$, but selects an interior endorsement $s_F^\ast \in (-m,m)$ if the friend led last.
            \end{itemize}
        Both $s_F^\ast$ and $y_E^\ast$ (weakly) decrease in $\theta$. A leading enemy's \eqref{Eq Agent DEC} always binds; a leading friend's \eqref{Eq Agent DEC} binds if and only if $\theta>\underline{\theta}$ and the contract is in the embracing phase.

        Deviations are punished by unconditional full endorsement for the non-deviator and full polarization in policy choices.
    \end{proposition}

   The exclusion phase and its intuition are unchanged compared to the example. The friend is willing to comply, because he receives a higher continuation value during the exclusion phase than during the embracing phase---the enemy has no agency, so his incentives are irrelevant.

  \paragraph{Opportunism.} We now turn to $\theta\geq\overline{\theta}$. At $\theta=\overline{\theta}$, the reasoning of \cref{prop:embraceGenCom} continues to apply, but the principal has exhausted all her resources: she needs to fully endorse whoever led last to uphold that period's incentives for moderation. She promises the friend full endorsement after leading, $s_F=-m$, and, in exchange, he moderates exactly to $y_F^\ast=\overline{\theta}$.  By the symmetry of the principal's strategy, the enemy chooses $y_E^\ast=1-\overline{\theta}$ in exchange for $s_E=m$.  Since the principal is now \emph{opportunistic}, the exclusion and embracing phases become almost indistinguishable. The principal follows her genuine preference for the friend only in the very beginning, before any party acts.

  For $\theta>\overline{\theta}$, the principal is incapable of incentivizing any party to choose her bliss point $\theta$.  As a result, the parameter $\theta$ has no payoff relevance to the parties and the optimal contract is identical for all $\theta \in [\overline{\theta},\frac{1}{2})$. By the familiar logic from above, since the enemy is at his \eqref{Eq Agent DEC}, there is no point in decreasing $s^\ast_E=m$ to provide incentives to the friend. Hence, the principal is opportunistic. Parties moderate as much as they can be moderated, which is $y_F^\ast=\overline{\theta}$ and $y_E^\ast=1-\overline{\theta}$, independently of $\theta$.

  \begin{proposition}\label{res:OEndorsement}
            Suppose $\theta\geq \overline{\theta}$. After an initial endorsement of the friend, $s^0=-m$ in the first period, the contract is stationary. On-path, the principal fully endorses whoever led last, i.e., $s^\ast_F=-m$ and $s^\ast_E=m$. A leading friend chooses $y_F^\ast=\overline{\theta}$ and a leading enemy chooses $y^\ast_E = 1 {-} \overline{\theta}$. Any leading party's \eqref{Eq Agent DEC} binds.

            Deviations are punished by unconditional full endorsement for the non-deviator and full polarization in policy choices.
        \end{proposition}

 \paragraph{First-Best Contracts.} To close the characterization, we briefly discuss first-best contracts. Because the friend's bliss point is closer to the principal's than the enemy's, a first-best contract is, intuitively, possible only if the principal can ensure $y_E^\ast=\theta$.

 From our previous discussion, we can derive an immediate necessary condition for the first best: $\theta\geq 1-\overline{\theta}$. As we know from \cref{res:OEndorsement}, $1-\overline{\theta}$ is the furthest we can moderate the enemy's (average) action by putting him to his \eqref{Eq Agent DEC} while maximizing his chances to lead next period, and promising him the best possible continuation value if he is not selected.

 However, the condition is only necessary, not sufficient. The reason is that to promise the enemy the best continuation value conditional on not leading, the friend may need to choose a policy $y_F>\theta$, which violates the first-best presumption.

 Thus, there is a second, necessary condition: $\theta\geq 1-\breve{\theta}$, and
 \[\breve{\theta}:= \beta\left( \frac{1}{2} + m(1+2b) \right),\]
 where $\breve{\theta}$ is the symmetric version of $\overline{\theta}$: the furthest distance the friend and the enemy are willing to symmetrically move in policy from their respective bliss points toward each other. Our second condition ensures that even if at $y_F=\theta$ the friend has slack on his \eqref{Eq Agent DEC}, the enemy is willing to choose $y_E=\theta$. Jointly, the two conditions are sufficient to achieve first best.

 \begin{proposition}\label{result:firstbest}
    The principal achieves her first best if and only if $\theta \geq \max \{1-\overline{\theta},1-\breve{\theta}\}.$
 \end{proposition}

 The comparative statics are intuitive. As $b$ grows large, the enemy becomes primarily power hungry. He is responsive to the threat of losing endorsement and willing to concede on policy to the principal; both first-best thresholds converge to $-\infty$. As $m$ grows, the principal's endorsement becomes more valuable and punishment more severe. So, provided $\beta$ and $b$ are not too small, the first best becomes feasible. However, even in the limit $m\to 1/2$, the principal can obtain the first best only if the friend is sufficiently close. Otherwise, even the promise of an everlasting exclusion phase cannot induce the friend to moderate up to $\theta$. Finally, as $\beta$ increases, more outcomes become implementable by the standard logic of repeated games. Even so, note in the example of $\theta=0$ that, as $\beta\to 1$, it is not guaranteed the first best is always implementable---we need both $m$ and $b$ to be sufficiently large.

 \paragraph{Polarization.} The next proposition shows that a more balanced principal is more effective at achieving moderation.
    \begin{proposition}[\textbf{Polarization}]\label{prop:polarization}
        The distance between the parties' equilibrium policies, $y^\ast_E-y^\ast_F$, decreases as the principal's preferences become more balanced.
    \end{proposition}
This result highlights the moderating role of the centrist principal. It introduces a novel channel in the existing literature on repeated policy trading \citep[see,][]{dixit2000dynamics,acemoglu2011power,invernizzi2024institutions,mcclellan2023intraparty}. These papers consider settings without a principal but assume curvature in parties' utility. By contrast, our parties face a constant-sum game between them, which rules out bilateral policy trading. Their mechanism and ours are complementary: introducing utility curvature would reinforce our channel, implying that the principal's embracing strategy generates even greater moderation.\footnote{In \cref{sub:curvature}, we sketch the formal argument underlying this conclusion.}

 \paragraph{The Symmetric Case. }
Before turning to the model without commitment, it is worth discussing the case of a fully balanced principal, i.e., $\theta=1/2$. Naturally, in this case, the meaning of ``friend'' and ``enemy'' is lost. However, the results are (almost) identical to those of the limit case $\theta \to 1/2$. To see this, recall from \cref{res:OEndorsement} that the maximum moderation the principal can induce from either party is a distance $\overline{\theta}$ away from their respective bliss points. This cutoff is, importantly, independent of the principal's bliss point $\theta$. Now, if parameters are such that $\overline{\theta}\geq 1/2$, then a fully balanced principal can achieve first best. Otherwise, she offers the opportunistic contract described in \cref{res:OEndorsement}, which---after the first period---is fully symmetric.

The only difference arises in this first period: while a biased principal has a \emph{strict} incentive to endorse the friend, a fully balanced principal is indifferent.

\begin{proposition}\label{prop:symmetry}
    In the symmetric case, $\theta=1/2$, either a first-best contract is optimal, or the opportunistic contract described in \cref{res:OEndorsement} is optimal. In either case, the principal is indifferent whom to endorse in the first period.
\end{proposition}

\subsection{General Model: No Commitment}\label{sec: nocommitment}

    We now remove the principal's commitment power. Recall the principal's dynamic enforcement constraint,
    \begin{equation*}
        w_P(C,h)\geq \underline{w}_P, \tag{\ref{Eq:PrincipalDEC}}
    \end{equation*}
    which implies that, at any history $h$, the principal prefers to continue with contract $C$ over switching to her worst continuation contract.

    As in the example of $\theta=0$, whether our results extend to the no-commitment case depends on two conditions: (i) the principal's desire not to renege on a given \emph{on-path} promise, and (ii) her ability to promise a harsh punishment if the game moves \emph{off path} because a party deviates. On the equilibrium path, \eqref{Eq:PrincipalDEC} becomes relevant when the contract requires the principal to endorse the enemy; off path, when she wants to punish a deviating friend. The reason is straightforward: the principal genuinely prefers the friend to lead, so endorsing the enemy strains her incentive constraint.

    \subsubsection{Punishments}

    How strained those incentives become depends, of course, on the ability of the two parties to punish the principal. Thus, we need to determine both the principal's and the friend's punishment jointly. That interconnectedness of punishments often makes it hard to derive explicit, behavioral \emph{penal codes} \`a la \citet{abreu1988theory} in multiplayer games.\footnote{Formally, a player's penal code is the continuation game played after they deviated. An \emph{optimal} penal code is the harshest such punishment.} In this case, however, the structure of our problem---in particular the constant-sum property between enemy and friend---allows us to explicitly derive each player's penal code.

    \paragraph{Punishing the Enemy.} The enemy's grim-trigger punishment from the commitment solution corresponds to the NES contract, which poses no incentive problem for the principal.

     \paragraph{Punishing the Friend.} Punishing the friend is considerably more difficult. A literal implementation of the commitment punishment requires parties to polarize while the principal fully endorses the enemy. Generically, such behavior is not incentive compatible for the principal.\footnote{It is feasible when $\theta=1/2$, because under symmetry, friend and enemy are the same to the principal.}  Consequently, the friend's optimal penal code must take a different structure. To construct it, we use the \emph{back-to-business property}: after an initial punishment phase, play returns to the embracing phase of the principal-optimal contract.

The intuition behind this property connects directly to the logic of the embracing phase itself: When the enemy first gains the lead, the principal moderates him by offering a desirable continuation game in exchange for maximal policy concessions. By the constant-sum property, a continuation game that benefits the enemy is harmful to the friend. In fact, part of what makes the promised continuation game attractive for the enemy is that the friend, upon leading, is held to his binding \eqref{Eq Agent DEC} and asked to moderate. This feature carries over to the friend's penal code: once the friend regains the lead after a deviation, his optimal punishment prescribes the same continuation play he faces on path. The back-to-business property serves a dual purpose. For the friend, it ensures a low continuation payoff upon returning to lead. For the principal, it guarantees a high continuation payoff because it coincides with the optimal contract---this, in turn, incentivizes the principal during the initial punishment phase.

Now, consider the punishment phase. The principal endorses the enemy, who moderates just enough for the principal's \eqref{Eq:PrincipalDEC} to hold. We get a first insight into how the penal codes of principal and friend are interrelated: the more the principal fears her own penal code, the greater the polarization she is willing to tolerate during the friend's punishment phase, and hence the harsher the penalty she can credibly impose on the friend.

    \paragraph{Punishing the Principal.} Finally, we determine the principal's optimal penal code. Although its concrete structure depends on parameters, two features are always present: (i) the principal ``defends'' by fully endorsing the friend---after all, she prefers the friend at all stages and has, within the penal code, nothing to lose; (ii) the enemy fully polarizes, choosing $y^P_E=1$.

    The only remaining object to characterize is the friend's action, which depends on the principal's bliss point, $\theta$.  When  $\theta$ is close enough to $0$, the logic from our special case $\theta=0$ applies. To punish the principal, the friend chooses a policy to the right of $\theta$. How far the friend is willing to go depends on the friend's own penal code, implying a second linkage between the two penal codes.  However, if we consider a large $\theta$, the friend would need to choose a policy that is significantly to the right. If the friend is unwilling to choose $y_F^P\geq 2 \theta$, then he instead reverts to his bliss point, $y_F^P = 0$. Hence, for sufficiently large $\theta$, the penal code becomes the NES contract.

Because the principal's penal code has these two regimes, her punishment value is non-monotone in her bliss point, $\theta$. If the punishment involves $y_k^P\geq\theta$, its severity diminishes as $\theta$ approaches these policy choices. By contrast, the NES punishment becomes more severe as $\theta$ increases. She moves away from the friend's policy, $y^P_F = 0$, and closer to the enemy's policy, $y^P_E =1$, but, because $s=-m$, the former effect dominates. We see, thus, that parties can coordinate to punish the principal harshly when she is either extreme ($\theta$ close to $0$) or centrist ($\theta$ close to $1/2$). But a principal that is partially aligned ($\theta$ interior) has an attenuated punishment. This observation, of course, feeds back into the principal's ability to promise an embracing strategy.

\subsubsection{Optimal contract}

    \begin{figure}
    \centering
    \includegraphics[width=0.49\textwidth,valign=t]{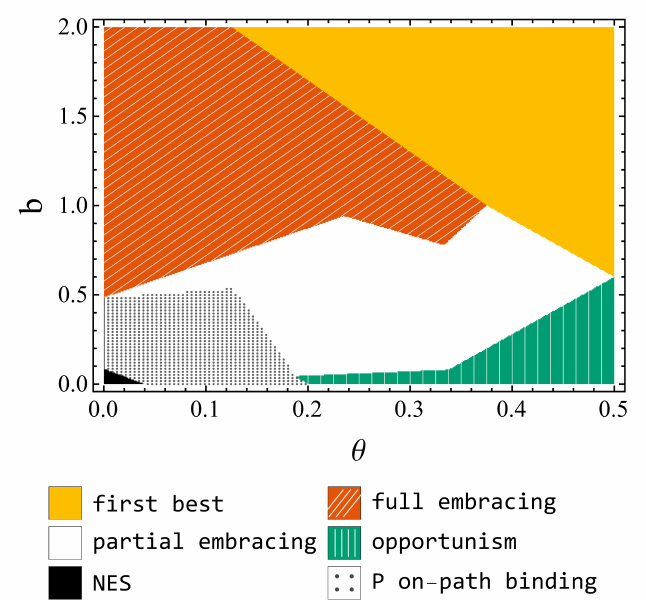}
    \includegraphics[width=0.49\textwidth,valign=t]{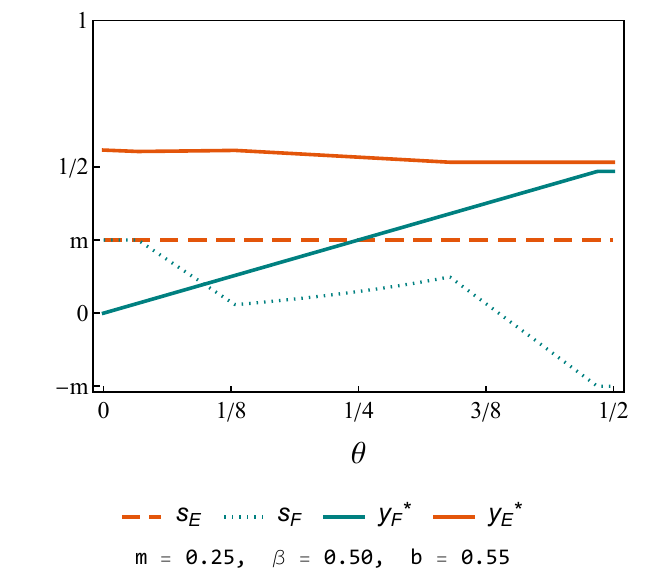}
    \caption{\textbf{Embracing Phase Without Commitment.} \emph{Left:} Regimes in the $(\theta,b)$ space. For low $(\theta,b)$, only the NES contract is feasible. Except in the dotted region, the binding constraint for the principal is mainly her ability to punish the friend, not to uphold her on-path promises. Commitment is irrelevant for large $(\theta,b)$. The full embracing region may be non-convex because the principal's punishment value is non-monotone in $\theta$.\\
    \emph{Right:} Equilibrium strategies as a function of $\theta$ for given $(m,\beta,b)$. Relative to the commitment case, the friend requires more endorsement, which reduces the enemy's moderation. Kinks in $s_F$ arise (in order) as $F$'s \eqref{Eq Agent DEC} binds on-path, $P$'s penal code switches to NES, and $P$'s commitment becomes irrelevant. The non-monotonicity in $s_F$ reflects the non-monotonicity of the principal's punishment value.}\label{fig:Nocommitment}
    \end{figure}

We now (behaviorally) characterize features of the optimal non-commitment contract. To highlight the role that commitment plays in our model, we relate our findings directly to those of the commitment case.

We begin by showing that the logic on the leadership rent, $b$, from \cref{prop:theta0nocommitment} extends to the general setting: when $b$ is large, whether the principal can(not) commit does not matter. The same holds for \emph{any} $b$ if the principal's bliss point is sufficiently close to $1/2$.

\begin{proposition}\label{prop:thresholdforcommitment}
    For any $(\beta,m)$, there exist thresholds $\hat{\theta}(b)<1/2$  and  $\overline{b}(\theta)>0$ such that the optimal non-commitment contract is (observationally) identical to the optimal commitment contract if $\theta > \hat{\theta}(b)$ or $b > \overline{b}(\theta)$.
\end{proposition}

The result hinges on the credibility of punishments: First, if the principal is balanced, she has almost no intrinsic preference between parties and can therefore credibly punish both. Second, if $b$ is large, the parties are highly responsive to the threat of losing endorsement and thus moderate their policies on the equilibrium path, benefiting the principal. Through the back-to-business property, these benefits provide the necessary carrot that makes the principal willing to incur the cost of a severe punishment phase against a deviating friend. In turn, precisely because a harsh punishment against the friend is credible, a harsh punishment for the principal---under which the friend is willing to choose policies to the right of the principal's bliss point---is also credible. Taken together, these harsh off-path punishments ensure the principal's \eqref{Eq:PrincipalDEC} has slack on the equilibrium path, even in periods when she endorses the enemy. Comparing the right panel of Figure \ref{fig:Nocommitment} to that of Figure \ref{fig:commitment} illustrates that for $\theta>3/8$, indeed the commitment solution is feasible.

This result does not tell us, however, whether the qualitative predictions of the commitment case---namely, the four \emph{regimes} of full embracing, partial embracing, opportunism, and first best---survive for interior $\theta\in (0,1/2)$. The next result states that for each regime, there exists a (non-knife-edge) region in which commitment plays no role. This is not a folk-theorem type of result in which direct incentives become irrelevant, and any outcome can be sustained regardless of commitment.  Instead, for each regime, there is a non-trivial set of parameters for which the same incentives discussed in the commitment case are at play, and the same contract is optimal.

\begin{proposition}\label{prop:regimesnocommitment}
    Fix any $(\beta,m)$. For any regime that is optimal under commitment, there exists a region of parameters $\tilde{\Theta}\times \tilde{B}$, with a non-empty interior relative to the parameter space, $\Theta\times B$, such that the optimal non-commitment contract is observationally identical to the optimal commitment contract if $(\theta,b) \in \tilde{\Theta}\times \tilde{B}$.
\end{proposition}

The intuition behind this result comes from the analysis at the boundaries of the $\Theta$ space. At $\theta=0$, \cref{prop:theta0nocommitment} establishes that, generically, either the principal has slack on her (on- and off-path) constraints, or only the NES contract is feasible. Likewise, \cref{prop:thresholdforcommitment} applies close to $\theta=1/2$, where these constraints have slack, too. Continuity of payoffs then implies that, for $b> \overline{b}_0$, around $\theta=0$ the optimal contract close to these boundaries is unaffected by the principal's ability to commit.

However, as we move toward the interior of the $\Theta$ space, lack of commitment becomes more consequential. Although moving in from either side can increase the principal's payoffs under the commitment contract (see also \cref{prop:non-monotoneWelfare}), the principal's optimal penal code also becomes less severe. This weakens the principal's ability both (i) to punish the friend, and (ii) to embrace the enemy on path. Each problem jeopardizes the sustainability of the commitment contract, and both recur throughout the analysis.

Having discussed cases where the commitment contract survives entirely, a logical next step is to ask when our regimes survive qualitatively. For instance, we may wonder whether in the no-commitment case sharp cutoffs exist that partition the $\Theta$ space into different regimes, similar to $\underline{\theta}$ and $\overline{\theta}$.

It turns out that such cutoffs exist and generate dynamics resembling those of the commitment contract. Broadly, we can still identify the first-best region, as well as our three second-best regimes (full embracing, partial embracing, and opportunism). In addition, we have the NES region, where cooperation breaks down entirely. The left panel of Figure \ref{fig:Nocommitment} illustrates our qualitative regions.

For full embracing, the intuition is identical to that of the commitment case, but the region shrinks because the friend's punishment is milder. Moreover, the frontier of the full embracing region comes from the friend's (\ref{Eq Agent DEC}), and not from the principal's (\ref{Eq:PrincipalDEC}). Hence, as one leaves full embracing, the principal begins to partially endorse the friend, providing him with more incentives to moderate. We see then that, absent first best, the optimal contract transitions from full to partial embracing.

The logic of partial embracing is more subtle. The extent and the reason why the principal partially embraces the enemy depend on which of her constraints is first-order: the off-path constraint when punishing the friend or the on-path constraint when embracing the enemy. If \eqref{Eq:PrincipalDEC} binds only off-path, the logic is as in the commitment case. Expected moderation by the enemy alone is insufficient to induce the friend to moderate up to the principal's bliss point, so the principal supplements it with just enough endorsement to satisfy his (\ref{Eq Agent DEC}) with equality (i.e., $s_F^\ast < m$ while $s_E^\ast = m$). This corresponds to the white region in the left panel of Figure \ref{fig:Nocommitment}. However, in the dotted region of that panel, the principal's \eqref{Eq:PrincipalDEC} also binds on the equilibrium path. The principal is constrained when embracing the enemy---after all, an enemy in power is always short-term worse than a friend in power. In this case, partial embracement occurs \emph{after} the enemy leads, not to accommodate the friend's incentive problem, but to accommodate the principal's own. This implies that there may be situations in which both $s_E^\ast$ and $s_F^\ast$ are interior, so we need to broaden our definition of the partial embracing regime when thinking about the no-commitment case. Comparing the right panels of Figures \ref{fig:commitment} and \ref{fig:Nocommitment} provides a visual illustration of how the optimal contract changes in that region.

The logic of opportunism is unchanged from that of the commitment case. The only difference is that, because the friend's punishment may be milder than the enemy's, parties may moderate asymmetrically. In the interior of the opportunism region, the principal's \eqref{Eq:PrincipalDEC} (generally) has slack along the equilibrium path, even in periods in which she endorses the enemy. As $b$ decreases, and so does equilibrium moderation, the principal's \eqref{Eq:PrincipalDEC} may eventually start binding, at which point the contract transitions to the dotted region of partial embracing. To satisfy her \eqref{Eq:PrincipalDEC}, the principal can no longer credibly promise full endorsement to the enemy
 after he led. Instead, she can offer at most partial endorsement, so that $s_E^\ast < m$ and $s_F^\ast = -m$. This weakens the cross-subsidization effect, putting additional strain on the principal's incentives. As $b$ falls further, her endorsement becomes increasingly tilted toward the friend, until NES is her only remaining option. Because contract adjustments are marginal throughout, if the optimal contract ever reaches the NES region, it smoothly converges to it as $b$ decreases.

While it is possible to characterize the cutoffs defining each regime, these thresholds are less informative than under commitment. One central issue is that they are sensitive to the precise structure of the optimal penal codes at each point and thus involve a set of case distinctions. We therefore refrain from providing explicit formulas for them.\footnote{We provide code that generates a complete symbolic characterization in the supplementary material. Online Appendix \ref{sec:MoreNoCommitment} provides some technical results resembling our discussion here.}

To illustrate the additional complexities introduced by non-commitment, it is useful to consider the full embracing regime. We can show that, relative to the commitment benchmark, the full embracing region (weakly) shrinks. Moreover, for any given $(m,\beta,\theta)$ under which full embracing arises with commitment, there exists a threshold $b_\theta$ such that full embracing survives whenever $b \geq b_\theta$. Both results can be easily seen by comparing the left panels of Figures \ref{fig:commitment} and \ref{fig:Nocommitment}. Importantly, however, this threshold $b_\theta$ is non-monotone in $\theta$: Full embracing may be optimal up to some $\theta'$, partial embracing may then emerge for $\theta\in(\theta',\theta'')$, and full embracing may reappear for $\theta\in[\theta'',\underline{\theta}]$. The reason is the non-monotonicity of the principal's punishment value, which makes her punishment milder for intermediate values, in turn weakening the punishment of the friend. As a result, there may exist an interval $\theta\in(\theta',\theta'')$ where the friend is unwilling to moderate up to the principal's bliss point without receiving endorsement. In contrast, for $\theta<\theta'$ and $\theta>\theta''$, punishments are sufficiently harsh to discipline the friend under full embracing.

  To conclude the analysis, comparing our results with \cref{prop:theta0nocommitment} gives a clear intuition: the comparative statics on $b$ are less drastic when the principal is balanced. There are two reasons for this: First, because the friend is often at his \eqref{Eq Agent DEC}, the optimal contract adjusts even if the principal has no on-path commitment problem. Second, even when the principal has a commitment problem both on- and off-path, cross-subsidization may still allow for a better contract than NES.

\section{Implications}\label{sec:implications}

In this section, we discuss the empirical predictions our model delivers for political economics. Although our model is deliberately stylized to isolate the core drivers of our mechanism, it captures key features of parliamentary coalition formation, or the behavior of interest groups. We organize the section as follows: we first address our main modeling choices. Second, we formulate three main empirical predictions, each illustrated with direct evidence from historical cases. Lastly, we turn to broader, more indirect, empirical predictions of our model.

\subsection{Discussion of the Model}

\paragraph{Community of Fate.} A defining feature of our model is that players form a community of fate: They are long-lived and tied together through an organization.  Policy decisions are payoff-relevant to everyone at any time, regardless of who takes them, and no one can be fully excluded. This differs from classical contracting problems, where replacement from a pool of identical players is feasible and the replaced player exits receiving an outside option. In a political system, replacement is difficult. While individual stakeholders may be replaced, the ideas behind parties, factions, or lobby organizations are there to last.

Absent the community-of-fate assumption, our results would be less interesting. If, say, the principal could (at no cost) replace any party with a (closer) friend, she would naturally do so. An interpretation of our discount factor is that with probability $\beta$ the principal is stuck with the current parties and needs to find a way to work with them. A weaker version of replacement is to somehow eliminate the enemy, e.g., by banning the party through a constitutional court. We will show (\cref{prop:competition}) that, often, it is not in the principal's interest to do that.

\subparagraph{Delegation and Limited Power.} Delegation of policy choices is central to our setting. A centrist party selects coalition partners who will then shape policy from their cabinet posts, while an interest group provides electoral support to a party that then approves legislation.\footnote{Delegation problems in coalitions are commonly described as \textit{ministerial drift} \citep[see][for an introduction to the literature]{MartinVanberg2004}.}

However, exogenous forces curb the authority of these power brokers. Unlike in standard delegation problems, our principal can influence the selection process but cannot fully control it. As a result, the principal needs a plan to deal with an enemy in power. Our results show that it is in her best interest to address that issue proactively.

We believe this assumption captures a feature common to most competitive political systems. A centrist party in a multi-party system, for instance, sometimes acts as a kingmaker, determining whether the left or right bloc enters the government. Other times, the voters have given power directly to a certain bloc.\footnote{In \cref{sec:kingmaker} we discuss a literal interpretation of the kingmaker model.} Similarly, interest groups may support parties that ultimately lose at the polls.

\subsection{The Cordon Sanitaire}\label{subsec:predictions}

A particular application of our setting is the entry of radical newcomers into parliament. Our results provide a theory of the dynamics of the so-called cordon sanitaire that mainstream parties may set up around radical newcomers by ostracizing them politically.

\paragraph{The Ally Principle vs. the Cordon Sanitaire.} Our embracing the enemy result stands, at first glance, in stark contradiction to the Ally Principle, which posits that a principal should hand power to her closest ally whenever possible. Considering the dynamics of our model, what we obtain is a temporary Ally Principle---the cordon sanitaire. It is optimal to collaborate with your closest ally as long as the enemy is away from power. However, once the enemy acquires agency, embracing him is optimal.

The Ally Principle has been challenged using arguments of face washing, power sharing, and the use of delegation as a way of committing to a policy. Within economic theory, \citet{bendor2001theories} and \citet{callander2008theory} demonstrate, for example, how the Ally Principle can be invalidated if information acquisition is costly for the agents. Our mechanism differs. The principal starts to delegate power to an enemy because her power to exclude him is limited. As a result, our theory offers a clear-cut prediction of \textit{when} the Ally Principle holds, and \textit{when} it will be abandoned in favor of embracing.

\paragraph{The Cordon Sanitaire in Reality.}\label{subsec:cordon} The sustainability of cordons sanitaires is a central question in recent European politics. Although the empirical literature is still developing, existing studies of European populist movements are consistent with two of our main findings: First, parties that were initially ostracized often become regular coalition partners once they enter government and moderate their policies (see \citet{van2007party} and \citet{bichay2026converging}). Second, \citet{albarello2024coalition} documents that moderation is greater when formerly ostracized parties are incorporated by more centrist partners.

Our theory also yields clear predictions on the dynamics followed by the embrace of the enemy, addressing both \emph{when} a cordon sanitaire breaks and the \emph{extent} of the subsequent embrace. To illustrate the empirical relevance of these two additional findings, we discuss two historical cases. The first concerns the stabilization of the German Weimar Republic, showing how a cordon sanitaire ended at a particular juncture---namely, when the enemy gained agency for the first time.

\subparagraph{Case Study: The Stabilization of the Weimar Republic.} The proportionality of the Weimar electoral system and the highly fragmented electorate gave a persistent pivotal role to the centrist parties---the Zentrum and the German People's Party (\emph{DVP})---in the formation of all conceivable coalitions, either to their left with the Social Democratic Party (\emph{SPD}), or with the German National People's Party (\emph{DNVP}) to their right.\footnote{Our model abstracts from the fact that voters react to the parties' policy choices, thus affecting the principal's power in the future. During the 1920s, the centrist parties retained their pivotal role and faced a relatively stable menu of coalition options despite high electoral volatility.} The DNVP rejected the republican constitution of 1919 and the Treaty of Versailles, and was accordingly excluded from government throughout the early 1920s. The exclusion of the DNVP meant that the centrist parties had to rely on the SPD, which belonged to a very different political tradition and, although committed to the republican constitution, pursued markedly different economic policies.

In 1924, the constitutional amendments required to ratify the Dawes Plan made the DNVP's votes necessary. This necessity led to a bridge-building attitude from the centrist parties.  As \citet{peukert1993weimar} summarizes, the DVP leader Gustav Stresemann secured ratification of the Dawes Plan by \enquote{throwing out hints that there might be a place for the DNVP in the governing coalition.} After the plan was ratified, Stresemann
     \begin{quote}
       was conciliatory towards the DNVP, arguing that having been brought to accept realities, it should be allowed a share in government [\ldots] He argued that bringing the DNVP into government would be an act of statesmanship. No party, he said, had more reason than they to dislike the DNVP: ``But one cannot make domestic policy with sentimentality.'' \citep[296ff.]{wright2002gustav}
    \end{quote}

The ratification of the Dawes Plan marked the stabilization of the Weimar Republic, a period of DNVP moderation. The DNVP joined the government and, to maintain its position, implicitly recognized the Weimar constitution, renewing the \enquote{law on the protection of the Republic---originally passed as an emergency measure against `the enemy on the right'} \citep{maier2015recasting}. The DNVP leaders understood that participation in government allowed them to deliver benefits to their constituencies, particularly through protective tariffs for farmers. Until the collapse of the Weimar system, several power swings occurred in the Reichstag with the centrist parties alternating their collaboration with both the SPD and the DNVP. The Luther I cabinet (1925) and Marx IV cabinet (1927--28) included the DNVP, while the Marx III cabinet (1926--27) relied on SPD toleration, and the MÃ¼ller II cabinet (1928--1930) included the SPD.\footnote{\Cref{tab:weimar_coalitions} in Online Appendix \ref{app:coalitions} tracks the principal's options. It shows that the principal indeed exercised agency as both options were arithmetically feasible at times.}

The stabilization effectively ended largely due to forces exogenous to our setting. The Great Depression fueled the rise of the Nazi and Communist parties, leaving none of the above coalition options able to command a majority. Shortly before that, the DNVP grassroots radicalized, culminating in the ousting of the leadership that had presided over the moderation process. Within a year, Stresemann died.

\paragraph{The extent of the embrace.} While our prediction for \emph{when} the cordon sanitaire breaks down is independent of the parameterization, behavior within the embracing phase is not. To connect our qualitative results to observable behavior, we focus on the cleanest divide of our theory: whether the principal fully endorses the enemy during the embracing phase or not. The next proposition characterizes how the cutoff $\underline{\theta}$ varies with the parameters under commitment.

\begin{proposition}\label{prop:comparativeStatics}
The full-embracing region weakly expands in $b$, and $\beta$. With commitment, it also expands in $m$.
\end{proposition}

Therefore, conditional on the enemy having gained agency at least once, full-embracing behavior should be more likely in settings where office-holding is more valuable ($b$), the political environment supports longer relational horizons ($\beta$), and, at least for the commitment case, where the principal is stronger ($m$).

To relate these results to real-world settings, we briefly introduce a second historical example, drawn from postwar Italy. As we elaborate below, this example reflects a more stable environment, with a principal who is a stronger power broker and substantially closer to the friend.

\paragraph{The \emph{Apertura a Sinistra}.} In postwar Italy, the Christian Democratic Party (\emph{DC}) governed as the dominant centrist party alongside two smaller allies, the Republicans (\emph{PRI}) and Social Democrats (\emph{PSDI}). This centrist bloc faced a choice between two potential coalition partners: the Liberal Party (\emph{PLI}) to its right and the Socialist Party (\emph{PSI}) to its left. The ideological distance between the DC and the PSI was especially pronounced in the first postwar decade, when the Socialist leader Pietro Nenni maintained a unity-of-action pact with the Communists. Accordingly, throughout the 1950s, the DC excluded the PSI from government, forming a center-right coalition that commanded razor-thin majorities. When even this formula proved insufficient, DC minority governments relied on \emph{de facto} support from neo-fascist deputies of the Italian Social Movement (MSI)---a dependence that reached its crisis point with the Tambroni government. Tambroni's reliance on MSI votes triggered nationwide anti-fascist protests and violence, prompting the DC leadership to terminate the government. As one party leader put it, the party ``no longer had the choice'' of allying rightward \citep{leonardi1989italian}.\footnote{External changes also facilitated a leftward turn: the Kennedy administration reversed its opposition to Socialist participation, and Pope John XXIII adopted a stance of non-interference in coalition politics.}

The opening to the left proceeded in several steps. In 1960, the PSI abstained on a DC's minority government led by Amintore Fanfani, and in 1962, the PSI provided external support. In December 1963, Aldo Moro formed the first center-left cabinet with PSI ministers. The DC leadership understood the coalition shift in a way consistent with our embracing strategy. They intended to attract the Socialists with ``the trappings of office and the chance of carrying through some reforms,'' while retaining the threat of expulsion should the PSI refuse to moderate \citep{ginsborg1990history}. Nenni made significant concessions on NATO, and their initially ambitious reform program was progressively diluted. As the Christian Democrat Carlo Donat-Cattin observed, the DC leadership had turned the center-left governments into
\begin{quote}
a `power arrangement' in which the Socialists had been substituted for the Liberals, but in substance nothing had changed in the way that power was managed. \citep{leonardi1989italian}
\end{quote}
The Socialists became the regular coalition partner, and the Liberals did not return to government for almost a decade. This pattern is especially striking because parliamentary arithmetic made both formulas viable throughout the 1960s and early 1970s. Yet the DC consistently chose the center-left.\footnote{ \Cref{tab:italy_coalitions_1,tab:italy_coalitions_2,tab:italy_coalitions_3} in Online Appendix \ref{app:coalitions} track the principal's options.}

\paragraph{Comparing the Two Historical Examples.} Both examples exhibit embracing and moderation, but differ in the extent of the embrace. In the Weimar Republic, the centrist bloc alternated between the DNVP and the SPD within the space of three years. The Italian Christian Democrats, however, governed with the PSI for almost a decade before briefly restoring a coalition with the PLI. These contrasting patterns are consistent with our comparative statics from \cref{prop:comparativeStatics}. The first, and most important, distinction concerns the principal's relative ideological distance from the friend. In Weimar Germany, the SPD was also far from the centrist bloc, so the centrists needed to keep the incentives for moderation on both sides of the spectrum. In Italy, in contrast, the Liberals were significantly closer to the DC than the PSI. Second, the stability of the political environment and the influence of the principal differed. The Italian Republic, upheld by the American international order and anchored in the DC's solid electoral base, offered a longer effective time horizon and a more influential principal than the fragile Weimar system, which was buffeted by reparations, economic instability, and rapid partisan realignment. Third, the value of holding office was markedly different. Postwar Italian governments presided over the \emph{trente glorieuses}---a period of rapid economic growth in which office carried substantial rents, including widespread opportunities for patronage. In Weimar Germany, by contrast, office meant administering deeply unpopular austerity and reparations.

\subsection{Broader Implications}

We now address the broader implications of our model. We discuss how to identify relevant parameters or how, if the parameters are known, they map into economic predictions about behavior.

\paragraph{Principal's Welfare.} We show that a principal's payoff is highest when she exhibits some bias toward a party, but the bias is not extreme.
\begin{proposition}[\textbf{Principal's Payoff}]\label{prop:non-monotoneWelfare}
        Consider $(\beta,m,b)$ such that the first best is unattainable for any $\theta$. Then the principal's ex-ante payoff is non-monotone in $\theta$ with an interior maximum $\theta^{\max}\in (0,1/2)$. If the principal has commitment power that maximum is unique and at $\theta^{\max}=\overline{\theta}<\frac{1}{2}.$
\end{proposition}

The intuition for this result can be most easily seen in the commitment case. We start with an extreme principal, $\theta=0$, and gradually make her more balanced. For $\theta \leq \overline{\theta}$, the friend keeps choosing $y_F^\ast =\theta$. At the same time, the enemy both gets mechanically closer to the principal and---through the cross-subsidization effect---has stronger incentives to moderate. However, for $\theta>\overline{\theta}$, the optimal contract is opportunistic (\cref{res:OEndorsement}) and does not change as $\theta$ increases. Then moving the principal's bliss point toward the center makes her worse off when $F$ leads and better off when $E$ leads. However, because of the exclusion phase, alignment with the friend is more valuable for the principal---so her ex-ante payoff decreases in $\theta$.
Removing the principal's ability to commit does not change that intuition. The only difference is that single-peakedness is not guaranteed because the principal's punishment payoffs are non-monotone in $\theta$.

\paragraph{The Value of the Enemy.} We now ask whether the principal benefits from the existence of the enemy.  Absent the enemy, the principal loses the ability to moderate her friend.

To state our result formally, first let us define \[\Upsilon(m,b):=\left\{(\beta,\theta) \text{ such that } \theta<\max\left\{1-\overline{\theta},1-\breve{\theta}\right\}\right\},\]
which is the set of $(\beta,\theta)$ for a given $(m,b)$ such that the first best is not attainable (see \cref{result:firstbest}). We say that \emph{the enemy is valuable} if, ex ante, the principal prefers our baseline game to a benchmark without $E$, in which $F$ has full agency in every period.

        \begin{proposition}[\textbf{The Value of the Enemy}]\label{prop:competition}
            Fix the parameters $(m,b)$ such that $\Upsilon(m,b)$ is non-empty. The following holds:
            \begin{itemize}
                \item For any $\beta$, there exists a $\tilde{\theta}$ such that $(\beta,\tilde{\theta}) \in \Upsilon(m,b)$ and the enemy is valuable for the principal if $\theta\geq\tilde{\theta}$.
                \item If $2b\geq \frac{1-2m}{2m}$, then for any $\theta >0$, there exists a $\tilde{\beta}$ such that $(\tilde{\beta},\theta) \in \Upsilon(m,b)$ and the enemy is valuable for the principal if $\beta\geq\tilde{\beta}$.
            \end{itemize}
        \end{proposition}

The enemy's presence allows the principal to discipline both parties. The cost is, naturally, that the enemy occasionally chooses the policy. Intuitively, the more balanced the principal, the lower that cost, and the greater her need to discipline the friend. If players are patient enough, the principal's ability to moderate is strong. Therefore, she prefers a competitive setting, even when she is almost aligned with the friend.

\cref{prop:competition} offers a new perspective on why a polity may endure as a ``community of fate'', where factions fundamentally disagree over policy directions, yet none is permanently excluded. Conventional explanations are that exclusion is too costly---e.g., because it requires violence---or precluded by institutional constraints. Our result suggests an additional mechanism: the principal may preserve political rivalry as it induces both sides to moderate in equilibrium.

Removing commitment decreases the value of the enemy because the friend cannot be moderated as effectively. Yet, the sufficient condition in \cref{prop:competition} remains valid.

\section{Conclusion}\label{sec:conclusion}

	We provide a theory of why a weak power broker may want to embrace her enemy: Moderating the enemy's policy choices is better than trying to ensure her friend is selected more often. Moreover, the principal gains from having an enemy that challenges her friend, even if that implies some periods of worse policy decisions for her. The principal's bias against the enemy affects both her benefits from these relationships and the role commitment plays in them.

	There are, naturally, several aspects of reality our model ignores. Among these are evolving preferences, the monitoring of the principal's endorsement, and the potential role of asymmetric information. We leave those to future research.

\part*{Appendix}
\appendix

\paragraph{Notation: Histories.} A history $h$ is a vector describing past actions and realizations. The initial node is given by $h_0\equiv\emptyset$. We say $h \subseteq h'$ if history $h'$ is reached in a continuation game of $h$. If, in addition, $h \neq h'$, then $h \subset h'$. The set of histories $h$ in which party $k$ has been the only one leading is $\mathcal H_k$. The set of histories $h$ where $k$ leads for the first time at $h$ is $\mathcal H_k^\ast$.
To ensure readability, we abuse notation and suppress within stage game histories. For example, we write $y_k(h)$ meaning party $k$'s policy when selected in the stage game starting at history $h$ following ``on-path play'' by the principal in her choice $s(h)$. This allows us, for example, to compare the mutually exclusive $y_F(h)$ and $y_E(h)$ in a compact manner. When it is clear from the context what we mean, we drop the history argument for a more compact formulation describing on-path stationary (continuation) strategies.

\section[Appendix: Useful Lemmas]{Useful Lemmas}
    \begin{lemma}\label{Lem:ylsmalleryR}\addcontentsline{toc}{subsection}{Lemma \thelemma}
        Take any contract with ex-ante value $w_P(h_0)$ to the principal. Then there exists a contract with value at least $w_P(h_0)$ such that $y_F(\cdot)\leq y_E(\cdot)$.
    \end{lemma}

    \begin{proof}
        Consider a contract $C$ where at some history $h$, $y_F(h) > y_E(h)$. At this history, the principal's selection decision is $p(s(h))$, and the principal's continuation value before selection is $w_P(h)$.
        Suppose now that, in addition, either $0<y_F(h) < \theta$ or $1>y_E(h)>\theta$. Consider a contract $\widetilde{C}$ that is identical to $C$ except that at history $h$ it specifies \[\tilde{y}_F(h)=y_F(h) - \varepsilon \text{ and } \tilde{y}_E(h)=y_E(h) + \frac{1- p(s(h))}{ p(s(h))}\varepsilon\] for some $\varepsilon>0$. Because $|\tilde{y}_k-\theta_k|\leq|y_k-\theta_k|$, $\tilde{C}$ is a contract. Moreover, at history $h$ the principal's value is identical because
        \[
            \begin{split}
                \tilde{w}_P(h)-w_P(h) &= - \Big( p(s(h))(|\tilde{y}_E(h)-\theta| - |y_E(h)-\theta|) + (1-p(s(h)))(|\tilde{y}_F(h)-\theta| - |y_F(h)-\theta|) \Big) \\
                &= - \Big( (1-p(s(h)))\varepsilon-(1-p(s(h)))\varepsilon \Big) = 0.
            \end{split}
        \]
           because the net-present value to the principal of the game \emph{after} completing this period's stage game is constant by construction.\footnote{The ``present'' here is node $h$.} Now consider $ y_E(h) \leq \theta \leq y_F(h)$. As before, take a $\widetilde{C}$ that is identical to $C$ but with
        \[
            \tilde{y}_F(h)= \max\left\{\theta- \frac{p(s(h)) }{1- p(s(h))} (\theta - y_E(h) ),0\right\} \text{ and }
         \tilde{y}_E(h)=  \min\left\{\theta  + \frac{1-p(s(h)) }{ p(s(h))} (y_F(h)-\theta),1\right\}.
         \]
        Thus, $\tilde{C}$ is a contract with $\tilde{y}_F(h)\leq \tilde{y}_E(h)$, and $P$'s value (weakly) increases because
        \[
            -p(s(h))(\tilde{y}_E(h) - \theta) - (1-p(s(h)))(\theta - \tilde{y}_F(h))
            \geq -p(s(h))(\theta- y_E(h)) - (1-p(s(h)))(y_F(h) - \theta)
            \]
            which implies $\tilde{w}_P(h) {\geq} {w}_P(h)$. Unless either policy is at its extreme, $\tilde{w}_P(h)=w_P(h)$.
    \end{proof}

    \begin{lemma}\label{lem:DEC}\label{lem:noovershoot}\addcontentsline{toc}{subsection}{Lemma \thelemma}
        The optimal on-path policies satisfy $y_{E}(h)\geq\theta\geq y_{F}(h)~ \forall h$.
        It is without loss of generality to focus on contracts in which, if a party gets to choose $y$ on path, (i) either he chooses the principal's bliss point $y_{k}=\theta$, or (ii) \eqref{Eq Agent DEC} binds.
    \end{lemma}

    \begin{proof}
        \textbf{First Statement: $y_{E}\geq\theta\geq y_{F}$.}
        Fix a contract $C$ such that $y_{E}(h)<\theta$ for some on-path history $h$. We will augment the contract in (at most) two steps to construct an alternative that the principal prefers. First, consider an identical contract, apart from increasing $y_{E}(h)$ marginally. That contract improves the principal's payoff at node $h$ by $dy_{E}$, relaxes $E$'s \eqref{Eq Agent DEC} in all histories $h'\subseteq h$, and leaves them unchanged in all periods $h'' \supset h$. It is thus incentive compatible for $E$. Now consider the last history $\hat{h} \subset h$ in the sequence leading up to $h$ in which $F$ leads, if such a history exists. The increase of $y_{E}(h)$ has tightened $F$'s \eqref{Eq Agent DEC} by $\delta(h|\hat{h}) dy_{E}$, where $\delta(h|\hat{h})$ is the discounted probability of reaching $h$ from $\hat{h}$.\footnote{\label{fn:delta}Formally, if $h$ is a period-$t$ history and $h'$ a period-$t'$ history, then $\delta(h|h'):=\beta^{t-t'} Pr(h|h')$.} Thus, augmenting the contract once again by reducing the prescribed $y_{F}(\hat{h})$ by at most $\delta(h|\hat{h}) dy_{E}$ restores \eqref{Eq Agent DEC} of $F$ without violating $E$'s \eqref{Eq Agent DEC} anywhere. Note that if $y_F(\hat{h})>0$, the perturbation is always possible; if $y_F(\hat{h})=0$, we take the next ``earlier history'' at which $y_F>0$ (note that if there is any issue with F's \eqref{Eq Agent DEC} binding, such a history must exist). Now, observe that the first and second augmentation jointly are at worst ex-ante payoff neutral to $P$ because at node $\hat{h}$ her gains and losses equate. Hence, assuming $y_{E}(h) \geq \theta$ is without loss. The case for $\theta\geq y_{F}$ is analogous.\medskip

        \paragraph{Second Statement: Either \eqref{Eq Agent DEC} binds or $y_k=\theta$.}
        \begin{definition}\label{def:firstH}
            History $h$ with property $\Pi$ is a \emph{first history with property $\Pi$} if no history $h' {\subset} h$ has property $\Pi$. It is a \emph{last history with property $\Pi$} if no $h'\supset h$ has property $\Pi$.
        \end{definition}

        \subparagraph{Step 1. $k$'s dynamic enforcement constraint upon first selection.}

        Take an arbitrary contract $C$ with $y_E(\cdot)\geq y_F(\cdot)$ and assume there is a first history in which $k$ leads, $h \in \mathcal{H}_k^\ast$, and $k$'s \eqref{Eq Agent DEC} has slack. Then consider a contract $\widetilde{C}$ identical to $C$ apart from $\tilde{y}_k(h)$ which is chosen such that $|\tilde{y}_k(h)-\theta| < |y_k(h)-\theta|$ and $k$'s \eqref{Eq Agent DEC} at $h$ is not violated. Since $\tilde{y}_k$ implies greater moderation, $\tilde{C}$ is strictly preferred by $P$ and $-k$. Now, either $\tilde{C}$ exists for $\tilde{y}_k(h)=\theta$, or there is a $\tilde{y}_k(h)\neq \theta$ such that $k$'s \eqref{Eq Agent DEC} binds.
        We iterate this procedure for any $h \in \mathcal{H}_k^\ast$ and obtain that an optimal contract exists in the class $\mathcal{C}^1$ in which the second statement holds for any $k$ and $h \in \mathcal H_K^\ast$.
        \subparagraph{Step 2. $k$'s dynamic enforcement constraint upon re-selection.}

        Consider a contract $C^1$ in the class of $\mathcal{C}^1$ and assume that there is at least one history $h$ such that $k$'s \eqref{Eq Agent DEC} has slack. Take the first on-path history $h$ with that property.

        Consider then a candidate contract $C^2$ which is identical to $C^1$ with the exception that at history $h$, we pick the associated $\tilde{y}_k(h)$ such that either $\tilde{y}_k(h)=\theta$ or $k$'s \eqref{Eq Agent DEC} binds. Using the same arguments as in step 1, $C^2$ is incentive compatible for $P$ and $-k$. However, it may not be incentive-compatible for $k$ at a previous selection.

        To construct a contract incentive compatible for $k$, take the last on-path history $h'\subset h$ at which $k$ led before $h$. Consider now a contract $C^3$ identical to $C^2$, but with the difference that at history $h'$, party $k$ chooses $\tilde{y}_{k}(h')$ such that $|\tilde{y}_{k}(h')-y_{k}(h')|=\delta(h|h')|\tilde{y}_{k}(h)-y_k(h)|$ and $|\tilde{y}_{k}(h')-\theta_k|<|y_{k}(h')-\theta_k|$.\footnote{Recall the definition of $\delta(\cdot|\cdot)$ from Footnote \ref{fn:delta}. As in the proof of the first statement, if $|\tilde{y}(h')-\theta|$ cannot be chosen because $y_k(h')$ is too close to $\theta$, we can move to an earlier history $h''$ with the same argument and move $\tilde{y}(h'')$ at that history.} Such a $\tilde{y}_{k}(h')$ exists by construction and implies that $k$ decreases his moderation at history $h'$ by exactly the net-present value of his own increase in moderation at history $h$. Thus, $k$ is indifferent at history $h'$ between contracts $C^3$ and $C^1$ and, by construction, $k$'s \eqref{Eq Agent DEC} at node $h'\subset h$ holds in contract $C^3$. Further, since these concessions are equivalent in history-$h'$ net-present values, they are also equivalent at any $h'' \subset h'$. Two implications follow from this feature: first, $C^3$ is ex-ante payoff equivalent to $C^1$ for $P$ and second, $C^3$ is incentive compatible if the initial $C^1$ is.

        Applying Step 2 iteratively at any history in which $k$ leads implies the statement.
    \end{proof}

    We now derive the ex-ante worst payoff for a party.
     \begin{definition}[Grim Trigger punishment]\label{def:grim}
        The contract of party $k$ is said to be \emph{grim trigger} if it prescribes that the principal fully endorses $j\neq k$ in any continuation game and both parties choose their respective bliss points whenever leading, $y_F=0$ and $y_E=1$.
    \end{definition}

\begin{lemma}\label{lem:grimtrigger}\addcontentsline{toc}{subsection}{Lemma \thelemma}
In a contract with commitment, a party's worst ex-ante continuation payoff is attained by
grim-trigger punishment and equals
\[
\alpha:=\frac{(1/2-m)b-(1/2+m)}{1-\beta}.
\]
\end{lemma}

\begin{proof}
    Let $\underline{w}_k$ be party $k$'s worst ex-ante continuation payoff.
        Under grim-trigger punishment, the principal and the other party ensure that the probability of achieving the lower bound of the stage payoff, $-1$, is maximized. Moreover, in a punishment contract, when leading, the punished party's payoff is at least $b+\beta \underline{w}_k$.
    Thus, the lowest ex-ante payoff must satisfy
    \[\underline{w}_k\geq  (1/2-m)(b+\beta\underline{w}_k)+(1/2+m)(-1+\beta \underline{w}_k) ~\Leftrightarrow~ \underline{w}_k\geq \frac{(1/2-m)b-(1/2+m)}{1-\beta}=:\alpha.\]
    Grim trigger implies a payoff of $\alpha$ and therefore $\underline{w}_k=\alpha$.\qedhere
\end{proof}

    \begin{lemma}\label{lem:FullEndorsment}\addcontentsline{toc}{subsection}{Lemma \thelemma}
        In any optimal contract with commitment in which for some $h$, $y_E(h)\neq \theta$, the principal either fully endorses $E$ at any history $\eta \supseteq h$ or $F$'s \eqref{Eq Agent DEC} binds at some $\eta \supseteq h$. In any optimal contract without commitment, the same holds unless $P$'s \eqref{Eq:PrincipalDEC} binds.
    \end{lemma}

\begin{proof}
    Consider some optimal contract $C$ and consider a first history $\hat{h}$ of that contract at which $E$'s \eqref{Eq Agent DEC} holds with equality and $y_E(\hat{h})\neq \theta$. Now assume the contract specifies $s(\check{h})\neq m$ for some history $\check{h}\supset \hat{h}$.

    We first show that unless $F$'s \eqref{Eq Agent DEC} binds at some $\eta \supset \hat{h}$, the principal has a strict incentive to increase $s(\check{h})$ contradicting optimality. This proves the commitment case.

    To see this, observe that a \emph{ceteris paribus} increase in $s(\check{h})$ increases $E$'s continuation payoff at $\check{h}$ by a \emph{policy} term (a change in the average policy $-1/(1-\beta)$ allocated to $E$ moving forward from $\check{h}$) and an \emph{leadership} term (a change in the fraction of $b/(1-\beta)$ allocated to $E$ moving forward from $\check{h}$). That means, the previous payoff $w_E(\check{h})$ increases to $w_E(\check{h})+ \gamma+\omega$ where $\gamma> 0$ is the policy term and $\omega>0$ the leadership term. Moreover, by \eqref{lem:noovershoot} and the fact that $P$ therefore always prefers a leading $F$, $P$'s continuation payoff at $\check{h}$ decreases at most by $E$'s policy term to $w_P(\check{h})-\gamma$.

    The realization of $\check{h}$ is uncertain, and parties discount the future by $\beta$. However, both parties face the same uncertainty. Thus, from the perspective of $\hat{h}$, both discount changes in the continuation payoffs at a future history $\check{h}$ in the same way from their history $\hat{h}$ perspective, implying history-$\hat{h}$ continuation payoffs of $v_E(\hat{h};E)+\hat{\gamma}+\hat{\omega}$ and $v_P(\hat{h};E)-\hat{\gamma}$, respectively. If, in addition, $\check{h}$ occurs with positive probability and in finite time given $\hat{h}$, both $\hat{\gamma}$ and $\hat{\omega}$ are positive.

    Being promised a strictly higher continuation payoff, $E$'s \eqref{Eq Agent DEC} no longer binds. He is thus willing to decrease $y_E(\hat{h})$ by $\hat{\gamma}+\hat{\omega}$---a pure policy effect translating to an increase in $P$'s payoff to $v_P(\hat{h};E)+\hat{\omega}$. Thus, we have constructed a contract that keeps $E$'s incentives in place, but is strictly preferred by $P$ at any history $h \subseteq \hat{h}$. An immediate contradiction to optimality of $C$ unless our construction violates $F$'s incentives.

    To see the result for the no-commitment case, observe that the only difference is that an increase in $s(\check{h})$ may now violate $P$'s \eqref{Eq:PrincipalDEC}. Unless it does, the argument above applies. That is, without commitment there cannot be any history $\check{h}$ such that $s(\check{h})<m$ unless either $F$'s \eqref{Eq Agent DEC} or $P$'s \eqref{Eq:PrincipalDEC} binds along the path from $\hat{h}$ to $\check{h}$.
\end{proof}

\section[Appendix: Example with Full Alignment]{Example with Full Alignment}

\subsection{Proof of Proposition \ref{prop:theta0}}

\begin{proof}
    Since $\theta= 0$, in any optimal contract $y_F=\theta$ with $F$'s \eqref{Eq Agent DEC} trivially satisfied. Before $E$ leads for the first time, i.e., for $h \in \mathcal{H}_F$, $P$'s endorsement has no effect on $E$'s incentives, so $s(h)=-m$. After the first time $E$ leads, i.e., for $h \not\in \mathcal{H}_F$, by \cref{lem:DEC}, either $E$ chooses $P$'s preferred policy $\theta$ or $E$'s \eqref{Eq Agent DEC} binds. Applying \cref{lem:FullEndorsment}, this implies $s(h)=m$. Solving for $y_E(h)$ using inequality \eqref{Eq Agent DEC} yields $y^\ast_E$.
\end{proof}
\subsection{Proof of Proposition \ref{prop:theta0nocommitment}}\label{sec:proof_theta0_nocommitment}
\begin{proof}
    When the commitment contract yields the first best, the result holds trivially. For the remainder, we prove the non-first-best case.
    \paragraph{First Statement} Consider the following contract:
    \begin{description}
        \item[On the equilibrium path.] The contract is identical to the optimal commitment contract described in \cref{prop:theta0}.
        \item[If $E$ deviated.] The continuation contract is the NES outcome, $(y_F=0,y_E=1,s=-m)$.
        \item[If $P$ deviated.] The continuation contract is stationary and given by $(y_F=\hat{y}_F,y_E=1,s=-m)$; $\hat{y}_F>0$ and makes $F$'s \eqref{Eq Agent DEC} bind.
        \item[If $F$ deviated.] The continuation contract has two phases: (i) punishment phase and (ii) back-to-business phase. The punishment phase lasts until the friend is selected for the first time (after his initial deviation from the on-path contract). During the punishment phase, every period's action profile is $(y_E{=}\hat{y}_E,s{=}m)$, where $\hat{y}_E$ makes $P$'s \eqref{Eq:PrincipalDEC} bind. Once we enter the back-to-business phase, the contract is identical to the on-path contract's embracing phase.
    \end{description}
The optimality of these penal codes is proved for the enemy and principal by \cref{lem:grimtrigger} and \cref{Lemma P Punishment 2}, respectively, and for the friend by \cref{lem:b2bFullEmbaracing} if the optimal on-path contract features full embracing.
    By construction, $P$ is indifferent in the punishment phase of $F$'s penal code between choosing $s=m$ and deviating to $s=-m$. Thus, we have that
    \begin{equation}\label{ProofProp2_underP}
        - p(m) \hat{y}_E+ \beta \underline{w}_P = \underline{w}_P=-\frac{p(m)\hat{y}_F+ (1-p(m))}{1-\beta}.
    \end{equation}
    The first equality comes from $P$'s indifference in $F$'s punishment, the second by the definition of $P$'s punishment payoff. Likewise, $F$ is indifferent in $P$'s penal code between choosing $y_F=\hat{y}_F$ or deviating to $y_F=0$ when leading. That is,
    \begin{equation}\label{ProofProp2_underF}
        -\hat{y}_F + \beta w_F^P = \beta \underline{w}_F.
    \end{equation}
    Next, observe that because all actions are taken (because $\theta=0$) to the right of the principal's bliss point, the friend's continuation payoff at any history is equal to that of the principal plus the friend's expected leadership rent. In particular, this implies
    \begin{equation}\label{ProofProp2_FriendPayoff}
       \underline{w}_F= \underline{w}_P+ \frac{1-p(m)}{1-\beta}b= \frac{-p(m) \hat{y}_E + (1-p(m))b}{1-\beta}
    \end{equation}
    with the last equality coming from substituting for $\underline{w}_P$ via \eqref{ProofProp2_underP}. This yields four linear equations: two from \eqref{ProofProp2_underP} and those from \eqref{ProofProp2_underF} and \eqref{ProofProp2_FriendPayoff}. If these equations have a solution, it is unique and yields, in particular,
     \begin{equation*}\hat{y}_E= 1- \frac{(2p(m)-1)(1-\beta(1+  p(m) b))}{p(m)(1-\beta)}.\label{eq:hatyr}\end{equation*}
   Now, recall that in the exclusion phase of the optimal contract, the principal has no commitment problem because she supports the friend anyhow. In the embracing phase, she aims to implement a contract that is identical to $F$'s penal code, apart from $E$ selecting $y_E^\ast$ instead of $\hat{y}_E$ in the punishment phase. We can find $\hat{y}_E\geq y_E^\ast$ if and only if
   \[b\geq \frac{(1 - \beta)^2}{\beta p(m)  (2 - \beta  (2-p(m)))  }=:\overline{b}_0.\]
     \paragraph{Second statement.} First, by \cref{lem:noovershoot}, we see that $F$'s \eqref{Eq Agent DEC} never binds on the equilibrium path (at best it holds with equality). This implies, invoking \cref{lem:FullEndorsment}, that either $s=m$ or $P$'s \eqref{Eq:PrincipalDEC} binds at some history on the equilibrium path. Second, since $F$ needs no incentives to select his bliss point, we have that $y^\ast_F = 0$. Third, by optimality, $E$'s \eqref{Eq Agent DEC} binds at least in his first action node in the embracing phase.

    We now want to show that it is without loss of generality to assume that $E$'s \eqref{Eq Agent DEC} binds at all of $E$'s decision nodes. To do that, take an optimal contract at which $E$'s \eqref{Eq Agent DEC} binds in his first action node, $h_t \in \mathcal{H}_E^\ast$, but has slack at history $h_{t+1}$ with $h_{t+1} \supset h_t$.
    Note that there exists an alternative strategy profile that is equivalent in everything except that it lowers $y_E(h_{t+1})$ and raises $y_E(h_t)$ in a way such that (i) $E$'s \eqref{Eq Agent DEC} binds both at $h_t$ and $h_{t+1}$; and (ii) $P$'s payoff before $h_t$ is unchanged. This strategy profile is a contract because, by reshuffling $E$'s policies, we have weakly improved $P$'s payoff at $w_P(h_t)$, implying that $P$'s \eqref{Eq:PrincipalDEC} holds. Since $F$ needs no incentives to choose $y^\ast_F = 0$, the same argument generalizes to the case in which $F$ is selected for some number of periods before $E$ is selected again. Reiterating this same argument across every node in which $E$ acts, we obtain an optimal contract such that $E$'s \eqref{Eq Agent DEC} binds at every history, which is at least as good as the optimal contract we started out with.

    Consider the optimal contract we have constructed. Once the embracing phase starts, this contract is stationary such that it `restarts the embracing phase' at every history in which $E$ led. This implies that if $P$'s \eqref{Eq:PrincipalDEC} binds at some on-path history, then $P$'s \eqref{Eq:PrincipalDEC} binds at every on-path history. We can write $P$'s payoff for $h \in \mathcal{H}_E^\ast$ as:
    \[v^\ast_P(h \in \mathcal{H}_E^\ast) = \theta - y^\ast_E(h \in \mathcal{H}_E^\ast) + \beta \underline{w}_P. \]
    Because the contract is optimal for the principal, there exists no $y_E<y_E^\ast(h \in \mathcal{H}_E^\ast)$ that is incentive compatible for $E$. Thus, $E$ is at his \eqref{Eq Agent DEC} which implies
$$
v^\ast_E(h \in \mathcal{H}_E^\ast) = b + y^\ast_E(h \in \mathcal{H}_E^\ast)-1 + \beta w^\ast_E(h \in \mathcal{H}_E^\ast) = b + \beta \underline{w}_E.
$$
  Then, optimality for the principal guarantees that there is no $w_E(E)>w_E^\ast(h \in \mathcal{H}_E^\ast)$ that the principal can credibly offer $E$. If there were, she would do it in exchange for lowering $y^\ast_E(h \in \mathcal{H}_E^\ast)$. Hence, the continuation play of the optimal contract attains the maximum value for $E$ and, by the constant-sum property, the minimum value for $F$.
     Thus, an optimal penal code for $F$ consists of restarting the embracing phase of the optimal contract. We use this punishment to construct $P$'s penal code following \cref{Lemma P Punishment 2}, which has the following condition: \begin{equation}\label{eq:DECLwhenPunishingP}
    -\hat{y}_F + \beta w^P_F = \beta \underline{w}_F= \beta w^\ast_F(h \in \mathcal{H}_E^\ast).
    \end{equation}
     Similarly, $P$'s \eqref{Eq:PrincipalDEC} binding in the embracing phase of the optimal contract implies:
\begin{equation}\label{eq:LDECPunishment}
    \underline{w}_P = -\frac{(1- p(-m)) \hat{y}_F + p(-m) }{1-\beta} = w^{\ast}_P(h \in \mathcal{H}_E^\ast).
    \end{equation}
     And, because $y_F^\ast=0$ requires no incentives and $P$'s \eqref{Eq:PrincipalDEC} binds (at least) at the first node of the embracing phase, it is without loss to consider stationary choices in that phase, $(s^{\ast},y_E^\ast(s^{\ast}))$,\footnote{Recall from our description in the main text that (with some abuse of notation) $s^\ast$ denotes the principal's strategy during the embracing phase.} where the latter is chosen such that $E$'s \eqref{Eq Agent DEC} binds. Thereafter, we refer to $w_k^\ast(E)$ as $k$'s continuation value in the embracing phase. That is,
    \[-(1-y_E^\ast(s^{\ast})) + \beta w^{\ast}_E(E) = \beta \underline{w}_E= \beta \alpha,\]
    where
    \[ w^{\ast}_E(E)= \frac{p(s^{\ast}) (b-(1-y_E^\ast(s^{\ast}))) -(1-p(s^{\ast}))}{1-\beta}.\]

   Since $y_k \geq \theta=0$, we can express $F$'s payoffs as $P$'s payoffs plus his expected leadership rents:\footnote{This decomposition is more general and we use it beyond the $\theta=0$ case. In Online Appendix \ref{proof:commitmentOA}, we provide a general version of this decomposition.}
    \[{w^P_F} =  \frac{1 - p(-m) }{1-\beta} b+\underline{w}_P \quad \text{ and } \quad w^{\ast}_F(E) = \frac{ 1- p(s^{\ast} )}{1-\beta} b +\underline{w}_P,\]
    respectively. Plugging into \eqref{eq:DECLwhenPunishingP} and \eqref{eq:LDECPunishment} and solving for $\hat{y}_F$ and $s^{\ast}$, we obtain two solutions.

    One solution is $s^{\ast}=-m$ leading to $y_E^\ast(s^{\ast})=1$ and $\hat{y}_F=0$. This implies repetition of the NES outcome in the embracing phase of the optimal contract (and \emph{all} penal codes).  The other solution is given by $\hat{y}_F>0$ and
   \[s^{\ast}=\frac{\beta ^2 (b (2 m+3)+4)-4 \beta  (b m+b+2)+4}{2 b \beta  (\beta  (2 m-1)+2)},\]
    which is monotone in $b$ and leads to a $s^{\ast}\leq m$ only if $b\geq\overline{b}_0$. Thus, below $b<\overline{b}_0$, the only contract in which $P$'s \eqref{Eq:PrincipalDEC} holds with equality is the NES contract.\footnote{The case in which $s=m$ and $E$'s \eqref{Eq Agent DEC} binds is already considered in the proof of the first statement.}\qedhere
\end{proof}

\section[Appendix: Commitment Contract]{The Optimal Commitment Contract}\label{proof:commitment}
\subsection[Proof of Propositions \ref{prop:embraceGenCom},  \ref{res:OEndorsement}, and \ref{result:firstbest}]{Proof of \cref{res:OEndorsement,prop:embraceGenCom,result:firstbest}}
\begin{proof}
We provide the full closed-form characterization of the optimal contract in Online Appendix \ref{proof:commitmentOA}. We state the lemmas leading to these results here. Jointly, they prove \cref{res:OEndorsement,prop:embraceGenCom,result:firstbest}.

Specifically, \cref{lem:firstbestslack,lem:firstbestdec,lem:DEC1side} prove the if part of \cref{result:firstbest}.
\cref{lem:feasiblethetadec,lem:thetaDECoptimality,lem:optimalityDECDECisthetaDEC,lem:thetaDECDECDEC} prove \cref{prop:embraceGenCom}. \cref{lem:optimalitytopright} proves \cref{res:OEndorsement} and  the only-if part of \cref{result:firstbest}.\qedhere
 \end{proof}

    Many of the lemmas work the following way: We first impose a particular contract candidate (Step 1), then we check the parties' \eqref{Eq Agent DEC} to ensure it is indeed a contract (Step 2), and (if needed) we address the exclusion phase (Step 3). If the proof of a Lemma deviates from this structure, we comment on it here explicitly.

    We make use of the following shorthand expressions \(
    \hat{\theta}:=p(m)\beta. \)
\medskip

    \subsubsection*{Steps for the if part of First Best (\cref{result:firstbest})}
\begin{lemma}\label{lem:firstbestslack}
        Suppose $1-\breve{\theta} \leq \theta\leq \hat{\theta}$. Then there is an optimal contract $(s,y_F,y_E)=(m,\theta,\theta)$.
    \end{lemma}

        \begin{lemma}\label{lem:firstbestdec}
            Suppose $\theta \geq \max\{1-\overline{\theta},\underline{\theta}\}$. Then, the optimal contract implies $y_k=\theta$.
        \end{lemma}

    \begin{lemma}\label{lem:DEC1side}
        Suppose $\underline{\theta}>\theta>\hat{\theta}$. Then the optimal contract implies $y_k=\theta$.
    \end{lemma}\medskip

    \subsubsection*{For $\theta<\underline{\theta}$, Full embracing is optimal (Part of \cref{prop:embraceGenCom})}

    \begin{lemma}\label{lem:feasiblethetadec}
        If $\theta\leq \underline{\theta}$, there is an incentive-compatible contract with $y_F=\theta$ and $y_E\geq \theta$. If, in addition, $\theta<\hat{\theta}$ and first best is not feasible, agent $F$'s \eqref{Eq Agent DEC} has slack for $\theta<\underline{\theta}$.
        In that contract $s^\ast_0=-m$ until the first time $E$ leads. Thereafter, $s^\ast=m$.
    \end{lemma}

    \begin{lemma}\label{lem:thetaDECoptimality}
        The contract constructed in the proof of \cref{lem:feasiblethetadec} is optimal in its region.
    \end{lemma}

    To prove this lemma, we first make use of the preliminary results to simplify the class of contracts we need to consider. Then, we show that any contract in that class that does not coincide with the candidate from \cref{lem:feasiblethetadec} cannot be optimal.\medskip

    \subsubsection*{For $\theta \in (\underline{\theta},\overline{\theta})$ Partial Embracing is Optimal (Part of \cref{prop:embraceGenCom})}
    \begin{lemma}\label{lem:thetaDECDECDEC}
        If $\theta \in (\underline{\theta},\overline{\theta})$, there exists a contract in which \eqref{Eq Agent DEC} binds for both agents and $y_F=\theta$. In that contract, $s=-m$ until the first time $E$ leads. Thereafter, $s_E=m$ when $E$ led last and $s_F<m$ when $F$ led last.
    \end{lemma}

    \begin{lemma}\label{lem:optimalityDECDECisthetaDEC}
    The contract constructed in the proof of \cref{lem:thetaDECDECDEC} is optimal in its region if $\theta<\max\{1-\overline{\theta},1-\breve{\theta}\}$.
    \end{lemma}

    To prove optimality here, we exploit that players' \eqref{Eq Agent DEC} bind, leading to a well-defined set of Bellman equations.

    \subsubsection*{For $\theta>\overline{\theta}$ Opportunism is Optimal (\cref{res:OEndorsement})}

    \begin{lemma}\label{lem:optimalitytopright}
        If $\theta>\overline{\theta}$, an optimal contract is such that $y_i\neq \theta$ and $s_E=-s_F=m$.
    \end{lemma}
    To prove this lemma we combine the optimality result from the previous lemma with our usual construction approach.
    \subsubsection*{Only-if part of \cref{result:firstbest}}
    We have shown that for any region, other than those in \cref{lem:firstbestslack,lem:firstbestdec,lem:DEC1side}, the optimal contract does not achieve first best, which proves the only-if part.

    \subsection[Proof of Proposition \ref{prop:polarization}]{Proof of \cref{prop:polarization}}
    \begin{proof}
            \Cref{prop:polarization} follows from observing that $y^\ast_E$ ($y^\ast_F$) decreases (increases) in $\theta$ for $\theta{<}\overline{\theta}$ and remains constant thereafter.
    \end{proof}

\subsection[Proof of Proposition \ref{prop:symmetry}]{Proof of \cref{prop:symmetry}}
\begin{proof}
We prove the following lemma, which implies \cref{prop:symmetry}.

\begin{lemma}\label{lemma:FBorOPPat12}
    Fix any $(m,\beta)$ such that $2\beta m<  1-\beta$ and assume $\theta{=}1/2$. There exists a threshold $b^{FB}>0$ such that for $b<b^{FB}$, the optimal commitment contract is the opportunistic contract, while for $b\geq b^{FB}$ the optimal commitment contract is the first-best contract.
    If, instead, $2\beta m \geq 1-\beta$, first best is implementable for any $b>0$ if $\theta=1/2$.
\end{lemma}
\begin{proof}
We prove the results in steps. First, we prove that the optimal contract at $\theta=1/2$ is either the opportunistic contract or the first-best contract. Then we show that for $2 \beta m<1-\beta$, a cutoff $b^{FB}>0$ exists separating the two. We conclude by showing that for $2\beta m \geq 1-\beta$ the first best is possible for \emph{any} $b>0$.

    \paragraph{Step 1. $\overline{\theta}$ and $\breve{\theta}$ at $1/2$.} Recall that the first-best is implementable if and only if $\theta\geq \max\{1-\overline{\theta},1-\breve{\theta}\}$. We will now show that at $\theta=1/2$, for any $(m,b,\beta)$, the relevant first-best implementation constraint is $\theta\geq 1-\overline{\theta}$. We do this by showing that if $1/2\geq 1-\overline{\theta}$, then $1/2> 1-\breve{\theta}$. To do that, notice that both $\breve{\theta}$ and $\overline{\theta}$ are linearly increasing in $b$.
    Now fix an arbitrary $(m,\breve{\beta},b)$ such that $\breve{\theta}=1/2$; then, implicitly, it must hold that
    \[\breve{\beta}=\frac{1}{1 + 2m(2b+1))} \quad \Rightarrow\quad \frac{\breve{\beta}}{1-\breve{\beta}}= \frac{1}{2m(1+2b)}.\]
    Replacing $\breve{\beta}$ inside $\overline{\theta}$ gives us
    \[\overline{\theta}_{\beta=\breve{\beta}} = \frac{1+b}{2(1+2 b)}\frac{(1+2(1+ 4b)m)}{(1+2(1+2 b)m)}\]
    which is (weakly) larger than $1/2$ if and only if
    \[(1+b)(1+2(1+ 4b)m)\geq(1+2 b)(1+2(1+2 b)m) \Leftrightarrow m\geq 1/2.\]
    which is impossible.

    \paragraph{Step 2. Only Opportunistic or First-Best Optimal.} Fixing $\theta=1/2$ and selecting an arbitrary $(m,\beta)$, linearity in $b$ implies that there exists a threshold $\breve{b}$ such that $\theta<1-\breve{\theta}$ for $b<\breve{b}$ and $\theta>1-\breve{\theta}$ for $b>\breve{b}$. Notice further that \emph{at} $b=\breve{b}$, by construction, we have \[\theta=1-\breve{\theta}<1-\overline{\theta}.\] The inequality comes from the order for $\breve{\theta}=1/2$ that we derived above. Equivalently, if $\theta=1-\overline{\theta}=1/2$, then $b>\breve{b}$.
    Finally, at $\theta=1/2$ we must have either $\theta>\overline{\theta}$ or $\theta\geq 1-\overline{\theta}$. Now, recall from \cref{res:OEndorsement} that if $\theta>\overline{\theta}$, the optimal contract is the opportunistic contract. As we see in the proof of \cref{lem:optimalitytopright}, nothing in the argument relies on the fact that $\theta<1/2$ and therefore the result extends to the limit. Thus, at $\theta=1/2$, the optimal commitment contract is the opportunistic contract if $\overline{\theta}<1/2$ and the first-best otherwise.

    \paragraph{Step 3. Cutoff $b^{FB}.$} Recall from Step 1 that $\overline{\theta}$ is strictly (and linearly) increasing in $b$ and thus crosses $1/2$ from below once in $b$. Therefore, ignoring the non-negativity constraint for $b$, for any $(m,\beta)$, there exists some threshold $b' \in \mathbb{R}$ such that $\overline{\theta}$ is above $1/2$ if and only if $b>b'$.
    Moreover, if we fix $(m,\beta)$ such that  $2m \beta\geq 1-\beta$, then $\lim_{b\to 0} \overline{\theta} \geq 1/2$, which means $b'\leq 0$ and thus $\overline{\theta}$ is strictly larger than $1/2$ for any $b>0$ and first best is implementable for all positive $b$.
    If, instead, we fix $(m,\beta)$ such that $2m \beta < 1-\beta$, then $\lim_{b \to 0} \overline{\theta}<1/2$, which implies the cutoff from the lemma, ${b}^{FB}>0$, exists.
\end{proof}
What remains is to show the first-period behavior in the opportunistic contract, which follows directly from the observation that $|\overline{\theta}-1/2|=|(1-\overline{\theta})-1/2|$ and thus the principal is indifferent between endorsing either agent in the first period.
\end{proof}
\section[Appendix: Penal Codes]{Penal Codes}

Here, we provide further details on the optimal penal codes for the players. The results we state here provide the backbone of the punishment discussion in \cref{sec: nocommitment}. Due to space restrictions, we defer their proofs to Online Appendix \ref{proof:commitmentOA}.

Note first that \emph{simple penal codes} (as in \citet{abreu1988theory}, Proposition 5) suffice in our game due to perfect information. The culprit is always observed by other players, and so is their ``crime,'' which makes more complicated punishment schemes, as in, e.g., \citet{Mailath2017Punishment}, not relevant. Existence is guaranteed by the compactness of the equilibrium payoff set.

\emph{Note:} By self-generation of equilibrium payoff sets, penal codes are nothing other than those contracts that minimize the punished player's ex-ante value. We will therefore adjust the notation here to reflect this. Specifically, we consider $h_0$ to be the history at which the punishment started, i.e., the node at which the deviation occurred.\footnote{We sometimes use $\underline{w}_k(h_0)$ instead of $\underline{w}_k$ to emphasize that this is the \emph{ex-ante} payoff to $k$.}

\subsection{The Principal's Penal Code}\label{sec:principal_punishment}
We show that a stationary penal code for the principal exists.

\begin{lemma}\label{Lemma P Punishment 1}
        There is a penal code for the principal in which $y_E = 1$ whenever $E$ leads.
\end{lemma}
The intuition is straightforward: If there was a penal code in which $y_E<1$, then such a contract can only be an optimal penal code for the principal if the principal promises $E$ some endorsement in return. But then, using the logic of \cref{lem:DEC} in reverse, we can find a contract in which $E$ selects $y_E=1$. Having $y_E=1$ established, we can now construct the optimal penal code for the principal.
    \begin{lemma}\label{Lemma P Punishment 2}
        A stationary optimal penal code for $P$ exists. It is characterized by:
        \begin{itemize}
            \item the principal endorses $F$, $s=-m$,
            \item $E$ fully polarizes $y_{E}=1$
            \item $F$ either fully polarizes, $y_{F}=0$, or chooses $y_F=\hat{y}_F$, where $\hat{y}_F$ is such that his dynamic enforcement constraint binds.
        \end{itemize}
    \end{lemma}

    \begin{proof}
        From Lemma \ref{Lemma P Punishment 1}, we have $y_E=1$. Hence, irrespective of $y_F(\cdot)$, $P$ is worse off if $E$ leads because $\theta<1/2$. Thus, $P$ chooses $s=-m$.
        To conclude, we characterize $F$'s actions. Fixing $(s=-m, y_E=1)$, let $\hat{y}_F$ be $F$'s policy such that his \eqref{Eq Agent DEC} binds. Note further that if a certain $\hat{y}_F$ satisfies \eqref{Eq Agent DEC} for some $s$, it satisfies \eqref{Eq Agent DEC} for $s=-m$. It is straightforward to see that $F$'s action is either $y_F=0$, which is trivially incentive compatible, or $y_F=\hat{y}_F$, as $P$ prefers any intermediate $y_F\in (0, \hat{y}_F)$ to at least one of these two extremes. If $\hat{y}_F < 2\theta$, $y_F=0$ is worse for $P$, and otherwise, $y_F=\hat{y}_F$ is worse.
\end{proof}

\subsection{The Friend's Penal Code}

\paragraph{Note.} Compactness of the equilibrium payoff set and self-generation imply that for every implementable continuation value for $P$, $w_p$, there is a contract that minimizes $F$'s continuation value conditional on yielding $w_P$ to $P$.
The key to the construction of the Friend's penal code is the back-to-business property, which we now define formally.

\begin{definition}[\textbf{Back-to-business property}]
    A penal code for the friend, $C^F$, has the \emph{back-to-business property}, if any on-path history in that contract, $h \in \mathcal H_F^\ast(C^F)$ implies a continuation game identical to that of a history $h \notin \mathcal H_F(C^\ast)$ in the optimal contract $C^\ast$ that selects $F$ as the leader.
 \end{definition}

We now show that a penal code for the friend that sustains the optimal contract has the back-to-business property. The first case we address is when $F$'s on-path \eqref{Eq Agent DEC} binds. Since $F$'s \eqref{Eq Agent DEC} binds, a penal code that implements the optimal contract has to be optimal. We see that this optimal penal code has the back-to-business property.

\begin{lemma}\label{lem:backtobusinessDECDEC}
    Suppose $F$'s \eqref{Eq Agent DEC} binds in the embracing phase of the optimal contract and first best is infeasible. The optimal penal code for $F$ has the back-to-business property.
\end{lemma}
We prove \cref{lem:backtobusinessDECDEC} in Online Appendix \ref{proof:commitmentOA}. The construction admits \cref{cor:b2b}.

\begin{corollary}\label{cor:b2b}
    The optimal penal code for $F$ consists of two phases: (i) A punishment phase that ends when $F$ leads for the first time since the beginning of the penal code, and (ii) the back-to-business phase.

    \begin{enumerate}
        \item If the optimal contract is such that $P$ has slack in the embracing phase, then the punishment phase is such that the principal chooses a full embracing strategy $s^F_0=m$, and the enemy chooses $y_E^F=\min \{\hat{y}_E, 1\}$, where $\hat{y}_E$ is such that the principal's \eqref{Eq:PrincipalDEC} holds with equality.
        \item If the optimal contract is such that $P$'s \eqref{Eq:PrincipalDEC} holds with equality, the punishment phase replicates the on-path behavior for histories in which $E$ was the last to lead.
    \end{enumerate}
\end{corollary}

The idea is straightforward: because $F$'s \eqref{Eq Agent DEC} binds in the embracing phase, this contract is already the worst for $F$ conditional on leading. In the initial punishment phase, the result relies on the idea familiar from the proof of \cref{prop:theta0nocommitment}: If all policy choices $y_k\geq \theta$, we can decompose the friend's payoff into a policy payoff and a leadership rent.
 Thus, by ensuring that $P$'s \eqref{Eq:PrincipalDEC} binds, we ensure that indeed we are minimizing $F$'s ex-ante payoffs. If $P$'s \eqref{Eq:PrincipalDEC} has slack even for $y_E=1$, we have recreated the commitment punishment (in value) for $F$.
Our next result complements the above, showing that a penal code with back-to-business is always a valid penal code for implementing the full-embracing optimal contract.
 \begin{lemma}\label{lem:b2bFullEmbaracing}
    Whenever the optimal contract features full embracing in the embracing phase, it can be supported by a penal code for $F$, $C^F$, which consists of (i) a punishment phase that ends when $F$ leads for the first time since the beginning of the penal code, and (ii) the back-to-business phase. Moreover, in the punishment phase: \begin{itemize}
        \item $s = m$
        \item $y_E = \min\{\hat{y}_E, 1\}$, where $\hat{y}_E \in [0,\infty)$ is chosen such that $P$'s \eqref{Eq:PrincipalDEC} binds.
    \end{itemize}
\end{lemma}

The result relies on the decomposition (since $y_k\geq \theta$). The leadership rent is minimized under $s^F=m$. The policy payoff is minimized by making $P$'s \eqref{Eq:PrincipalDEC} bind at the initial node. If $P$ has slack even for $y_E =1$, either $F$'s \eqref{Eq Agent DEC} binds, in which case we have the commitment punishment, or it has slack, in which case the penal code is not necessarily optimal, but still valid to sustain the optimal contract.

\paragraph{Conclusion.} We demonstrate here that the penal codes for $P$ and $F$ are linked through $F$'s action when punishing $P$, and the back-to-business property of $F$'s penal code. Thus, they jointly form a system of equations that behaviorally characterizes both punishments and the optimal contract. However, depending on details, different constraints hold with equality, making a simple closed-form characterization difficult. In the supplementary material, we provide code that obtains the full symbolic characterization of all penal codes and the optimal contract for any arbitrary parameter constellation.
\section[Appendix: No Commitment Contracts]{The Optimal No-Commitment Contract}

\subsection[Proof of Proposition \ref{prop:thresholdforcommitment}]{Proof of \cref{prop:thresholdforcommitment}}
\begin{proof}
    For existence of $\overline{b}(\theta)$, observe that as $b \to \infty$, both $\breve{\theta},\overline{\theta} \to \infty$; thus, with commitment, $P$ could implement the first best, in particular via the following strategy: fully endorsing the agent who has led last.

    We will now show that the first-best can be implemented without commitment when $b$ is large. For that purpose, assume the following punishment: If an agent deviates, the principal fully endorses the non-deviator for one period before returning to the on-path contract. Agents continue to choose $y_k=\theta$. That punishment is sufficient to sustain $y_k=\theta$ if $2\beta m (b-|\theta_k-\theta|)\geq|\theta_k-\theta|$ which holds for $b\geq(|\theta_k-\theta|)(1+2\beta m)/(2\beta m)$. $P$'s \eqref{Eq:PrincipalDEC} holds on-path and off-path because all agents are expected to always select $\theta$.

    The existence of $\hat{\theta}$ follows from \cref{prop:balanceiscommitmentGeneral} which we state and prove below. \end{proof}

    \begin{lemma}\label{prop:balanceiscommitmentGeneral}
        Fix $(m,b,\beta)$. There exists a threshold $\hat{\theta}<1/2$ such that for any $\theta \in [\hat{\theta},1/2)$, even if the principal has no commitment power, she optimally implements a contract that is observationally equivalent to the optimal commitment contract.
    \end{lemma}

    \begin{proof}
        Note that, by construction, the principal prefers the on-path contract to the NES payoff. Moreover, the enemy's grim-trigger punishment is the NES contract, which is (trivially) implementable also absent principal commitment. Thus, if the principal can sustain a punishment that gives the friend the same off-path payoff as the grim-trigger punishment used in the construction of the commitment contract, the principal can implement a contract observationally equivalent to the commitment contract. We now construct such punishments.

        \paragraph{At the Limit.} At the limit $\theta=1/2$, either first best or the opportunistic contract is the commitment optimum (by \cref{lemma:FBorOPPat12}). Now, observe that the following class of contracts is also implementable at the limit: Parties choose their bliss points, $y_F=0$ and $y_E=1$, whenever they lead, and the principal chooses an arbitrary strategy $s \in [-m,m]$ at \emph{any} history. To see that, consider the static game: incentives for $F$ and $E$ hold trivially, and both policy choices lead to a payoff of $-1/2$ for the principal. That makes $P$ indifferent between \emph{any} $s$. Because the repetition of a stage Nash equilibrium is also an equilibrium of the repeated game, the claim follows. Note that this claim means in particular that the grim-trigger punishment for $F$ under commitment is implementable at the limit $\theta=1/2$.

        \paragraph{Outside the limit.} We now move slightly away from the limit $\theta=1/2$ and consider $\theta<1/2$ but close to it.
        We want to implement the optimal on-path commitment contract $(s_F^\ast,s_E^\ast,y_E^\ast,y_F^\ast)$ and consider the following punishment contract for the friend: If $F$ deviates, $P$ supports $E$, who does not moderate at all $(y_E^F=1)$ until the friend returns to the lead. From that point onward the continuation game follows the on-path commitment contract. If the principal or the enemy deviates, the continuation game is NES.

        The principal's value from punishing the friend is
            \[w_P^F=(1/2+m)(-(1-\theta) + \beta w_P^F)+(1/2-m)(v_P^\ast(F)) = -\frac{p(m)(1-\theta)-(1-p(m)) v_P^\ast(F)}{1-p(m)\beta}\]
        If the principal deviates her payoff is
        \[\underline{w}_P=w_P^E=-\frac{p(m)\theta+(1-p(m))(1-\theta)}{1-\beta}.\]

        Now at the limit $\underline{w}_P(\theta=1/2)=-1/(2(1-\beta))$, and $w_P^\ast>\underline{w}_P$ because if $\overline{\theta}\leq 1/2$ first best is implementable, and otherwise $y_F=\overline{\theta}, y_R=1-\overline{\theta}$ is implementable (\cref{lemma:FBorOPPat12}). Both are better than NES.
        Moreover, notice that \[
            v^\ast_P\left(F;\theta=\frac{1}{2}\right) \geq -\frac{1}{2} + \beta w_P^\ast \left(F;\theta=\frac{1}{2}\right)>  -\frac{1}{2} + \beta \underline{w}_P \left(\theta=\frac{1}{2}\right)=  -\frac{1}{2(1-\beta)} = \underline{w}_P\left(\theta=\frac{1}{2}\right).\]

        But then, we have that $w_P^F(\theta=1/2)>\underline{w}_P(\theta=1/2)$ which implies strict slack of the principal's incentives at the limit $\theta=1/2$. Since both $w_P^F(\theta)$ and $\underline{w}_P(\theta)$ are continuous in $\theta$, the principal's incentives are maintained even outside the limit.

        That implies that the commitment contract is sustainable for some $\theta$ outside the limit. Finally, if that punishment is sustainable for the principal, it is sustainable for both parties. The enemy gains from choosing $1$, and the friend is punished by being threatened to go through a punishment period of $y_E^P=1$ and $s_F^P=m$ instead of $s^\ast_F \leq m$ and $y_E^\ast<1$. Thus, both parties' \eqref{Eq Agent DEC} and the principal's \eqref{Eq:PrincipalDEC} hold and the no-commitment contract is observationally equivalent to the commitment contract for $\theta$ close to 1/2.
    \end{proof}

\subsection[Proof of Proposition \ref{prop:regimesnocommitment}]{Proof of \cref{prop:regimesnocommitment}}\label{proofregimes}
\begin{proof}
    To prove \cref{prop:regimesnocommitment}, we have to show that whenever a region exists with commitment, part of that region survives removing commitment. We do this by first showing that as $\theta\to 1/2$, full embracing vanishes for any $b$ if there ever is a partial embracing or opportunistic region. Moreover, as $\theta\to 1/2,$ partial embracing survives for intermediate $b$. We then use the ``centrism is commitment'' result of \cref{prop:thresholdforcommitment} to show robustness of parts of the first best, the partial embracing, and the opportunistic region.

    Then, we consider the other corner $\theta \to 0$ and show that full embracing survives in some neighborhood of $\theta=0$ if it survives at $\theta=0$.

    \begin{lemma}\label{lem:partialoutsidetheta12}
        If $1/2>\theta>\theta^\dagger:=\beta(1/2+m)$, there exist thresholds $b^\dagger<b^\ddagger$ such that the optimal commitment contract is the opportunistic contract for $b<b^\dagger$, the first best for $b>b^\ddagger$, and a partial embracing contract for $b\in (b^\dagger,b^\ddagger)$.
    \end{lemma}

    \begin{proof}
        We prove this statement by fixing $(m,\beta)$ and showing two steps separately.
        \begin{enumerate}
            \item showing that for $\theta=\theta^\dagger$, there is a unique $b^\diamond$ such that $1-\overline{\theta}=1-\breve{\theta}=\underline{\theta}=\theta^\dagger$,
            \item Fixing $(m,\beta)$ and $\theta \in(\theta^\dagger,1/2)$: \begin{enumerate}
                \item if $b$ is such that $\theta\leq\underline{\theta}$, then $\theta>\max\{1-\overline{\theta}, 1-\breve{\theta}\}$.
                \item if $b$ is such that $\theta\geq \max\{1-\overline{\theta},1-\breve{\theta}\}$ then $\theta>\overline{\theta}$
                \item if $b$ is such that $\theta\geq \overline{\theta}$, then $\theta<\max\{1-\overline{\theta}, 1-\breve{\theta},\underline{\theta}\}$.

            \end{enumerate}
        \end{enumerate}
        Result (1) implies that the three cutoffs coincide; then result (2) implies that above their intersection (in the $\theta$ space), a full embracing contract does not exist (2(a)), but a partial embracing contract (2(b)) and an opportunistic contract (2(c)) exist for some $b$.

        \paragraph{Result (1).} First, recall that all three cutoffs, $\underline{\theta}, \overline{\theta}, \breve{\theta}$, are linearly increasing in $b$. Thus any pair of $1-\overline{\theta}, 1-\breve{\theta},\underline{\theta}$ intersects at most once in $b$. The following Lemma states that they intersect at the same point. We prove it in \cref{proof:commitmentOA}.

        \begin{lemma}\label{lemma:daggertheta}
            Assume $2m\beta<1-\beta$. The following are equivalent for $\theta^\dagger=\beta(1/2+m):$  (1) $b=b^\diamond:=\frac{1-\beta}{2\beta m}-1$, (2) $\underline{\theta}=\theta^\dagger$, (3)  $1-\overline{\theta}=\theta^\dagger$, (4) $1-\breve{\theta}=\theta^\dagger$.
        \end{lemma}

        \paragraph{Result (2).} Note that
        \[0<\frac{\partial \breve{\theta}}{\partial b} = 2m\beta<2m\beta \underbrace{\frac{1-\beta(1/2+m)}{1-\beta}}_{>1} = \frac{\partial \overline{\theta}}{\partial b}\]
        which implies both $1-\breve{\theta}$ and $1-\overline{\theta}$ are decreasing in $b$ with $1-\overline{\theta}$ decreasing slower. That implies if $\theta=1-\overline{\theta}>\theta^\dagger$ then $\theta>1-\breve{\theta}$. Because $\underline{\theta}$ increases in $b$ and intersects with $1-\overline{\theta}$ at $(b^\diamond,\theta^\dagger)$, the 2(a) follows. The two other results, 2(b) and 2(c), follow because by construction $\overline{\theta}=1-\overline{\theta}$ at $(b^{FB},1/2)$ (see \cref{prop:symmetry}), and for $1/2>\theta=\overline{\theta}<1-\overline{\theta}$. The last result, 2(c), follows because $\overline{\theta}$ increases in $b$ and intersects with $1-\overline{\theta}$ at $(b^{FB},1/2)$, and for $1/2>\theta=\overline{\theta}<1-\overline{\theta}$.
    \end{proof}

    \cref{lem:partialoutsidetheta12} combined with \cref{prop:thresholdforcommitment} proves robustness of a partial embracing region, a first-best region, and an opportunistic region if they exist in the first place.\footnote{$\theta^\dagger{>}1/2 {\Leftrightarrow} 2m\beta{>}1-\beta$ which implies that only full embracing and first best are commitment optima.}

    \begin{lemma}\label{lem:FullEmbraceExistsOutsideTheta0}
        Fix $(m,\beta)$ and pick $b \in(\overline{b}_0,\breve{b})$.\footnote{These cutoffs are defined in \cref{prop:theta0nocommitment} and \cref{lemma:FBorOPPat12} respectively.} Then there exists $\theta^\P>0$ such that for any $\theta \in (0,\theta^\P)$, the optimal no-commitment contract features full embracing and is observationally equivalent to the optimal commitment contract.
    \end{lemma}

    \begin{proof}
         Because $\underline{\theta}>0$, for any $(m,\beta,b)$, there exists a region where full embracing is optimal under commitment. Also, if $b<\breve{b}$, then there exists a $\theta'>0$ such that for $\theta \in [0,\theta')$, the optimal commitment contract is not first best.\footnote{Details are found in the proof of \cref{lem:partialoutsidetheta12}.} Consider now the following results for $\theta =0$.
        First, for any $b \in (\overline{b}_0,\breve{b})$, the optimal commitment contract is implementable without commitment by \cref{prop:theta0nocommitment} and because $b>\overline{b}_0$, $P$'s \eqref{Eq:PrincipalDEC} has slack and $\hat{y}_E>y_E^\ast$, where $\hat{y}_E$ is $E$'s action during the punishment phase of $F$'s penal code (see the proof of \cref{prop:theta0nocommitment} for details). Second, because $\hat{y}_E>y^\ast_E$ and $F$'s punishment has the back-to-business property (see \cref{lem:b2bFullEmbaracing}), $F$'s \eqref{Eq Agent DEC} also has slack.

        Now, with these observations, the result follows from continuity in $\theta$ of the commitment continuation payoffs $w_P^\ast(h), w_F^\ast(h)$, and the penal codes for the non-commitment case, $\underline{w}_F$, and $\underline{w}_P$. Continuity in the on-path payoffs follows directly from the characterization of the optimal commitment contract (\cref{proof:commitment}), while continuity in $P$'s punishment payoffs follows from the construction in \cref{Lemma P Punishment 2,Lemma P Punishment 1}. Finally, continuity in $F$'s punishment payoff follows from the back-to-business property, the continuity of the on-path contract, and continuity of $\hat{y}_E$ in $\theta$, which is a consequence of the continuity of $\underline{w}_P$.
    \end{proof}
    \Cref{lem:FullEmbraceExistsOutsideTheta0} shows that if $b>\overline{b}_0$, then there is a region (with $\theta>0$) in which commitment plays no role and full embracing survives in the absence of commitment.
\end{proof}

\section[Appendix: Implications]{Implications}

\subsection[Proof of Proposition \ref{prop:comparativeStatics}]{Proof of \cref{prop:comparativeStatics}}

For the commitment case, the result follows from the observation that $\underline{\theta}$ is monotone in all three parameters.

For the no-commitment case, take $F$'s \eqref{Eq Agent DEC} binding in the full embracing contract:
\begin{equation}\label{eq F Fullembracing DEC}
    y_F^\ast= \theta = \beta( w^\ast_F - \underline{w}_F)
\end{equation}
Recall that $F$'s optimal penal code involves back-to-business (by Lemmas \ref{lem:b2bFullEmbaracing}-\ref{lem:backtobusinessDECDEC}), and thus both the optimal contract and $F$'s punishment have all $y_k \geq \theta$, which implies one can express $F$'s payoffs as $P$'s payoffs plus his expected leadership rents:
 \[w^\ast_F = \frac{1 - p(m) }{1-\beta} b+w^\ast_P - \frac{\theta}{1-\beta} \quad \text{ and } \quad {\underline{w}_F} =  \frac{1 - p(m) }{1-\beta} b + w_P^F - \frac{\theta}{1-\beta},\]
 where $w_P^F = \underline{w}_P$ in the relevant region, since otherwise $F$'s commitment punishment is attainable. Substituting in \eqref{eq F Fullembracing DEC}, we get $\theta = \beta( w^\ast_P - \underline{w}_P).$ As $\theta >0$, $P$'s \eqref{Eq:PrincipalDEC} has slack on path, and the frontier of the full embracing region is given by \eqref{eq F Fullembracing DEC}. The result follows from the fact that, for either penal code of $P$, $\beta( w^\ast_P - \underline{w}_P)$ increases both in $\beta$ and $b$.

\subsection[Proof of Proposition \ref{prop:non-monotoneWelfare}]{Proof of \cref{prop:non-monotoneWelfare}}

\begin{proof}
    For the commitment case, \Cref{prop:non-monotoneWelfare} follows for $\theta{<}\overline{\theta}$ because $y^\ast_F{=}\theta$ and $y^\ast_E$ (weakly) decreases in $\theta$. For $\theta \geq \overline{\theta}$, both $y^\ast_F$ and $y_E^\ast$ are constant, implying that the principal loses in the exclusion phase as $\theta$ increases; in the embracing phase, the changes cancel each other out.

    \cref{prop:balanceiscommitmentGeneral} implies that for $\theta$ close to $1/2$ commitment plays no role, thus the principal's ex-ante payoff decreases in $\theta$ if $\overline{\theta}<1/2$ also without commitment. This is straightforward for $b<\overline{b}_0$, in which case the no-commitment payoff is that of NES (by \cref{prop:theta0nocommitment}), which is dominated by the opportunistic contract under commitment for $\theta$ close to $1/2$.

    For the remaining case, the following lemma (which we prove in \cref{proof:commitmentOA}) implies the result in the no-commitment case for $b>\overline{b}_0$.
    \begin{lemma}\label{lem:nonMono_noCommit_blarge}
        There exists a $b_{\Delta}<\overline{b}_0$ such that for $b>b_{\Delta}$, the principal's payoff under commitment at $\theta=0$ is smaller than at $\theta \to 1/2$.
    \end{lemma}
Since the statement holds for $b\geq\overline{b}_0>b_{\Delta}$, this completes our proof.
\end{proof}

\subsection[Proof of Proposition \ref{prop:competition}]{Proof of \cref{prop:competition}}

\begin{proof}
        The first part of \cref{prop:competition} follows because when firing the enemy, the principal receives a per-period payoff of $-\theta$. In the optimal contract, the principal's payoff is continuous and she strictly prefers it over firing at the boundaries $\theta\in \{1/2,1-\overline{\theta},1-\breve{\theta}\}$, one of which is always admissible. Because $\Upsilon (b,m)$ is non-empty and payoffs are continuous, the result follows. For the second part of \cref{prop:competition}, observe that for a given $\theta$ the principal weakly prefers competition if \(y^\ast_E-\theta\leq \theta.\) Because commitment does not matter close to $\theta=1/2$, the result continues to hold.

        Now, if, for the chosen $(b,m,\theta)$, the first best is achieved for some $\hat{\beta}$ then, at $\hat{\beta}$, the left-hand side of the inequality is $0$. Continuity and monotonicity of $y^\ast_E$ in $\beta$ imply that there is a $\tilde{\beta}<\hat{\beta}$ such that competition is preferred. Observe that $\overline{\theta}\to \infty$ as $\beta\to 1$ and $\breve{\theta} \to 1/2+m(1+2b)\geq 1$ if and only if $2b\geq \frac{1-2m}{2m}$. Thus, under the condition, first best is achieved at least at the limit, which implies that, for every $\theta$, there is a $(\hat{\beta},\theta) \in \Upsilon(b,m)$ such that the principal does not want to fire the agent. Because the first best survives no commitment, the result follows in the no-commitment case.
\end{proof}
\part*{References}\addcontentsline{toc}{section}{References}
\singlespacing
\setlength\bibitemsep{0.5\itemsep}
\begin{refcontext}[sorting=nyt]
\printbibliography[heading=none]
\end{refcontext}

\newpage

\onehalfspacing
\setcounter{page}{1}
\renewcommand*{\thepage}{OA.\arabic{page}}
\begin{refsection}
\part*{Online Appendix}

\section[Online Appendix: Omitted Proofs]{Omitted Proofs}\label{proof:commitmentOA}

    We make frequent use of the following shorthand expressions
    \begin{align*}
    \hat{\theta}&:=p(m)\beta &
    \psi&:=-\beta \left(\frac{b+1-2\beta b}{1-\beta} +2 \beta \alpha\right).
    \end{align*}

    \subsection{Proof of Lemma \ref{lem:firstbestslack}}

    \begin{proof} \textbf{Step 1. A candidate.}
        Take the candidate $(s,y_F,y_E)=(m,\theta,\theta)$ with deviations punished via grim trigger.

        \paragraph{Step 2. The candidate is a contract.} Optimality follows because $y_k{=}\theta.$ $E$'s (\ref{Eq Agent DEC}) is
            \[b- (1-y_E) + \frac{\beta}{1-\beta} \bigg(p(m) (b-(1-y_E)) - (1 - p(m)) (1-\theta)\bigg) \geq b + \beta \alpha.\]
            Hence, $y_E=\theta$ satisfies \eqref{Eq Agent DEC} if
            \begin{align*}
            \theta &\geq 1-\frac{2 m (b+1)+\theta(1/2-m)}{1-\beta(1/2-m)} \beta &
            \Leftrightarrow \quad  \theta&\geq  1- \beta\left(\frac{1}{2} +m(1+2b)\right)=1-\breve{\theta}.
            \end{align*}

            Analogously, $F$'s (\ref{Eq Agent DEC}) is
            \[b- y_F + \frac{\beta}{1-\beta} \bigg( (1-p(m)) (b-y_F)- p(m)\theta \bigg) \geq b + \beta \alpha.\]
            Hence, $y_F=\theta$ satisfies \eqref{Eq Agent DEC} if
            \begin{align*}
            \theta&\leq  \frac{p(m)(1-\theta)\beta}{1-\beta p(m)} &
            \Leftrightarrow \quad  \theta &\leq  p(m)\beta=\hat{\theta}.\qedhere
            \end{align*}
        \end{proof}

        \subsection{Proof of Lemma \ref{lem:firstbestdec}}

          \begin{proof} \textbf{Step 1. A candidate.} Consider the following strategy profile with $y_k=\theta$ and on-path strategies, $(s_F,s_E)$, where $s_F$ and $s_E$ solve
        \[\begin{split}\theta &= \beta \left(\frac{b-1}{2(1-\beta)}-\alpha\right)\underbrace{-\beta \left(\frac{b+1-2\beta b}{1-\beta} +2 \beta \alpha\right)}_{=\psi} s_F,\\\theta&= 1- \beta \left(\frac{b-1}{2(1-\beta)}-\alpha\right) + \psi s_E.
        \end{split}\]
        The first-period endorsement strategy is arbitrary; deviations are punished by grim trigger.
        \paragraph{Step 2. The candidate is a contract.}

        The joint per-period utility of $F$ and $E$ is $u_F+u_E=b-1$. Assume, for now, that we are in a setting in which both players' \eqref{Eq Agent DEC} binds when leading. Thus, the leading party's continuation payoff is identical to deviating and entering the worst punishment, i.e., $b+\beta \alpha$. That implies that the non-leading party receives the residual $(b-1)/(1-\beta) - (b+\beta \alpha).$ But then, because both \eqref{Eq Agent DEC} bind,
        \begin{equation*}
            \begin{split}
          -y_F + \beta \left((1-p(s_F)) \left(b+\beta \alpha\right) +  p(s_F)\left(\frac{b-1}{(1-\beta)}-(b +\beta \alpha)\right)\right)&= \beta \alpha \\
            (y_E-1) + \beta \left( p(s_E) \left(b + \beta \alpha\right) + (1-p(s_E))\left(\frac{b -1}{(1-\beta)} -(b +\beta \alpha)\right)\right)&= \beta \alpha,
            \end{split}
        \end{equation*}
        we can solve for $y_F$ and $y_E$ as
        \begin{equation}\label{yrDECDEC}
            y_F(s_F) = \beta \left(\frac{b-1}{2(1-\beta)}-\alpha\right) + \psi s_F;
\text{ and } y_E(s_E) =1- \beta \left(\frac{b-1}{2(1-\beta)}-\alpha\right) + \psi s_E.
        \end{equation}

        Note that since both equations in \eqref{yrDECDEC} are expected to hold in all future periods, the leading party's choice is only a function of the next expected endorsement by $P$. Thus, $y_k=\theta$ satisfies both parties' \eqref{Eq Agent DEC} if and only if there are $s_k \in [-m,m]$ such that each RHS in \cref{yrDECDEC} is equal to $\theta$.
        Assume for now that $\psi<0$, which is a property that is necessary and sufficient for the interval we construct to be non-empty, and that we will verify later. Then, an $s_F$ that ensures $y_F=\theta$ exists if and only if
        \[\beta\left(\frac{b-1}{2(1-\beta)}-\alpha\right) + \psi m \leq \theta \leq  \beta \left(\frac{b-1}{2(1-\beta)}-\alpha\right) - \psi m,\]
        and rearranging,
        \[\underline{\theta}=2{\frac{\beta}{1-\beta} (b+1) m p(m)\beta} \leq \theta \leq{\frac{\beta}{1-\beta}(b+1)2 m\left(1- \beta p(m)\right)}=\overline{\theta},\]
        which is non-empty (and hence indeed $\psi<0$) if and only if $1-\beta>2m \beta$. Similarly, an $s_E$ that ensures $y_E=\theta$ exists if and only if
        \[1- \beta \left(\frac{b-1}{2(1-\beta)}-\alpha\right) + \psi m \leq \theta \leq  1- \beta \left(\frac{b-1}{2(1-\beta)}-\alpha\right) - \psi m,\]
        which is equivalent to $1-\overline{\theta} \leq \theta \leq 1-\underline{\theta}$. Using that $\theta \leq 1/2$ we can conclude $\theta>1-\overline{\theta} \Rightarrow \theta<\overline{\theta}$, and $\quad\theta>\underline{\theta} \Rightarrow \theta<1-\underline{\theta}.$
        Thus, a first-best contract exists if $\theta>\max\left\{1-\overline{\theta},\underline{\theta}\right\}$ conditional on $\psi<0$, which remains to be shown.

        Recall from above that $\psi<0$ iff $1-\beta>2m \beta$. If, instead, $1-\beta\leq 2m \beta$, then
        \[\underline{\theta}= \frac{1}{1-\beta}  \underbrace{m\beta}_{\geq \frac{1-\beta}{2}} \underbrace{(b+1)}_{>1}\underbrace{(1+2m)\beta}_{\geq 1}>1/2\]
        which implies that $\psi\geq 0 \Rightarrow \theta<\underline{\theta}$ making \cref{lem:firstbestdec} redundant.

        \paragraph{Step 3. Exclusion Phase.} Finally, since $y_i=\theta$ in any continuation game, $P$ is indifferent between any initial endorsement, which concludes the proof.
    \end{proof}

    \subsection{Proof of Lemma \ref{lem:DEC1side}}

      \begin{proof} \textbf{Step 1: A candidate.} Consider the following strategy profile with $y_k=\theta$ and on-path strategies, $s_F=m$ and
        \[s_E= m - \frac{\theta - \beta p(m)}{\beta \left(\beta p(m)(b+1) - \theta\right)}=: \overline{s}_E.\]
        The first-period endorsement strategy is arbitrary; deviations are punished via grim trigger.
        \paragraph{Step 2: The candidate is a contract.} We first show that the candidate satisfies $F$'s \eqref{Eq Agent DEC}. $F$'s value conditional on leading is
        \[v_F(F)= b-\theta + \beta \left(p(m) v_F(E) + (1 -p(m)) v_F(F)\right),\]
        and $F$'s value conditional on \emph{not} leading is
        \[v_F(E)= -\theta + \beta\left(p(s_E) v_F(E) + (1 -p(s_E)) v_F(F)\right).\]
        Solving this system for $v_F(F)$ and $v_F(E)$, we obtain
        \[v_F(F) = \frac{(1 - \beta p(s_E))b }{(1-\beta)(1+\beta(m-s_E))} - \frac{\theta }{(1-\beta)},\text{~and~}
        v_F(E) = \frac{(1-p(s_E))\beta b}{(1-\beta)(1+\beta(m-s_E))} - \frac{\theta}{1-\beta}.\]
        Hence, $F$'s \eqref{Eq Agent DEC} holds if and only if
        \(
         v_F(F) \geq b+ \beta \alpha \Leftrightarrow
          s_E\leq \overline{s}_E,
        \)
        which, in turn, holds by assumption. Observe that $\overline{s}_E<m$ if $\underline{\theta}>\theta > \hat{\theta}$ and $\overline{s}_E=m$ if $\theta=\hat{\theta}$.

        Consistency with $E$'s \eqref{Eq Agent DEC} follows from recalling that $u_F+u_E=b-1$ in every period, which implies that
         \[v_E(E)= \frac{b-1}{1-\beta} - v_F(E) = \frac{b+\theta-1}{1-\beta} - \frac{(1-p(s_E))\beta b}{(1-\beta)(1+\beta(m-s_E))}.\]

        Our candidate satisfies $E$'s \eqref{Eq Agent DEC} if
             \begin{align*}
             v_E(E) &\geq b + \beta \alpha &&
             \Leftrightarrow && s_E \geq m -\frac{\theta - (1-\beta p(m)) + 2bm\beta}{\beta \left((b+1) (1-\beta p(m)) - \theta \right)}=:\underline{s}_E.
             \end{align*}
        What remains is to show that $\overline{s}_E \geq \underline{s}_E$ for all $\theta \in (\hat{\theta}, \underline{\theta})$ which is equivalent to showing
            \[\frac{4b\left((b+1)m(1+2m) \beta^2 -(1-\beta)\theta\right)}{\beta\Big((1+b)^2(1+2m) \beta (2-\beta(1+2m)) - 4\theta \left((1+b)-\theta\right)\Big)}>0.\]
        Because $\theta<\underline{\theta}$ by assumption, the numerator is positive. What remains is to show a positive denominator. We do this by sequentially deriving bounds. First, notice that the relevant term (within the larger brackets) decreases in $\theta$ because $b>0$ and $\theta<1/2$. Thus, that term is positive for all $\theta \in (\hat{\theta}, \underline{\theta})$ if and only if it is non-negative for $\theta=\underline{\theta}$, that is, iff
        \begin{equation}\label{eq:pos}
            \frac{(1+b)^2 (1+2m)\beta}{(1-\beta)^2} \Bigg(2-\beta(1-2m)\Bigg)\Bigg(1-\beta(1+2m)\Bigg)^2\geq 0.
        \end{equation}
        All terms in \eqref{eq:pos} are non-negative; thus, the denominator is non-negative.
        \paragraph{Step 3. Exclusion Phase.} Since $y_k=\theta$ in any continuation game, $P$ is indifferent between any initial endorsement, which concludes the proof.
    \end{proof}

\subsection{Proof of Lemma \ref{lem:feasiblethetadec}}

 \begin{proof}
        The claim holds for all $\theta\geq \hat{\theta}$ by \cref{lem:DEC1side}. Observe that $\hat{\theta}$ and $\underline{\theta}$ intersect at most once for $\beta>0$ because both are increasing in $\beta$ with $\underline{\theta}$ convex, $\hat{\theta}$ linear, and $\underline{\theta} \rightarrow \hat{\theta}$ as $\beta \rightarrow 0$. This intersection occurs at \(\beta =\frac{1}{1+2m(1+b)}=:\hat{\beta},\)
        and $\hat{\theta}<\underline{\theta}$ if and only if $\beta>\hat{\beta}$.
        Thus, what remains is to show the statement for $\underline{\theta} \geq \theta< \hat{\theta}$.

        \paragraph{Step 1. A candidate.} Consider the following candidate with an on-path strategy $y_F=\theta$ whenever $F$ leads, and an on-path strategy
        \[\tilde{y}_E = 1- \frac{2 m (b+1)+\theta(1/2-m)}{1-\beta(1/2-m)}\beta.\]
        whenever $E$ leads. Until $E$ leads for the first time, $s_{0}=-m$, and thereafter, $s=m$ in any on-path node that follows. Deviations are punished via grim trigger.

        \paragraph{Step 2. The candidate is a contract.}

        Take $E$'s (\ref{Eq Agent DEC}) at the candidate. Observe that
        $\tilde{y}_E$ solves $E$'s \eqref{Eq Agent DEC} with equality. Because first-best is not feasible by assumption, the resulting $\tilde{y}_E>\theta$. Now consider $F$'s (\ref{Eq Agent DEC}) which is
        \[(b-\theta) + \frac{\beta }{1-\beta}\Big((1/2-m)(b-\theta) - (1/2+m) \tilde{y}_E\Big) \geq b + \beta \alpha,\]
        and substituting for $\tilde{y}_E$ and $\alpha$ implies
        \[\frac{2m(b+1)+\theta(1/2-m)}{1-\beta(1/2-m)}\beta \geq \frac{1-\beta(1/2+m)}{\beta(1/2+m)}\theta,\]
        which, in turn, is equivalent to
        \[\theta \leq \frac{\beta^2 m(2m+1)}{1-\beta} (b+1)=\underline{\theta}.\]
        \paragraph{Step 3. Initial periods.} $F$'s (\ref{Eq Agent DEC}) has slack when playing the principal's bliss point, $y_F=\theta$, and receiving no endorsement. Thus, more endorsement, $s<m$, leaves $F$'s incentives undistorted. Because $E$ optimizes dynamically, endorsement decisions prior to $E$'s first lead are inconsequential.
    \end{proof}

    \subsection{Proof of Lemma \ref{lem:thetaDECoptimality}}

    \begin{proof}
        In steps 1 and 2, we focus on histories $h \not\in \mathcal{H}_F$ and show that no better contract for $P$ exists; in step 3, we close by addressing $h \in \mathcal{H}_F$.

        \paragraph{Step 1. The relevant class of contracts.} Invoking \cref{lem:DEC}, we restrict attention, without loss, to contracts such that at any history the leading party either chooses $y_i=\theta$ or that party's (\ref{Eq Agent DEC}) binds and $y_F\leq \theta \leq y_E$. \cref{lem:FullEndorsment} implies that within the class of contracts such that $y_F=\theta$ and $y_E \neq \theta$ we only need to consider those in which $s=m$ unless either $F$'s (\ref{Eq Agent DEC}) binds or $E$ plays $y_{E}(h)=\theta$ at least at some on-path history $h$.

        \paragraph{Step 2. No other candidate in that class is optimal.}

        \subparagraph*{Step 2a. If $F$'s \eqref{Eq Agent DEC} binds, $E$'s \eqref{Eq Agent DEC} must have slack thereafter.} Consider a history in which a leading $F$'s \eqref{Eq Agent DEC} holds with equality. Then, a leading $E$'s \eqref{Eq Agent DEC} must have slack if $E$ leads the period immediately after (history $h'$). To see this, recall first that $u_F+u_E=b-1$ in each period. Thus, if $F$'s \eqref{Eq Agent DEC} binds in $h$ and $E$'s \eqref{Eq Agent DEC} binds in $h'$, then $F$'s value of the game when in the lead at $h$ is $b+\beta \alpha$ while $F$'s value of the game when $E$ leads in the next period is $(b-1)/(1-\beta) - b - \alpha$. For now, assume in addition that if $F$ leads in $h'$, his \eqref{Eq Agent DEC} also binds. Then, $F$'s \eqref{Eq Agent DEC} becomes
        \[ \begin{split}
            &b-y_F + \beta \left((1/2-s_F) (b+\beta \alpha) + (1/2+s_F) ((b-1)/(1-\beta) - b - \beta \alpha)\right) = b + \beta \alpha \\\Leftrightarrow &y_F = \beta \left(\frac{b-1}{2(1-\beta)}-\alpha\right) + \psi s_F.
        \end{split}\]
        If $\theta<\underline{\theta}$, there is no $s_F \in [-m,m]$ such that $y_F\leq \theta$ solves $\eqref{yrDECDEC}$ and hence, if $E$ leads in $h'$, $E$'s (\ref{Eq Agent DEC}) must have slack. Moreover, relaxing $F$'s \eqref{Eq Agent DEC} in $h'$ only makes matters worse. Invoking \cref{lem:DEC}, that argument implies $y_{E}(h')=\theta$ in $h'$ if $F$'s \eqref{Eq Agent DEC} binds in $h$.

        \subparagraph*{Step 2b. No optimal contract exists in which $y_E(h)=\theta$.} Finally, we show that there is no optimal contract in which, at $h'$, $y_{E}(h)=\theta$. To see this, recall that under the candidate from \cref{lem:feasiblethetadec} when $E$ leads he is promised the best possible continuation payoff within the relevant class of contracts in all nodes in which $E$ is not playing: $P$ fully endorses $E$, and $F$ chooses the highest policy, $y_F=\theta$, available within the relevant class of contracts. Leaving these actions unchanged to derive an upper bound of what is possible, it is only possible to let $E$ choose $y_{E}(h)=\theta$ if $E$ would choose $y_{E}(h')>\theta$ if leading at some future history $h'$. Once again by \cref{lem:DEC}, no such contract exists in the relevant class of contracts that is such that $E$'s (\ref{Eq Agent DEC}) holds with slack in $h'$. But then, from $E$'s perspective, such a promise is no different from promising to return to the candidate after $h$, ruling out that $y_{E}(h)=\theta$ satisfies $E$'s \eqref{Eq Agent DEC} if $\theta < 1- \frac{\beta}{2} - (1 + 2 b) m\beta$ and hence the principal optimally asks for $y_E>\theta$ such that (DEC) binds given $(s=m,y_F=\theta)$.

        \paragraph{Step 3. Exclusion Phase.} The contract delivers $P$'s first best until $E$ leads for the first time and $s=-m$ minimizes the chances of $E$ leading. That completes the proof.
    \end{proof}

    \subsection{Proof of Lemma \ref{lem:thetaDECDECDEC}}

     \begin{proof}
            We restrict attention to $\theta\leq 1-\overline{\theta}$. \cref{lem:firstbestdec} covers the complementary case.

            \paragraph{Step 1. A candidate.} When $F$ led last, the principal's on-path endorsement strategy is
            \[s_F=\frac{\alpha \beta + \theta - \frac{b-1}{2}\frac{\beta}{(1-\beta)}}{\psi}<m.\]
            When $E$ led last, it is $s_E=m$. On path, $y_F=\theta$ and
            \(y_E=1-\beta \left(\frac{b-1}{2(1-\beta)}-\alpha\right)+ \psi m.\)
            Until $E$ leads for the first time, $s_0=-m$; deviations are punished via grim trigger.

            \paragraph{Step 2. The candidate is a contract.} Suppose that both \eqref{Eq Agent DEC} bind. That is,
            \[y_E=1-\beta \left(\frac{b-1}{2(1-\beta)}-\alpha\right)+ \psi m \qquad
                y_F = \beta \left(\frac{b-1}{2(1-\beta)}-\alpha\right) + \psi s_F. \tag{\ref{yrDECDEC}} \]
            The proposed $s_F$ solves the second equation by construction for $y_F=\theta$. However, analogously to the proof of \cref{lem:firstbestdec}, $y_F=\theta$ can only be sustained if $\underline{\theta}\leq\theta\leq \overline{\theta}$ conditional on $\psi\leq 0$. To see that $\psi<0$, assume for a contradiction that $\psi>0$ and recall that this implies $\beta (1+2m)>1$ or $1-\beta<2\beta m$. Moreover, recall that this case is only relevant if
            \[1/2 \geq \underline{\theta} \quad
            \Leftrightarrow \quad \frac{\beta}{1-\beta} \beta  m(2m+1) (b+1) \leq 1/2 \quad
            \Leftrightarrow \quad  2\beta^2  m  (2m+1) (b+1) \leq 1-\beta.
            \]
            But then because $\psi>0$, we get that \(2\beta^2 m (2m+1)(b+1) > (1-\beta)(b+1)>(1-\beta),\)
            a contradiction. Hence, $\psi<0$ whenever $\theta\geq \underline{\theta}$; $s_F$ decreases in $\theta$ and obtains $s_F=m$ at $\underline{\theta}$ which proves $s_F<m$. By assumption, $\theta<1-\overline{\theta}$ and hence $y_E> \theta$ even if $s_E=m$.
        \paragraph{Step 3. Exclusion Phase.} The reasoning of step 3 in the proof of \cref{lem:feasiblethetadec} applies.
    \end{proof}

    \subsection{Proof of Lemma \ref{lem:optimalityDECDECisthetaDEC}}

    \begin{proof}
        In Steps 1-3, we focus on $h \not\in \mathcal{H}_F$. We turn to $h\in \mathcal{H}_F$ in Step 4.

        \paragraph{Step 1. The candidate is the optimal contract with binding \eqref{Eq Agent DEC}.} Consider $P$'s problem
        \begin{equation*}
            \begin{split}
            v_P(s_E;s_F)&= \max_{s_E} -(1-\beta)|\theta-y_E(s_E)| + \beta\left(p(s_E) v_P(s_E;s_F) + (1-p(s_E)) v_P(s_F;s_E)\right)\\
            v_P(s_F;s_E)&= \max_{s_F} -(1-\beta)|\theta-y_F(s_F)| + \beta\left(p(s_F) v_P(s_E;s_F) + (1-p(s_F)) v_P(s_F;s_E)\right),
            \end{split}
        \end{equation*}
        where parties' policies are given by their binding \eqref{Eq Agent DEC}, which give the equations \eqref{yrDECDEC}. Using the two Bellman equations above, we can solve for $P$'s value of selecting the optimal strategy $s_F$ in all periods in which $F$ leads, taking the choice $s_E$ in periods in which $E$ leads as given. Thus, $P$'s objective becomes
        \begin{equation}\label{eq:VPRsolved}\tilde{v}_P(s_E;s_F) = (\theta-y_E(s_E))\rho_E(s_E;s_F)) + (y_F(s_F)-\theta) (1-\rho_E(s_E;s_F)) \tag{VPR}\end{equation}
        and
        \begin{equation}\label{eq:VPLsolved}\tilde{v}_P(s_F;s_E) = (\theta-y_E(s_E))\rho_F(s_F;s_E)+(y_F(s_F)-\theta)(1-\rho_F(s_F;s_E)).\tag{VPL}\end{equation}
        with
         \[\rho_E(s_E;s_F)=\frac{1-\beta(1-p(s_F))}{1{-}\beta(1- p(s_F))+\beta(1- p(s_E))} \text{ and } \rho_F(s_F;s_E)=\frac{\beta p(s_F)}{1{-}\beta p(s_E){+}\beta p(s_F)}.\]
        Observe that in objective $\tilde{v}_P(s_i;s_j)$, $s_i$ is the choice whereas $s_j$ is assumed to be chosen optimally the next time $j$ leads. Taking derivatives of $P$'s objectives yields
        \begin{equation}\label{eq:FOCL}
            \frac{\partial \tilde{v}_p(s_F;s_E)}{\partial s_F}{=} \frac{\left(2{-}\beta(1{+} 2 s_E)\right)}{2(1{-}\beta(s_E{-}s_F))^2}\Bigg(2 \beta \theta {-} \beta \left(y_F(s_F){+} y_E(s_E)\right) {+}  y_F'(s_F)\left(1{+}\beta(s_F{-}s_E)\right)\Bigg)
        \end{equation}
        and
        \begin{equation}\label{eq:FOCR}
            \frac{\partial \tilde{v}_p(s_E;s_F)}{\partial s_E}= \frac{\left(2{-}\beta(1{-} 2 s_F)\right)}{2(1{-}\beta(s_E{-}s_F))^2}\Bigg(2\beta \theta {-} \beta(y_F(s_F){+}y_E(s_E)) {-} y_E'(s_E)\left(1{+}\beta(s_F{-}s_E)\right)\Bigg)
        \end{equation}
        which, after replacing via \cref{yrDECDEC,yrDECDEC}, can be written as
        \begin{equation}\label{eq:FOCL'}
            \begin{split}
                \frac{\partial \tilde{v}_p(s_F;s_E)}{\partial s_F}&{=} \frac{\left(2{-}\beta(1{+} 2 s_E)\right)}{2(1{-}\beta(s_E{-}s_F))^2}\Bigg(2 \beta \theta {-} \beta \left(1+ \psi(s_F+s_E)\right) +  \psi + \psi\beta(s_F-s_E)\Bigg)\\
                &= \frac{\left(2-\beta(1+ 2 s_E)\right)}{(1-\beta(s_E-s_F))^2}\Bigg(- \beta\left(\frac{1}{2}-\theta\right) + \left(\frac{1}{2}-\beta s_E\right)\psi \Bigg)
            \end{split}
        \end{equation}
        and
        \begin{equation}\label{eq:FOCR'}
            \begin{split}
                \frac{\partial \tilde{v}_p(s_E;s_F)}{\partial s_E}&= \frac{\left(2{-}\beta(1{-} 2 s_F)\right)}{2(1{-}\beta(s_E{-}s_F))^2}\Bigg(2\beta \theta {-} \beta(1+\psi(s_F+s_E)) {-} \psi - \psi \beta(s_F{-}s_E)\Bigg)\\
                &=\frac{\left(2{-}\beta(1{-} 2 s_F)\right)}{(1{-}\beta(s_E{-}s_F))^2}\Bigg( - \beta\left(\frac{1}{2}-\theta\right) - \left(\frac{1}{2}+\beta s_F\right) \psi\Bigg).
            \end{split}
        \end{equation}
        Recall that the relevant case here is one in which \(2\frac{\beta}{1-\beta} (b+1) m \beta\left(\frac{1}{2} + m \right)=\underline{\theta}<1/2\), as otherwise our case is not relevant. Thus, relevance implicitly gives a constraint on the parameter space. But then, with some algebra, we can show that \eqref{eq:FOCL'} is always negative, implying that the principal will choose the smallest $s_F$ that guarantees $y_F\leq \theta$ \emph{and} \eqref{Eq Agent DEC} binding for $F$ no matter the principal's choice of $s_E$. But then, our candidate contract precisely describes the smallest $s_F$ such that $y_F\leq \theta$ for a binding \eqref{Eq Agent DEC}.\footnote{Observe that without \eqref{Eq Agent DEC} binding for $F$, we must have $y_F=\theta$ by \cref{lem:DEC} which is ruled out because $\theta>\underline{\theta}$ (see proof of \cref{lem:DEC1side} for the relevant inequality).}
        Replacing $s_F$ with $s_F^\ast$ in \eqref{eq:FOCR'} and rearranging implies that $\eqref{eq:FOCR'}$ increases in $s_E$. But then, no constraint is violated even for the corner solution $s_E=m$, which is then optimal.
        Thus, if each party's (\ref{Eq Agent DEC}) binds, the candidate contract is optimal.

        \paragraph{Step 2. $E$'s \eqref{Eq Agent DEC} binds at the optimum.} Suppose that there exists an optimal contract in which $E$'s (\ref{Eq Agent DEC}) has slack when leading at some history $h$. Then, because that contract is optimal and $E$'s (\ref{Eq Agent DEC}) holds with slack, there is an optimal contract in which $E$'s (\ref{Eq Agent DEC}) has slack again in the period immediately thereafter, conditional on leading again. Using \cref{lem:DEC}, it is without loss to assume that $y_{E}(h')=\theta$ for any $h' \supset h$ such that between $h$ and $h'$, $F$ did not lead. Moreover, if such a contract is feasible, it is also feasible assuming that whenever $F$ leads the next time, $F$'s (\ref{Eq Agent DEC}) holds with equality. Because $u_F+u_E=b-1$ in every period, such a continuation game implies $E$'s \emph{best} possible continuation game conditional on $F$ leading, i.e., $\overline{v}^E_E(F)$.
        But then, when $\theta<1-\overline{\theta}$ which holds whenever $\overline{\theta}<1/2$, all contracts that satisfy $E$'s (\ref{Eq Agent DEC}) with equality for a fixed $y_E$ until the next time $F$ leads imply $y_E>\theta$. By monotonicity of the (\ref{Eq Agent DEC}), no contract with $y_E\leq \theta$ exists that satisfies $E$'s (\ref{Eq Agent DEC}). A contradiction to the premise that $E$'s (\ref{Eq Agent DEC}) has slack at some history $h$.

        \paragraph{Step 3. No other candidate is optimal.} Observe that by \cref{lem:DEC}, whenever $F$ is not at his (\ref{Eq Agent DEC}) we need $y_F=\theta$. Moreover, we know from the previous argument that $E$'s (\ref{Eq Agent DEC}) holds with equality but $y_E=1-\overline{\theta}>\theta$ where the last step follows because we assume $\theta>\underline{\theta} \Rightarrow 1-\overline{\theta}>1-\breve{\theta}$.\footnote{This relation is purely algebraic. We discuss it in greater detail in the proof of \cref{lemma:daggertheta} below and that of \cref{prop:regimesnocommitment} in \cref{proofregimes}.} Invoking \cref{lem:FullEndorsment} and noting that $(s=m, y_F=\theta)$ again does not satisfy $F$'s (\ref{Eq Agent DEC}) because $\theta>\underline{\theta}$, we conclude that the candidate contract is optimal.

        \paragraph{Step 4. Exclusion Phase.} The reasoning of step 3 in the proof of \cref{lem:feasiblethetadec} applies.
    \end{proof}

    \subsection{Proof of Lemma \ref{lem:optimalitytopright}}

    \begin{proof}\textbf{Step 1. A candidate.} Principal chooses $s_E{=}-s_F{=}m$ and parties:
          \begin{equation*}
                \begin{split}
                    \tilde{y}_F(s_F)&=  \beta \left(\frac{b-1}{2(1-\beta)}-\alpha\right)-\psi m,\\
                    \tilde{y}_E(s_E)&=1-\beta \left(\frac{b-1}{2(1-\beta)}-\alpha\right)+\psi m,
                \end{split}
         \end{equation*}
         which make their \eqref{Eq Agent DEC} bind; deviations are punished via grim trigger.
         \paragraph{Step 2. The candidate is a contract.} Because $1/2\geq \theta>\underline{\theta}$, it follows that $\psi<0$ and thus, $y_F(s_F= -m)$ and $y_E(s_E= m)$ satisfy \eqref{Eq Agent DEC}---as shown in \eqref{yrDECDEC}.

         \paragraph{Step 3. The candidate is optimal.} Imagine party $i$ is currently leading. By $u_F+ u_E = b -1$, promising that party $j$ is brought to his (\ref{Eq Agent DEC}) the next time he leads implies $i$'s best payoff conditional on any future selection of $j$, i.e., $\overline{v}^i_i(j)$. By monotonicity of \eqref{yrDECDEC} in $s_i$ and the fact that the party who led last receives $P$'s full endorsement, no other contract satisfying parties' \eqref{Eq Agent DEC} implies a $y_i$ that is (on average) closer to $\theta$. \qedhere
    \end{proof}

\subsection{Proof of Lemma \ref{Lemma P Punishment 1}}

\begin{proof}

        A principal's optimal penal code exists because the equilibrium payoff set is compact.

        Observe first that the principal's punishment starts mid-period after the principal deviated at the beginning of the period. Second, observe that in that period, if $E$ leads, $y_E=1$ because this minimizes $P$'s myopic payoff and no incentives have to be taken into account at such an initial node. Moreover, if, after an initial lead by $E$ and $P$'s subsequent endorsement decision, $E$ takes the lead again, the same argument holds and $y_E=1$.\footnote{If $P$'s \eqref{Eq:PrincipalDEC} were binding, she would be indifferent between choosing whatever $s$ is prescribed at that node and $s=-m$. It is thus without loss to assume $s=-m$ in that case and restart the punishment.}

        Now assume $F$ leads after $P$'s deviation. Then, either $y_F<\theta$ or $y_F\geq \theta$. In either case, we can minimize the principal's continuation value by promising $y_E=1$ should $E$ lead in the next period. To see this, observe that the only reason to have $y_E<1$ would be to provide incentives for $F$ to increase his initial $y_F$. If $y_F<\theta$, a marginal increase is counterproductive because it benefits the principal only (and so does $E$'s promised moderation). If $y_F \geq \theta$, to make the principal worse off we would want to increase $y_F$. We could do it by then decreasing $y_E$ thereafter. However, because $y_F$ and $y_E$ are already to the right of $\theta$, $F$ and $P$ have perfectly aligned interests over marginal policy shifts. Decreasing $y_E$ then provides the same policy benefits to $F$ as it provides to $P$. Thus, if the reduction in $y_E$ is then counteracted by an increase in $y_F$ to offset $F$'s continuation payoffs, that change also exactly offsets $P$'s payoffs, making such a construction no worse than the one with $y_E=1$. Because $y_F\leq y_E$, $P$ has, in addition, no change in incentives through this trade. Therefore, it is without loss to have $y_E=1$. Iterating forward on this argument proves \cref{Lemma P Punishment 1}.
        \end{proof}

\subsection{Decomposition in policy and leadership values} \label{subs:decomposition}

To carry out some of the non-commitment proofs, it is useful to separate the continuation values of the friend into two parts. To that end, we introduce two pieces of additional notation: the \emph{policy part} of party $F$'s continuation value, $\Gamma_F$, and the \emph{leadership part}, $\Omega_F b$, such that, for example, $w^\ast_F = \Gamma^\ast_F + \Omega^\ast_F b$.

More rigorously, fix a strategy profile $\sigma$. Then, $\sigma$ induces a probability law $\mathbb{P}^\sigma$ on the path of outcomes $(k_t,y_t)_{t\geq0}$ starting from the initial public history. Let $\mathbb{E}^\sigma$ be the associated expectation. Define \[\Gamma^\sigma_F:= \mathbb{E}^\sigma\left[-\sum\limits_{t=0}^\infty \beta^t y_t\right],\] the net-present value of future policy choices, and \[\Omega^\sigma_F:=\sum_{t=0}^\infty \mathbb{P}^\sigma[k_t=F] \beta^t,\] the friend's discounted expected time in the lead. When we look at an equilibrium, then $\Omega^\sigma_F$ is governed by $s$ itself. Therefore, it is convenient to abuse notation in such cases and write $\Omega_F(s)$ for the discounted expected leadership time the principal allocates to the friend under (equilibrium) strategy $s$.

Consider now a contract such that every on-path policy is to the right of $\theta$. We can then write $F$'s policy value in terms of the principal's value:
$$
\Gamma_F^{\ast} = w_P^\ast - \frac{\theta}{1-\beta},
$$
and we can rewrite $F$'s on-path value as:
 \begin{align}\label{Eq F's onpath value decomposition}
w_F^\ast = w_P^\ast - \frac{\theta}{1-\beta} + \Omega_F(s^\ast)b.
\end{align}

\subsection{Proof of Lemma \ref{lem:backtobusinessDECDEC}}
\begin{proof}
    We prove \cref{lem:backtobusinessDECDEC} constructively, which also proves \cref{cor:b2b}.

    Consider the optimal contract, and assume (without loss of generality for our argument) that $E$ is selected at the initial period, i.e., the embracing phase starts right away at $h_0$. Fixing this particular realization at $h_0$ allows us to use the following notation: $\mathcal{H}_E$ is the set of histories following $h_0$ where only $E$ has been selected, and $\mathcal{H}^\ast_F$ is the set of histories in which $F$ is selected for the first time.

    \paragraph{Part 1.} Suppose first that the optimal contract is such that $P$'s \eqref{Eq:PrincipalDEC} binds at $h_1 \in \mathcal{H}_E$. That means that $w_P(h_1) = \underline{w}_P$. We can write $P$'s payoff for $h_0$ after $E$ is selected as:
    $$
    v^\ast_P(h_0, E) = \theta - y^\ast_E(h_0) + \beta \underline{w}_P.
    $$
    Because the contract is optimal for the principal, there exists no $y_E<y_E^\ast(h_0)$ that is incentive compatible for $E$. Thus, $E$ is at his \eqref{Eq Agent DEC} which implies
    $$
    v^\ast_E(h_0,E) = b + y^\ast_E(h_0)-1 + \beta w^\ast_E(E) = b + \beta \underline{w}_E.
    $$

    In addition, optimality for the principal guarantees that there is no $w_E(E)>w_E^\ast(E)$ that the principal can credibly offer $E$. If there were, she would do so in exchange for lowering $y^\ast_E(h_0)$.\footnote{Because we assume that \eqref{Eq:PrincipalDEC} \emph{binds}, $y_E^\ast(E)\geq \theta$.} Hence, the continuation play of the optimal contract attains the maximum value for $E$ and, by the constant-sum property, the minimum value for $F$.

    Therefore, an optimal penal code for $F$ is the continuation play of the optimal contract as if $E$ had been the one leading last.

    \paragraph{Part 2.} Suppose now that the optimal contract, $C^\ast$, is such that the principal's \eqref{Eq:PrincipalDEC} has slack at $h_1 \in \mathcal{H}_E$.
    By \cref{lem:DEC}, we can assume $E$'s \eqref{Eq Agent DEC} binds whenever he leads.

        \subparagraph{Claim.} \emph{If the optimal contract is such that the principal has slack after $E$'s lead in $h_0$, it is without loss to assume $s^\ast(h_1)=m$.}
        \begin{proof}
            Note first that since $F$ has not yet led, his incentives are unaffected by $s(h_1)$. Now, suppose there exists a contract $C$ such that $s(h_1)=s_1<m$ and $y_E(h_0)=y_E>\theta$, which is without loss because the first best is not feasible by assumption. Then, there exists an alternative contract $C'$ identical to $C$ except for two actions:
            \begin{itemize}
            \item[(i)] The principal chooses $s(h_1)=s_1'>s_1$;
            \item[(ii)] the enemy chooses $y_E(h_0)=y'_E<y_E$ such that $E$'s \eqref{Eq Agent DEC} holds with equality.
            \end{itemize}
            Note that there exists an incentive-compatible $s_1'$ because, by assumption, the principal's \eqref{Eq:PrincipalDEC} has slack under $s_1$. Moreover, $s'_1>s_1$ implies a larger continuation payoff to $E$, so $y'_E<y_E$ exists too. Lastly, compare the principal's payoff in $C'$ relative to $C$: As $s_1'$ provides benefits in both the office and policy dimension for the enemy, and then $y'_E$ makes \eqref{Eq Agent DEC} bind, any loss in future policy choices from the point of view of the principal created by $s_1'$ is (at least) offset by the decrease in $y'_E$. Thus, $v_P(h_0; C') \geq v_P(h_0; C)$. Iterating over this argument, we either have a principal-optimal contract at which \eqref{Eq:PrincipalDEC} binds at $h_1$ or $s(h_1)=m$.
        \end{proof}

        Since $E$'s \eqref{Eq Agent DEC} holds at every on-path history with $E$ leading, the above claim implies that the principal also has slack at $h_t \in \mathcal{H}_E$ and $s^\ast(h_t) = m$.

        Having established the properties of $C^\ast$, we will now construct $F$'s penal code based on that contract. To that end, consider the following alternative contract following $h_0$, $\hat{C}$, identical to $C^\ast$ apart from the following difference:
        \begin{itemize}
            \item For every history $h \in \mathcal H_E$ and $h \neq h_0$, $\hat{y}_E(h)=\hat{y}_E>y_E^\ast(h)=y_E^\ast$ is chosen either such that $P$'s \eqref{Eq:PrincipalDEC} holds with equality, or, if that is not possible, such that $\hat{y}_E=1$.
        \end{itemize}

        By assumption, the principal's \eqref{Eq:PrincipalDEC} has slack in $C^\ast$, so $\hat{y}_E>y_E^\ast$ exists.

        Now, suppose that $\hat{y}_E=1$. Consider that $F$ is at his \eqref{Eq Agent DEC} whenever taking the lead and $\hat{y}_E=1$. Then, there exists a continuation play that implements the same payoff as the friend's grim-trigger punishment under commitment. That contract thus constitutes an optimal penal code for $F$ and possesses the back-to-business property.

        Suppose now that $\hat{y}_E<1$. Next, denote $\varepsilon:= \hat{y}_E- y_E^\ast$, and let $w_P^\ast(E)$ be the principal's continuation payoff in the optimal contract conditional on $E$ having led last. The principal's payoff from $\hat{C}$ at $h_0$ is then:
        \[\begin{split}
            \hat{v}_P(h_0,E)&= \theta - y_E^\ast + \beta \underbrace{ \big(w^\ast_P(E) - \frac{ p(m)}{1-\beta p(m)}\varepsilon \big)}_{=\underline{w}_P}  \\&=v^\ast_P(h_0,E) - \beta\frac{ p(m)}{1-\beta p(m)}\varepsilon;
        \end{split}\]
        and the enemy's payoff from $\hat{C}$ at $h_0$ is:
        \[\begin{split}
            \hat{v}_E(h_0,E)&=b -(1 - y_E^\ast) + \beta \big(w^\ast_E + \frac{ p(m)}{1-\beta p(m)}\varepsilon \big)  \\
            &=v^\ast_E(h_0,E) + \beta\frac{ p(m)}{1-\beta p(m)}\varepsilon.
        \end{split}\]

        Because the enemy is at his \eqref{Eq Agent DEC} in $C^\ast$, he now has an \emph{initial} slack of $\beta\frac{ p(m)}{1-\beta p(m)}\varepsilon$ in $\hat{C}$. Thus, in principle we can enforce a reduction in $E$'s payoff by $\beta\frac{ p(m)}{1-\beta p(m)}\varepsilon$ without violating $E$'s \eqref{Eq Agent DEC}.

        Now consider a contract $\hat{C}'$ identical to $\hat{C}$ with the only exception that at $h_0$ it prescribes:
        $y_E(h_0)= \hat{y}'_E=y_E^\ast - \beta \frac{p(m)}{1-\beta p(m)}\varepsilon$.
        If $\hat{y}'_E \geq \theta$, then this contract is incentive compatible for all players and \emph{optimal} for the principal. Moreover, no other contract that is incentive compatible for all players can yield a higher payoff to $E$ \emph{after} the initial choice $\hat{y}'_E$. Since the continuation contract is an equilibrium, it is an $E$-optimal contract, and thus an optimal penal code for $F$ with the back-to-business property.

        What remains is the case in which $\hat{y}'_E < \theta$. In that case, the contract $\hat{C}'$ is not optimal because the enemy takes too low an action. However, this is irrelevant for the continuation contract which still gives $E$ the most slack feasible in the initial period. Thus, also then, the continuation play of $\hat{C}$ from $h_1$ is an optimal penal code for $F$ with the back-to-business property.
\end{proof}

\subsection{Proof of Lemma \ref{lem:b2bFullEmbaracing}}

\begin{proof}
    We first establish that $C^F$ satisfies the parties' enforcement constraints. In the punishment phase, this follows for the case of the enemy from the fact that he needs to moderate less than in the optimal contract, while the remaining behavior is unchanged; and for the principal, directly from the definition of $\hat{y}_E$. Naturally, in the back-to-business phase, this follows immediately from the optimal contract.

    Now, we show that if $\hat{y}_E \leq 1$, $C^F$ is indeed a friend's optimal penal code. Note first that every policy choice, $y_i$, is to the right of $\theta$. Therefore, we can decompose $F$'s value at the beginning of the penal code into a policy and a leadership component as follows: \[w_F^F(h_0) = w_P^F(h_0) - \frac{\theta}{1-\beta} + \Omega_F(s^F)b.\] Since $P$'s \eqref{Eq:PrincipalDEC} binds at the beginning of the penal code, $w_P^F(h_0) =\underline{w}_P$, and hence the policy component is minimized. Second, since $s = m$, the leadership component is also minimized. Thus, $C^F$ provides $F$ with both the lowest feasible policy payoff and leadership rent, making it a worst contract for $F$ and thus an optimal penal code for $F$.

    Lastly, if $\hat{y}_E >1$, it may be that $F$'s \eqref{Eq Agent DEC} binds at the embracing phase, in which case the optimality of the penal code follows from Lemma \ref{lem:backtobusinessDECDEC}. Alternatively, $F$'s \eqref{Eq Agent DEC} can have slack, in which case this penal code is not necessarily optimal but still valid to implement the optimal full-embracing contract by construction.
\end{proof}

\subsection{Proof of Lemma \ref{lemma:daggertheta}}

\begin{proof}
    To see that, observe from their respective definitions that
        \begin{equation}\label{eq:thetadagger}
            \overline{\theta} = \frac{1-\beta(1/2+m)}{\beta(1/2+m)}\underline{\theta} \quad \Rightarrow\quad 1-\overline{\theta}= 1- \frac{1-\beta(1/2+m)}{\beta(1/2+m)}\underline{\theta}=\underline{\theta} \Leftrightarrow \underline{\theta}=\theta^\dagger.
        \end{equation}
        Observe that $\overline{\theta}=\breve{\theta}$ implies
        \begin{equation}\label{eq:breveisoverline}
            \breve{\theta}= \beta(1/2+m)+2 m b \beta=\underbrace{2 \frac{\beta}{1-\beta} m \left(1-\beta(1/2+m)\right)}_{=:\kappa} (b+1)=\overline{\theta}
        \end{equation}
        Let $b^\diamond$ be the unique solution to \eqref{eq:breveisoverline}, that is
        \[b^\diamond+1= \frac{\beta(1/2- m)}{ \kappa-2 m \beta} = \frac{1-\beta}{2\beta m}.\]
        Plugging this back into \eqref{eq:breveisoverline} implies that when $\overline{\theta}=\breve{\theta}$,
        \[1-\overline{\theta}=1-\kappa \frac{1-\beta}{2\beta m} = \beta(1/2+m)=\theta^\dagger.\]
        Thus, all three cutoffs coincide at $(\theta^\dagger, b^\diamond)$.
\end{proof}

\subsection{Proof of Lemma \ref{lem:nonMono_noCommit_blarge}}

    \begin{proof}
         For $b>\overline{b}_0$, we can use \cref{prop:theta0,res:OEndorsement} to compute

    \[w_P(\theta \to 1/2)-w_{P}(\theta=0) = \frac{2 p(m) - 1}{2(1-\beta)^2(1-\beta p(m))}\left( 2 \beta H(m,\beta)b - (1-\beta p(m))(2\beta^2 p(m)-3\beta+1)\right)\]

    with $H(m,\beta):= \beta^2 p(m)^2 \beta(3 p(m)-1)+p(m)$.

    Note that this difference is linearly increasing in $b$ and equal to zero for a cutoff \[b=b_{\Delta}:= (1-\beta p(m))(2 \beta^2 p(m)-3\beta+1)/(2 \beta H(m,\beta)).\]

    Now $\overline{b}_0= \frac{(1-\beta)^2}{\beta p(m) (2-\beta(2-p(m)))}>b_{\Delta}$ which can be verified by cross-multiplying the denominators and looking at the sign of the numerator, which becomes:

    \[2(1-\beta)^2 H(m,\beta) - p(m)(2-\beta(2-p(m)))(1-\beta p(m))(2\beta^2 p(m)-3\beta+1)>0\]

    Substituting $q=1-p(m) \in (0,1/2)$ and $r=1-\beta \in (0,1)$ and simplifying, we get that this holds if

    \[2r^2(1+(1-r)^3(1-q)(2-3q))>(1-q+r(1+q))(q+r-rq)\Bigg(r(2r-1)-2q(1-r)^2\Bigg)\]

    which can be verified by direct computation to hold.
\end{proof}

\section{Construction of the Non-commitment Results}\label{sec:MoreNoCommitment}

In this section, we present the structure behind the full non-commitment characterization. Our supplementary Mathematica code uses this structure to solve the systems of equations and select the relevant optimal branch. In the embracing phase, the on-path Bellman equations are, for each \(\ell\in\{F,E\}\),
\[
w_i^*(\ell)
=
(1-p(s^\ast_\ell))v_i^*(F)+p(s^\ast_\ell)v_i^*(E).
\]

The post-selection payoffs are
\begin{align*}
    v_F^*(F)&=b-y^\ast_F+\beta w_F^*(F),  &
v_F^*(E)&=-y^\ast_E+\beta w_F^*(E), \\
v_E^*(F)&=-(1-y^\ast_F)+\beta w_E^*(F), &
v_E^*(E)&=b-(1-y^\ast_E)+\beta w_E^*(E),
\end{align*}
and since \(y^\ast_F\leq \theta\leq y^\ast_E\),
\[
v_P^*(F)=-(\theta-y^\ast_F)+\beta w_P^*(F),
\qquad
v_P^*(E)=-(y^\ast_E-\theta)+\beta w_P^*(E).
\]

\subsection{Punishment values}

\paragraph{Enemy punishment.}
The enemy is punished by the Nash equilibrium of the stage game:
\[
s^E=-m,\qquad y_F^E=0,\qquad y_E^E=1.
\]
Hence
\[
w_E^E
=
\frac{(1/2-m)b-(1/2+m)}{1-\beta}.
\]

\paragraph{Principal punishment.}
The principal's penal code has
\[
s^P=-m,\qquad y_E^P=1,
\]
and either \(y_F^P=0\), corresponding to the NES punishment contract, or
\[
y_F^P=\widehat y_F^P>0,
\]
corresponding to what we refer to as the \emph{sophisticated} punishment contract. The latter is determined by $F$'s dynamic enforcement constraint inside $P$'s penal code, which is
\[
v_F^P(F)=b+\beta w_F^F.
\]
Substituting the right-hand side yields:
\[
b-\widehat y_F^P+\beta w_F^P
=
b+\beta w_F^F,
\]
so that
\[
\widehat y_F^P
=
\beta\bigl(w_F^P-w_F^F\bigr).
\]
The relevant punishment payoff for $P$ is
\[
w_P^P=\min\{w_P^{P,NES},w_P^{P,S}\},
\]
where $S$ denotes the sophisticated punishment, which is used only if it is admissible and yields a lower payoff for $P$.

\paragraph{Friend punishment.}
The friend's penal code features the back-to-business property: the punishment phase lasts until $F$ leads
for the first time. Once $F$ leads, continuation returns to the continuation play of the embracing phase of the optimal contract after $F$ led last.

Let $s_0^F$ be the endorsement used in the punishment phase, and let $y_E^F$ be
$E$'s policy during that phase. Then,
\[
w_i^F(E)
=
(1-p(s_0^F))v_i^F(F)+p(s_0^F)v_i^F(E).
\]
If $F$ is selected, play returns to the embracing phase of the optimal contract, so $v_i^F(F) = v_i^\ast(F)$. For this reason, we will solve for the friend's penal code together with the optimal contract.
If $E$ is selected, the punishment phase continues:
\begin{align*}
    v_F^F(E)&=-y_E^F+\beta w_F^F(E), \\
    v_E^F(E)&=b-(1-y_E^F)+\beta w_E^F(E), \\
    v_P^F(E)&=-(y_E^F-\theta)+\beta w_P^F(E).
\end{align*}
The principal's \eqref{Eq:PrincipalDEC} in $F$'s penal code is:
\[
w_P^F(E)\geq w_P^P.
\]
When this constraint binds, \(y_E^F\) solves $P$'s \eqref{Eq:PrincipalDEC}:
\[
w_P^F(E)=w_P^P.
\]
If the cap \(y_E^F=1\) is reached first, then the friend punishment coincides with the commitment punishment in the punishment phase.

\subsection{Enforcement Constraints}

The enemy's on-path \eqref{Eq Agent DEC} is:
\[
v_E^*(E)\geq b+\beta w_E^E,
\]
which always binds except in the first-best case.

The friend's on-path \eqref{Eq Agent DEC} is:
\[
v_F^*(F)\geq b+\beta w_F^F.
\]
This constraint binds whenever the friend must be given direct incentives (through $s$), namely in
partial embracing (with and without the principal's \eqref{Eq:PrincipalDEC} binding on-path) and opportunism (with and without the principal's \eqref{Eq:PrincipalDEC} binding on-path).

The principal's on-path \eqref{Eq:PrincipalDEC} is:
\[
w_P^*(\ell)\geq w_P^P,
\qquad
\ell\in\{F,E\},
\]
where the relevant constraint is typically
\[
w_P^*(E)\geq w_P^P,
\]
because the principal is tempted to renege after promising support to the enemy.

\subsection{Candidate systems}

For fixed $(m,\beta,\theta)$, the algorithm solves candidate systems along the $b$
dimension. Conceptually, as $b$ decreases, as in the commitment case, we move from \emph{first-best} to \emph{full embracing} to \emph{partial embracing}, and then either to \emph{opportunism} or directly to \emph{NES}. It also holds that $P$'s \eqref{Eq:PrincipalDEC} may only hold with equality on the equilibrium path in the partial embracing setting and in the opportunism setting, yet never close to the boundary of \emph{full embracing} in the partial embracing case.

\paragraph{First best.}
Impose
\[
y^\ast_F=y^\ast_E=\theta.
\]
The Bellman equations are solved together with the feasibility constraints
\[
v_E^*(E)\geq b+\beta w_E^E,\qquad
v_F^*(F)\geq b+\beta w_F^F,
\]
\[
w_P^*(F)\geq w_P^P,
\qquad
w_P^*(E)\geq w_P^P,\quad
\text{and}\quad
w_P^F(E)\geq w_P^P.
\]

\paragraph{Full embracing.}
Impose
\[
s^\ast_F=s^\ast_E=m,
\qquad
y_F^\ast=\theta,
\qquad
y_E^\ast\in[\theta,1),
\]
where $y_E^\ast$ solves:
\[
v_E^*(E)=b+\beta w_E^E.
\]
The friend's punishment features back-to-business. Hence, in the punishment phase
\[
s_0^F=m,
\]
and \(y_E^F\) solves:
\[
w_P^F(E)=w_P^P,
\]
unless the cap $y_E^F\leq1$ binds first, in which case $y_E^F=1$.

Feasibility requires
\[
v_F^*(F)\geq b+\beta w_F^F\quad
\text{and}\quad
w_P^*(E)\geq w_P^P.
\]
In this regime, $F$ is disciplined by the threat of losing $E$'s moderation. No direct endorsement is required to incentivize him.

\paragraph{Partial embracing with $P$'s \eqref{Eq:PrincipalDEC} slack on path.}
Impose
\[
s_E^\ast=m,
\qquad
-m<s_F^\ast<m,
\qquad
y_F^\ast=\theta,
\qquad
y_E^\ast\in[\theta,1).
\],
where $y_E^\ast$ and $s_F^\ast$ solve:
\[
v_E^*(E)=b+\beta w_E^E,
\quad\text{and}\quad
v_F^*(F)=b+\beta w_F^F.
\]
The friend's punishment again satisfies back-to-business, with $y_E^F$ in the punishment phase solving:
\[
w_P^F(E)=w_P^P,
\]
unless the cap \(y_E^F\leq1\) binds first, in which case $y_E^F=1$. Feasibility requires $P$'s \eqref{Eq:PrincipalDEC} to have slack on path:
\[
w_P^*(E)>w_P^P.
\]
In this regime, $P$ can still fully endorse $E$ after $E$ led, but must lower \(s^\ast_F\) after $F$ led
to induce him to choose \(y^\ast_F=\theta\). Note that at the boundary of the full-embracing region, lowering $b$ marginally always implies that we enter this regime. The reason is that at the boundary of the full-embracing regime, $P$'s on-path \eqref{Eq:PrincipalDEC} always holds with slack.

\paragraph{Opportunistic boundary with $P$'s \eqref{Eq:PrincipalDEC} slack on path.}
If solving the partial-embracing system yields \(s^\ast_F<-m\), impose
\[
s^\ast_F=-m,
\qquad
s^\ast_E=m.
\]
Then, $y_F^\ast$ and $y_E^\ast$, satisfying $y^\ast_F\leq \theta\leq y^\ast_E$, solve:
\[
v_E^*(E)=b+\beta w_E^E,
\quad\text{and}\quad
v_F^*(F)=b+\beta w_F^F.
\]
In this regime, the principal's \eqref{Eq:PrincipalDEC} has slack on path:
\[
w_P^\ast(E)>w_P^P.
\]
Here, the principal has exhausted her endorsement incentives for $F$. Further adjustment takes the form
of weaker policy moderation.

\paragraph{Partial embracing with $P$'s \eqref{Eq:PrincipalDEC} binding on path.}
When $P$ can no longer credibly promise full endorsement to $E$ after $E$ led, \(s^\ast_E=m\) is
relaxed and $P$'s on-path constraint binds:
\[
w_P^\ast(E)=w_P^P.
\]
The core system is
\[
v_E^*(E)=b+\beta w_E^E,
\qquad
v_F^*(F)=b+\beta w_F^F,
\quad\text{and}\quad
w_P^*(E)=w_P^P.
\]
Because $P$'s \eqref{Eq:PrincipalDEC} binds on-path, the friend-punishment phase replicates the continuation of the optimal contract after $E$ led:
\[
w_i^F(E)=w_i^*(E).
\]
There are two subcases: if $F$ can still be induced to choose $P$'s bliss point, impose
\[
y_F^\ast=\theta,
\qquad
y_E^\ast \in [\theta, 1),
\qquad
-m\leq s_F^\ast\leq s_E^\ast <m,
\]
and hence use the above system of equations to solve $s_F^\ast$, $s_E^\ast$, and $y_E^\ast$.
If, instead, $y_F=\theta$ is no longer feasible, impose
\[
y_F^\ast \in (0, \theta),
\qquad
y_E^\ast \in [\theta, 1),
\qquad
-m= s_F^\ast\leq s_E^\ast <m,
\]
and hence use the above system of equations to solve $s_E^\ast$, $y_F^\ast$, and $y_E^\ast$.

In this regime, partial embracing is driven by $P$'s own incentive constraint. $P$ would like to promise more support to $E$, but cannot credibly do so. We can transition to this regime either from Partial equilibrium or from Opportunism, both with $P$'s \eqref{Eq:PrincipalDEC} binding on path. As we said before, we cannot transition directly from Full Embracing to it, since at the boundary of the Full Embracing region $P$'s \eqref{Eq:PrincipalDEC} has slack. The same holds for the next regime, the NES regime.

\paragraph{Nash equilibrium of the stage game (NES).}
The fallback candidate is
\[
s_F=s_E=-m,
\qquad
y_F=0,
\qquad
y_E=1.
\]
It is always feasible and is selected whenever no cooperative candidate survives the
feasibility filters or whenever it gives the highest payoff to $P$ among feasible candidates.

\subsection{Selection}

For each candidate family and punishment branch, our code performs the following steps.
First, it solves the candidate's Bellman equations and binding constraints. Second, it
filters roots by admissibility:
\[
-m\leq s_F,s_E,s_0^F\leq m,
\]
\[
0\leq y_F\leq \theta\leq y_E\leq 1,
\]
\[
0\leq y_E^F\leq 1,
\qquad
0\leq y_F^P\leq 1.
\]
Third, it imposes all regime-specific slack inequalities. Finally, among all feasible roots,
it selects the one maximizing the principal's value, with NES included as the fallback.

The construction does not rely on closed-form cutoffs. Instead, the cutoffs are generated endogenously by solving the relevant systems, discarding inadmissible roots, and selecting the principal-preferred feasible continuation. It provides a full, symbolic characterization of the equilibrium.

\section[Online Appendix: Non-Linear Preferences]{Non-Linear Policy Preferences}
\label{sub:curvature}

In this section, we illustrate why linear policy preferences bias our setting toward \emph{less embracing} by sketching the mechanics in an example with curved utilities. These mechanics also address a claim we make in \cref{sec: Analysis} in relation to previous literature that curved utilities lead to greater moderation. They show that both effects work in the same direction. We do this within the special case of \cref{sec: Example}, because the contrast is clearest there.

Formally, we assume that all players have a policy payoff of
\begin{equation*}
    u_{i,t}(y)= -(y-\theta_i)^2.
\end{equation*}
The key observation is that promising endorsement to the enemy at some future date provides him with larger gains (in payoffs) than the principal loses. Thus, promising embracing allows the principal to extract further rents from the enemy even ignoring the leadership rents. This is in contrast to the baseline case in which those policy trades are (by the linear assumptions) always one-to-one.

To see this contrast, assume an optimal contract in which the enemy \emph{does not} receive full endorsement at some history $h$, $s(h)<m$. Now take an identical contract with the only difference that at $h$ we have $\hat{s}(h)=s(h)+\varepsilon$ for some small $\varepsilon$.

Now, the principal's policy gain from that change is given (in history $h$-present terms):

\[\gamma_E= \varepsilon \left(-(1-y^\ast_E)^2 + (1-y^\ast_F)^2\right)= \varepsilon (2 y^\ast_E-(y^\ast_E)^2)\]
where the second equality follows from $y^\ast_F=0$ because we assume $\theta=0$.

Now consider the respective policy loss by the principal which is simply
\[\gamma_P= -\varepsilon (y^\ast_E)^2 \]
again, because $y^\ast_F=0$. Comparing gains and losses we see that
\[\gamma_P+\gamma_E= \left(2 y^\ast_E - 2 (y^\ast_E)^2\right)\varepsilon\]
which is non-negative if $y^\ast_E\leq 1$.

But then, even ignoring the leadership dimension, the enemy is willing to give up strictly more utility to the principal at any stage prior to history $h$ in which the enemy leads to gain that endorsement (unless in the boundary case of $y_E=1$). Thus, unless the principal already gets her first best, she always has an incentive to promise endorsement in the future even absent any leadership term and even in the special case $\theta=0$ where the cross-subsidization effect is shut down.

\section[Online Appendix: Monetary Transfers]{Monetary Transfers}
\label{sub:transfers}

In this part, we consider a model in which the parties can directly transfer utility to each other. We will first discuss when and how utility transfers improve in the commitment case. Then we will briefly discuss the no-commitment case.

\begin{proposition}\label{prop:transferCommitment}
Suppose the principal has commitment power and utility transfers are allowed. If $\theta \leq \underline{\theta}$, the optimal contract from the no-transfer case remains optimal. If $\theta > \underline{\theta}$, the principal can improve the contract by transferring utility to the enemy when the enemy leads.
\end{proposition}

\begin{proof}
First, consider the case in which $\theta \leq \underline{\theta}$. In this case, if the principal pays a transfer to $F$, she does not benefit from it, since $F$, whenever leading, chooses the principal's bliss point despite not being promised any endorsement.

Second, if the principal pays a transfer to $E$, she does not benefit either. The only way for $E$ to repay is by moderating his policy, which in utility terms is no larger than the utility transferred to $E$ in the first place. Any such change implies an externality on $F$, whose \eqref{Eq Agent DEC} already holds with slack at $y_F^\ast=\theta$. Thus, there are no benefits.

However, if $\theta > \underline{\theta}$, the principal can improve the contract by transferring utility to the enemy when the enemy leads. In this case, the principal can pay a transfer to $E$ when $E$ leads. This transfer will be repaid by $E$ moderating his policy. However, it also relaxes $F$'s \cref{Eq Agent DEC} in periods prior to the transfer. In such periods, if they exist, the principal can then increase $s_F$ until $F$'s \cref{Eq Agent DEC} binds. By \cref{lem:FullEndorsment}, this increases the principal's ex-ante utility. Thus, transfers benefit the principal.
\end{proof}

As we can see from \cref{prop:transferCommitment}, transfers are a sort of money pump for the principal. By exchanging favors with one party (which is zero-sum), the implied externality on the other party allows for a better contract. That contract involves more embracing of the enemy. However, such pumping is only possible if the principal is not yet at a corner solution.

In the no-commitment case, there is a second role for transfers. Here, that effect becomes relevant if the principal desires to punish her friend. The optimal punishment in this case may involve not full polarization, but some moderation by the enemy---enough to ensure the principal indeed wants to endorse the enemy for a bit. In this case, instead of moderating, the enemy could polarize and pay the principal a transfer. Then, the principal would still be willing to endorse the enemy in expectation of that transfer, while the friend would suffer more, as he does not benefit from moderation. Punishment is worse, and more can be implemented.

\section[Online Appendix: A Kingmaker Model]{A Kingmaker Model} \label{sec:kingmaker}

In this section, we consider a model in which the current leader is sometimes chosen exogenously (probability $1-2m$) and sometimes by the principal (probability $2m$), who then becomes the \emph{kingmaker}. We show here (for the commitment case) that the two models are isomorphic.

\paragraph{Model.} The model is the same as in \cref{sec:model}, except that the principal can choose the leader with probability $2m$ and nature chooses the leader with probability $1-2m$. Nature is assumed to use an unbiased coin to select the leader. The principal has access to a uniform public randomization device with realization $x \in [0,1]$ and chooses a cutoff $s$ such that if $x\leq s$, the principal picks the enemy if asked, and the friend otherwise.

Because the principal has commitment power, she can choose any strategy $s$ resulting in an unconditional probability of $p(s)= 1/2(1-2m) + 2 m s = 1/2 + 2m (s-1/2) \in [1/2-m,1/2+m]$. That is enough to show the isomorphism between the two models.

\section[Online Appendix: Roman Republic Example]{Julius Caesar and the late Roman Republic}\label{app:other}

In this section, we discuss an example from the late Roman Republic to illustrate how the mechanics of our model map onto real-world behavior. Since some important details of Roman politics in the late Republic remain obscure, we treat this example primarily as suggestive.

The mechanism we identify for when a cordon sanitaire is breached can also be traced back to the late Roman Republic. In particular, it can be traced to Marcus T. Cicero and the political situation created by Julius Caesar's military victories in Gaul.

The Roman Republic lacked fully coherent political parties, but Cicero and Caesar were often on different sides of the political divide. Cicero was a moderate \textit{optimate}---a supporter of the more aristocratic elements of the constitution, in particular, the authority of the Senate---whereas Caesar was inclined to appeal to the popular assemblies and favor some redistributive legislation like land reforms. The more staunch optimates led by Marcus P. Cato sabotaged Caesar in his early years. They filibustered Caesar's request for a triumph in 60 BC and then blocked his legislation in the Senate. However, Caesar's success in Gaul produced what \citet{morstein2021julius} refers to as the \enquote{Senate's embrace of Caesar's agenda in 56 BC}. The Senate passed a series of decrees in his favor, including an extension of Caesar's governorship of Transalpine Gaul and the authorization for him to run for the consulship immediately after. It is in these debates that Cicero presents the logic of the embracement (cited from \citet{morstein2021julius}):
\begin{quote}
    Never was there anyone who could be a leader here [i.e., in the Senate] who preferred to be a ``Friend of the People.'' But some men [\ldots] have often almost been forced to throw themselves into those heavy seas. If they turn their gaze away from that kind of tossing about on a popular course back to the Senate after serving the Republic well and desire to be considered worthy of the highest prestige, which belongs to this body [i.e., the Senate], they should not merely not be pushed away but even be courted.
\end{quote}
As \citet{morstein2021julius} reasons, \enquote{even if most senators did not really believe in redemption, Caesar's stock was now extremely high at all levels of society, and it must have seemed to those senators [\ldots] to make accommodations to political reality, especially if one could convince oneself, as Cicero claims to have done, that Caesar for his part was also prepared to return to a more harmonious relationship with the Senate}.

The Senate decrees in 56 BC brought seven years of stability for the Republic. It remains unclear (and subject to a heated debate among historians) why the embracing operation broke down in 49 BC, when the civil war started. The year 49 BC was crucial because Caesar's governorship ended, and the second part of the agreement had to be fulfilled: Caesar was supposed to run for the consulship immediately. However, running for the consulship required Caesar to travel to Rome and leave his army behind. The temptation for the Senatorial faction to break their promise was hence much greater than back in 56 BC, when Caesar had his army with him. The situation required greater credibility of promises then than before.

At the same time, several elements exogenous to our model point to a structural break. There are signals of a growing polarization in Rome: popular agitation and violence in the streets, as well as the growing radicalism of the most conservative faction in the Senate, led by Cato the Younger. Also, a particularly obscure event that some historians, like \citet{canfora2007julius}, give nonetheless significant centrality is the murder of Publius Clodius, who was a popular agitator, but also may have been an agent of Caesar. At the same time, two important political figures had died during this period: Marcus Licinius Crassus, a colleague of Caesar and Pompey in the first triumvirate who could have balanced between them; and Caesar's daughter, who was Pompey's wife, and whose death destroyed an important linkage between the two leaders.
\pagestyle{empty}
\section[Online Appendix: Supplementary Tables for Section \ref{sec:implications}]{Supplementary Tables for Section \ref{sec:implications}} \label{app:coalitions}
Here we provide timeline tables showing the feasibility of coalitions in our two historical examples.
\begin{table}[htbp]
\centering

\footnotesize
\renewcommand{\arraystretch}{1.16}
\setlength{\tabcolsep}{4pt}

\caption{Coalition feasibility in the Weimar Republic, 1920--1930}
\label{tab:weimar_coalitions}

\begin{tabularx}{1.1\textwidth}{@{}p{3cm}Yp{1.8cm}p{1.8cm}p{4.2cm}@{}}
\toprule
\textbf{Period} & \textbf{Government} &
\makecell[l]{\textbf{Left option}\\[-2pt]\scriptsize \textcolor{cred}{\textbf{SPD}} + pivot} &
\makecell[l]{\textbf{Right option}\\[-2pt]\scriptsize \textcolor{cblue}{\textbf{DNVP}} + pivot} &
\\
\midrule

\elecrow{Election --- Jun 1920 (459 seats, maj.\ = 230). Weimar Coalition loses majority; SPD collapses to 103.}

Jun 1920--May 1921
& \textbf{Fehrenbach}\newline{\scriptsize Z + DVP + DDP\newline(minority; \textcolor{cred}{\textbf{SPD}} toleration)}
& \yes{292}
& \yes{260}
& \sideneither\newline{\scriptsize DNVP excluded}
\\

May--Nov 1921
& \textbf{Wirth\,I}\newline{\scriptsize \textcolor{cred}{\textbf{SPD}} + Z + DDP\newline(Weimar Coal.\ revived; min.)}
& \chosenL{292}
& \yes{260}
& \sidelean\newline{\scriptsize Post-reparations ultimatum}
\\

Nov 1921--Nov 1922
& \textbf{Wirth\,II}\newline{\scriptsize \textcolor{cred}{\textbf{SPD}} + Z + DDP\newline(Weimar Coal.; minority)}
& \chosenL{292}
& \yes{260}
& \sidelean\newline{\scriptsize Rathenau assassination}
\\

Nov 1922--Aug 1923
& \textbf{Cuno}\newline{\scriptsize Z + DVP + DDP + BVP\newline(``business cabinet''; minority)}
& \yes{292}
& \yes{260}
& \sideneither\newline{\scriptsize Ruhr occupation}
\\

Aug--Nov 1923
& \textbf{Stresemann\,I--II}\newline{\scriptsize \textcolor{cred}{\textbf{SPD}} + Z + DVP + DDP\newline(grand coalition)}
& \chosenL{292}
& \yes{260}
& \sideleft\newline{\scriptsize Hyperinflation crisis}
\\

Nov 1923--May 1924
& \textbf{Marx\,I}\newline{\scriptsize Z + DVP + DDP + BVP\newline(minority)}
& \yes{292}
& \yes{260}
& \sideneither\newline{\scriptsize \textcolor{cred}{SPD} left after Stresemann ouster}
\\

\elecrow{Election --- May 1924 (472 seats, maj.\ = 237). Extremes surge; \textcolor{cblue}{\textbf{DNVP}} = 95 seats.}

Jun--Dec 1924
& \textbf{Marx\,II}\newline{\scriptsize Z + DVP + DDP\newline(minority; SPD toleration)}
& \yes{254}
& \yes{249}
& \sideneither\newline{\scriptsize Dawes Plan vote (Aug):\newline \textcolor{cblue}{\textbf{DNVP}} first gets agency}
\\

\elecrow{Election --- Dec 1924 (493 seats, maj.\ = 247). Moderates recover; \textcolor{cblue}{\textbf{DNVP}} = 103.}

Jan--Dec 1925
& \textbf{Luther\,I}\newline{\scriptsize Z + \textcolor{cblue}{\textbf{DNVP}} + DVP + DDP + BVP}\newline{\scriptsize\textcolor{cblue}{\textsc{Cordon broken}}}
& \yes{302}
& \chosenR{274}
& \sideright\newline{\scriptsize \textcolor{cblue}{\textbf{DNVP}} enters gov't for 1st time}
\\

Jan--May 1926
& \textbf{Luther\,II}\newline{\scriptsize Z + DVP + DDP + BVP\newline(minority; \textcolor{cblue}{\textbf{DNVP}} left, Locarno)}
& \yes{302}
& \yes{274}
& \sideneither\newline{\scriptsize \textcolor{cblue}{\textbf{DNVP}} withdrew}
\\

May 1926--Jan 1927
& \textbf{Marx\,III}\newline{\scriptsize Z + DVP + DDP + BVP\newline(minority; \textcolor{cred}{\textbf{SPD}} ext.\ support)}
& \yes{302}
& \yes{274}
& \sidelean\newline{\scriptsize \textcolor{cred}{\textbf{SPD}} toleration}
\\

Jan 1927--Jun 1928
& \textbf{Marx\,IV}\newline{\scriptsize Z + \textcolor{cblue}{\textbf{DNVP}} + DVP + BVP\newline(near-majority; DDP out)}
& \yes{302}
& \chosenR{274}
& \sideright\newline{\scriptsize \textcolor{cblue}{\textbf{DNVP}} returns; moderation}
\\

\elecrow{Election --- May 1928 (491 seats, maj.\ = 246). \textcolor{cblue}{\textbf{DNVP}} collapses to 73; Hugenberg takes over Oct.}

Jun 1928--Mar 1930
& \textbf{M\"uller\,II}\newline{\scriptsize \textcolor{cred}{\textbf{SPD}} + Z + DVP + DDP + BVP\newline(grand coalition)}
& \chosenL{302}
& \no{222}
& \sideleft\newline{\scriptsize Right option no longer feasible}
\\

\elecrow{Election --- Sep 1930 (577 seats, maj.\ = 289). NSDAP surges to 107; \textcolor{cblue}{\textbf{DNVP}} = 41.}

Mar 1930$\to$
& \textbf{Br\"uning\,I} \newline{\scriptsize Z + DVP + DDP + BVP\newline(presidential cab.; \textcolor{cred}{\textbf{SPD}} toler.)}\newline{\scriptsize\textcolor{camber}{\textsc{End of parliamentary gov't}}}
& \no{280}
& \no{178}
& \sideend\newline{\scriptsize Article 48 decree rule}
\\

\bottomrule
\end{tabularx}

\medskip
\parbox{\textwidth}{\scriptsize\textbf{Notes:} ``Pivot bloc'' = Z + DVP + DDP + BVP.  \colorbox{cgreenlight}{\textcolor{cgreen}{$\checkmark$}} = arithmetically feasible majority. \colorbox{credlight}{$\times$} = falls short. \colorbox{credlight}{\textcolor{cred}{$\leftarrow$}} / \colorbox{cbluelight}{\textcolor{cblue}{$\leftarrow$}} = option chosen. Br\"uning I took office under the 1928 Reichstag. His austerity budget failed to be approved despite using an emergency decree, rejected both by parties to his left (\textcolor{cred}{\textbf{SPD}} + KPD) and right (\textcolor{cblue}{\textbf{DNVP}} + NSDAP). This led to the 1930 election. }

\end{table}

\begin{table}[htbp]
\centering
\caption{Coalition feasibility in the Italian Republic (Part I: Centrist Era, 1947--1958)}
\label{tab:italy_coalitions_1}
\footnotesize
\renewcommand{\arraystretch}{1.16}
\setlength{\tabcolsep}{4pt}

\begin{tabularx}{\textwidth}{@{}p{3cm}Yp{1.8cm}p{1.8cm}p{4.2cm}@{}}
\toprule
\textbf{Period} & \textbf{Government} &
\makecell[l]{\textbf{Left option}\\[-2pt]\scriptsize \textcolor{cred}{\textbf{PSI}} + pivot} &
\makecell[l]{\textbf{Right option}\\[-2pt]\scriptsize \textcolor{cblue}{\textbf{PLI}} + pivot} &
 \\
\midrule

\elecrow{May 1947: De Gasperi expels PCI and \textcolor{cred}{\textbf{PSI}} from cabinet. Centrist formula established.}

Jun 1947--Apr 1948
& \textbf{De Gasperi\,IV}\newline{\scriptsize DC + PSLI + \textcolor{cblue}{\textbf{PLI}} + PRI\newline(\textcolor{cred}{\textbf{PSI}} in pact with PCI)}
& \yes{378}
& \chosenR{323}
& \sideright\newline{\scriptsize Cordon established}
\\

\elecrow{Election --- Apr 1948 (574, maj.\ = 288). DC wins absolute majority. \textcolor{cred}{\textbf{PSI}} ran in joint list with PCI.}

May 1948--Jan 1950
& \textbf{De Gasperi\,V}\newline{\scriptsize DC + PSDI + \textcolor{cblue}{\textbf{PLI}} + PRI}
& \yes{399}
& \chosenR{366}
& \sideright\newline{\scriptsize DC has own majority;\newline allies for breadth}
\\

Jan 1950--Jul 1951
& \textbf{De Gasperi\,VI}\newline{\scriptsize DC + PSDI + PRI\newline(\textcolor{cblue}{\textbf{PLI}} leaves)}
& \yes{399}
& \yes{366}
& \sideneither\newline{\scriptsize Centrist core only}
\\

Jul 1951--Jun 1953
& \textbf{De Gasperi\,VII}\newline{\scriptsize DC + PRI}
& \yes{399}
& \yes{366}
& \sideneither\newline{\scriptsize DC majority suffices}
\\

\elecrow{Election --- Jun 1953 (590, maj.\ = 296). DC loses absolute majority.}

Jul--Aug 1953
& \textbf{De Gasperi\,VIII}\newline{\scriptsize DC monocolore}
& \yes{362}
& \yes{300}
& \sideneither\newline{\scriptsize Failed confidence vote}
\\

Aug 1953--Feb 1954
& \textbf{Pella / Fanfani\,I}\newline{\scriptsize DC monocolore (both fail)}
& \yes{362}
& \yes{300}
& \sideneither\newline{\scriptsize Failed confidence vote}
\\

Feb 1954--Jul 1955
& \textbf{Scelba}\newline{\scriptsize DC + PSDI + \textcolor{cblue}{\textbf{PLI}} \newline(PRI external support)}
& \yes{362}
& \chosenR{300}
& \sideright\newline{\scriptsize Majority by 4 seats}
\\

Jul 1955--May 1957
& \textbf{Segni\,I}\newline{\scriptsize DC + PSDI + \textcolor{cblue}{\textbf{PLI}} \newline(PRI external support)}
& \yes{362}
& \chosenR{300}
& \sideright\newline{\scriptsize Same razor-thin formula}
\\

May 1957--Jun 1958
& \textbf{Zoli}\newline{\scriptsize DC monocolore\newline(minority; MSI + monarchist votes)}
& \yes{362}
& \yes{300}
& \sidefarright\newline{\scriptsize Survives on neo-fascist votes}
\\

\bottomrule
\end{tabularx}

\medskip
\parbox{\textwidth}{\scriptsize\textbf{Notes:} ``Pivot'' = DC + PSDI + PRI. ``Left option'' adds \textcolor{cred}{\textbf{PSI}}; ``Right option'' adds \textcolor{cblue}{\textbf{PLI}}. \colorbox{cgreenlight}{\textcolor{cgreen}{$\checkmark$}} = feasible majority. \colorbox{credlight}{\textcolor{cred}{$\leftarrow$}} / \colorbox{cbluelight}{\textcolor{cblue}{$\leftarrow$}} = option chosen. From 1947 to 1963, \textcolor{cred}{\textbf{PSI}} was excluded due to its alliance with PCI. The right option was always arithmetically tighter (margins of 4--19 seats vs.\ 50--86 for the left).}
\end{table}

\begin{table}[htbp]
\centering
\caption{Coalition feasibility in the Italian Republic (Part II: Opening to the Left, 1958--1968)}
\label{tab:italy_coalitions_2}
\footnotesize
\renewcommand{\arraystretch}{1.16}
\setlength{\tabcolsep}{4pt}

\begin{tabularx}{\textwidth}{@{}p{3cm}Yp{1.75cm}p{1.75cm}p{4.2cm}@{}}
\toprule
\textbf{Period} & \textbf{Government} &
\makecell[l]{\textbf{Left option}\\[-2pt]\scriptsize \textcolor{cred}{\textbf{PSI}} + pivot} &
\makecell[l]{\textbf{Right option}\\[-2pt]\scriptsize \textcolor{cblue}{\textbf{PLI}} + pivot} &
 \\
\midrule

\elecrow{Election --- May 1958 (596, maj.\ = 299). }

Jul 1958--Feb 1959
& \textbf{Fanfani\,II}\newline{\scriptsize DC + PSDI\newline(near-minority)}
& \yes{385}
& \yes{318}
& \sideneither\newline{\scriptsize Neither ally in cabinet}
\\

Feb 1959--Mar 1960
& \textbf{Segni\,II}\newline{\scriptsize DC monocolore\newline(minority; MSI abstention)}
& \yes{385}
& \yes{318}
& \sideneither\newline{\scriptsize Rightward drift}
\\

Mar--Jul 1960
& \textbf{Tambroni}\newline{\scriptsize DC monocolore\newline(MSI votes decisive)}\newline{\scriptsize\textcolor{camber}{\textsc{Tambroni crisis}}}
& \yes{385}
& \yes{318}
& \sidefarright\newline{\scriptsize Anti-fascist protests;\newline 5 killed. Gov't falls.}
\\[4pt]

Jul 1960--Feb 1962
& \textbf{Fanfani\,III}\newline{\scriptsize DC monocolore\newline(minority; \textcolor{cred}{\textbf{PSI}} \emph{abstains})}
& \yes{385}
& \yes{318}
& \sidelean\newline{\scriptsize First \textcolor{cred}{\textbf{PSI}} opening}
\\

Feb 1962--Jun 1963
& \textbf{Fanfani\,IV}\newline{\scriptsize DC + PSDI + PRI\newline(\textcolor{cred}{\textbf{PSI}} external support)}
& \yes{385}
& \yes{318}
& \sidelean\newline{\scriptsize \textcolor{cred}{\textbf{PSI}} supports from outside;\newline ENEL nationalized}
\\

\elecrow{Election --- Apr 1963 (630, maj.\ = 316). DC drops to 260. \textcolor{cblue}{\textbf{PLI}} surges to 39. }

Jun--Dec 1963
& \textbf{Leone\,I}\newline{\scriptsize DC monocolore (caretaker)}
& \yes{386}
& \yes{338}
& \sideneither\newline{\scriptsize Transition}
\\

Dec 1963--Jul 1964
& \textbf{Moro\,I}\newline{\scriptsize DC + \textcolor{cred}{\textbf{PSI}} + PSDI + PRI}\newline{\scriptsize\textcolor{cred}{\textsc{Cordon broken}}}
& \chosenL{386}
& \yes{338}
& \sideleft\newline{\scriptsize \textcolor{cred}{\textbf{PSI}} enters gov't;\newline Nenni Deputy PM}
\\

Jul 1964--Feb 1966
& \textbf{Moro\,II}\newline{\scriptsize DC + \textcolor{cred}{\textbf{PSI}} + PSDI + PRI}
& \chosenL{${\sim}$361}
& \yes{338}
& \sideleft\newline{\scriptsize PSIUP split costs \textcolor{cred}{\textbf{PSI}} seats}
\\

Feb 1966--Jun 1968
& \textbf{Moro\,III}\newline{\scriptsize DC + PSU + PRI\newline(\textcolor{cred}{\textbf{PSI}}--PSDI merge into \textcolor{cred}{\textbf{PSU}})}
& \chosenL{${\sim}$361}
& \yes{338}
& \sideleft\newline{\scriptsize Centre-left consolidated}
\\

\bottomrule
\end{tabularx}

\medskip
\parbox{\textwidth}{\scriptsize\textbf{Key transition:} After the Tambroni crisis (1960), DC pivoted left in three steps: \textcolor{cred}{\textbf{PSI}} abstention (1960), external support (1962), and full cabinet participation (1963). Both options remained arithmetically feasible. Note the \textcolor{cblue}{\textbf{PLI}} surge to 39 seats in 1963: voters punished DC's leftward turn by shifting right, yet the center-left formula still commanded a larger majority.}
\end{table}

\begin{table}[htbp]
\centering
\caption{Coalition feasibility in the Italian Republic (Part III: Centre-Left as Norm, 1968--1979)}
\label{tab:italy_coalitions_3}
\footnotesize
\renewcommand{\arraystretch}{1.16}
\setlength{\tabcolsep}{4pt}

\begin{tabularx}{1.1\textwidth}{@{}p{3cm}Yp{1.75cm}p{1.75cm}p{4.2cm}@{}}
\toprule
\textbf{Period} & \textbf{Government} &
\makecell[l]{\textbf{Left option}\\[-2pt]\scriptsize \textcolor{cred}{\textbf{PSI}} + pivot} &
\makecell[l]{\textbf{Right option}\\[-2pt]\scriptsize \textcolor{cblue}{\textbf{PLI}} + pivot} &
 \\
\midrule

\elecrow{Election --- May 1968 (630, maj.\ = 316). }

Jun--Dec 1968
& \textbf{Leone\,II}\newline{\scriptsize DC monocolore (caretaker)}
& \yes{366}
& \yes{335}
& \sideneither\newline{\scriptsize Transition}
\\

Dec 1968--Aug 1969
& \textbf{Rumor\,I}\newline{\scriptsize DC + \textcolor{cred}{\textbf{PSU}} + PRI}
& \chosenL{366}
& \yes{335}
& \sideleft\newline{\scriptsize Centre-left continues}
\\

Aug 1969--Mar 1970
& \textbf{Rumor\,II}\newline{\scriptsize DC monocolore\newline(minority; \textcolor{cred}{\textbf{PSI}}/PSDI ext.)}
& \yes{366}
& \yes{335}
& \sidelean\newline{\scriptsize Hot Autumn; PSU splits}
\\

Mar--Aug 1970
& \textbf{Rumor\,III}\newline{\scriptsize DC + \textcolor{cred}{\textbf{PSI}} + PSDI + PRI}
& \chosenL{366}
& \yes{335}
& \sideleft\newline{\scriptsize Quadripartite restored}
\\

Aug 1970--Feb 1972
& \textbf{Colombo}\newline{\scriptsize DC + \textcolor{cred}{\textbf{PSI}} + PSDI + PRI}
& \chosenL{366}
& \yes{335}
& \sideleft\newline{\scriptsize Workers' Statute enacted}
\\

Feb--Jun 1972
& \textbf{Andreotti\,I}\newline{\scriptsize DC monocolore}
& \yes{366}
& \yes{335}
& \sideneither\newline{\scriptsize \textbf{Failed confidence vote}}
\\

\elecrow{Election --- May 1972 (630, maj.\ = 316). \textcolor{cblue}{\textbf{PLI}} falls to 20. MSI-DN surges to 56. }

Jun 1972--Jul 1973
& \textbf{Andreotti\,II}\newline{\scriptsize DC + PSDI + \textcolor{cblue}{\textbf{PLI}} + PRI\textsuperscript{ext}}
& \yes{371}
& \chosenR{330}
& \sideright\newline{\scriptsize \emph{Only} post-1963 return\newline to right; lasts 13 months}
\\

Jul 1973--Mar 1974
& \textbf{Rumor\,IV}\newline{\scriptsize DC + \textcolor{cred}{\textbf{PSI}} + PSDI + PRI}
& \chosenL{371}
& \yes{330}
& \sideleft\newline{\scriptsize Centre-left restored}
\\

Mar--Nov 1974
& \textbf{Rumor\,V}\newline{\scriptsize DC + \textcolor{cred}{\textbf{PSI}} + PSDI\newline(PRI external)}
& \chosenL{371}
& \yes{330}
& \sideleft
\\

Nov 1974--Jul 1976
& \textbf{Moro\,IV--V}\newline{\scriptsize DC + PRI\newline(\textcolor{cred}{\textbf{PSI}} external support)}
& \yes{371}
& \yes{330}
& \sidelean\newline{\scriptsize Minority; \textcolor{cred}{\textbf{PSI}} outside}
\\

\elecrow{Election --- Jun 1976 (630, maj.\ = 316). \textcolor{cblue}{\textbf{PLI}} collapses to 5. Right option no longer feasible.}

Jul 1976--Mar 1978
& \textbf{Andreotti\,III}\newline{\scriptsize DC monocolore\newline(``national solidarity''; all\newline parties abstain incl.\ PCI)}
& \yes{348}
& \no{296}
& \sideneither\newline{\scriptsize PCI in \emph{non-sfiducia};\newline right option gone}
\\

Mar 1978--Jan 1979
& \textbf{Andreotti\,IV}\newline{\scriptsize DC + PSDI + PRI\newline(PCI + \textcolor{cred}{\textbf{PSI}} in parl.\ majority)}
& \yes{348}
& \no{296}
& \sidelean\newline{\scriptsize \textcolor{cred}{\textbf{PSI}} supports;\newline Moro kidnapped \& killed}
\\

Jan--Aug 1979
& \textbf{Andreotti\,V}\newline{\scriptsize DC + PSDI + PRI\newline(PCI withdraws $\to$ falls)}
& \yes{348}
& \no{296}
& \sideend\newline{\scriptsize PCI exits majority;\newline legislature ends}
\\

\bottomrule
\end{tabularx}

\medskip
\parbox{\textwidth}{\scriptsize\textbf{Notes:} Andreotti\,II (1972--73) was the \emph{sole} post-1963 government to include \textcolor{cblue}{\textbf{PLI}} instead of \textcolor{cred}{\textbf{PSI}}: it collapsed within 13 months. After the 1976 election, \textcolor{cblue}{\textbf{PLI}}'s collapse to 5 seats eliminated the right option entirely. Both options were feasible throughout 1953--1976; only the left option survived after 1976.}
\end{table}

\clearpage\pagestyle{plain}
\setlength\bibitemsep{0.5\itemsep}
\printbibliography
\end{refsection}
\end{document}